\documentclass[aps,twocolumn,amsmath,amssymb,prd]{revtex4}

\def\al{\alpha}
\def\be{\beta}
\def\ga{\gamma}
\def\de{\delta}
\def\ep{\epsilon}
\def\ve{\varepsilon}
\def\ze{\zeta}
\def\et{\eta}
\def\th{\theta}

\def\la{\lambda}

\def\si{\sigma}

\def\ph{\phi}

\def\ch{\chi}
\def\ps{\psi}
\def\om{\omega}

\def\De{\Delta}

\def\La{\Lambda}

\def\Ps{\Psi}
\def\Om{\Omega}

\def\cl{{\cal L}}

\def\cO{{\cal O}}

\def\fr#1#2{{{#1}\over{#2}}}
\def\frac#1#2{{\textstyle{{#1}\over{#2}}}}
\def\half{{\textstyle{1\over 2}}}
\def\ol{\overline}
\def\prt{\partial}

\def\Re{\hbox{Re}\,}
\def\Im{\hbox{Im}\,}

\def\lsim{\mathrel{\rlap{\lower4pt\hbox{\hskip1pt$\sim$}}
    \raise1pt\hbox{$<$}}}
\def\gsim{\mathrel{\rlap{\lower4pt\hbox{\hskip1pt$\sim$}}
    \raise1pt\hbox{$>$}}}

\def\etal{{\it et al.}}

\def\vev#1{\langle {#1}\rangle}

\def\ket#1{|{#1}\rangle}

\def\eff{{\rm eff}}

\def\sqr#1#2{{\vcenter{\vbox{\hrule height.#2pt
         \hbox{\vrule width.#2pt height#1pt \kern#1pt
         \vrule width.#2pt}
         \hrule height.#2pt}}}}

\newcommand{\beq}{\begin{equation}}
\newcommand{\eeq}{\end{equation}}
\newcommand{\bea}{\begin{eqnarray}}
\newcommand{\eea}{\end{eqnarray}}
\newcommand{\rf}[1]{(\ref{#1})}

\newcommand{\bM}{\begin{pmatrix}}
\newcommand{\eM}{\end{pmatrix}}

\def\nn{\nonumber}

\def\w{w}
\def\f{w}

\def\psfb{\ol{\ps_\f}{}}

\def\mbf#1{\boldsymbol #1}

\def\syjm#1#2{{}_{#1}Y_{#2}}

\def\O{\mathcal O}
\def\Q{\mathcal Q}

\def\V{\mathcal V}

\def\T{\mathcal T}

\def\K{\mathcal K}

\def\pvec{\mbf p}

\def\sivec{\mbf\si}

\def\bevec{\mbf\be}

\def\pmag{|\pvec|}

\def\punit{\hat p}

\def\xunit{\hat x}
\def\yunit{\hat y}
\def\zunit{\hat z}
\def\epunit{\hat\ep}
\def\thunit{\hat\th}
\def\phunit{\hat\ph}

\def\phat{\mbf\punit}

\def\xhat{\mbf\xunit}
\def\yhat{\mbf\yunit}
\def\zhat{\mbf\zunit}
\def\ephat{\mbf\epunit}
\def\thhat{\mbf\thunit}
\def\phhat{\mbf\phunit}

\def\widebar{\overline}

\def\Qhat{\widehat\Q}

\def\Tdual{\widetilde{\widehat\T}\phantom{}}

\def\gt{\widetilde g}
\def\Ht{\widetilde H}

\def\Htf#1#2#3{{{\tilde{H}}_{#1}}\,\hspace{-1 pt}^{(#2)#3}_\eff}
\def\gtf#1#2#3{{{\tilde{g}}_{#1}}\,\hspace{-1 pt}^{(#2)#3}_\eff}

\def\X{X}
\def\Y{Y}
\def\Z{Z}
\def\Xhat{\widehat\X}
\def\Yhat{\widehat\Y}
\def\Zhat{\widehat\Z}

\def\nr{{\rm NR}}

\def\nrtemplate#1#2#3{#1^{\nr#3}_{#2}}

\def\cs133{\rm Cs}

\def\Vnr#1{\nrtemplate{\V}{#1}{}}
\def\TzBnr#1{\nrtemplate{\T}{#1}{(0B)}}
\def\ToBnr#1{\nrtemplate{\T}{#1}{(1B)}}

\def\Vnrf#1#2{\nrtemplate{{\V_{#1}}}{#2}{}}
\def\TzBnrf#1#2{\nrtemplate{{\T_{#1}}}{#2}{(0B)}}
\def\ToEnrf#1#2{\nrtemplate{{\T_{#1}}}{#2}{(1E)}}
\def\ToBnrf#1#2{\nrtemplate{{\T_{#1}}}{#2}{(1B)}}

\def\anrf#1#2{\nrtemplate{{a_{#1}}}{#2}{}}
\def\cnrf#1#2{\nrtemplate{{c_{#1}}}{#2}{}}
\def\gzBnrf#1#2{\nrtemplate{{g_{#1}}}{#2}{(0B)}}
\def\goBnrf#1#2{\nrtemplate{{g_{#1}}}{#2}{(1B)}}

\def\HzBnrf#1#2{\nrtemplate{{H_{#1}}}{#2}{(0B)}}
\def\HoBnrf#1#2{\nrtemplate{{H_{#1}}}{#2}{(1B)}}

\def\sVnrf#1#2{\nrtemplate{{\V_{#1}}}{#2}{,{\rm Sun}}}

\def\sgzBnrf#1#2{\nrtemplate{{g_{#1}}}{#2}{(0B), {\rm Sun}}}
\def\sgoBnrf#1#2{\nrtemplate{{g_{#1}}}{#2}{(1B),{\rm Sun}}}
\def\sHzBnrf#1#2{\nrtemplate{{H_{#1}}}{#2}{(0B),{\rm Sun}}}
\def\sHoBnrf#1#2{\nrtemplate{{H_{#1}}}{#2}{(1B),\rm Sun}}
\def\sTzBnrf#1#2{\nrtemplate{{\T_{#1}}}{#2}{(0B),{\rm Sun}}}
\def\sToBnrf#1#2{\nrtemplate{{\T_{#1}}}{#2}{(1B),\rm Sun}}

\def\m{m_\ps}

\def\widecheck#1{\hskip#1pt\huge$\check{}$}
\def\bighacek#1#2{\vbox{\ialign{##\crcr\widecheck#2\crcr
  \noalign{\kern-9.5pt\nointerlineskip}
   $\hfil\displaystyle{#1}\hfil$\crcr}}}

\def\mr{\ol{m}_{\rm r}}

\def\k{k}

\def\atm{{\rm atom}}

\def\AzB#1#2{{\La_{#1}}^{(0B)}_{#2}}
\def\AoB#1#2{{\La_{#1}}^{(1B)}_{#2}}
\def\AzE#1#2{{\La_{#1}}^{(0E)}_{#2}}
\def\Agen#1{{\La_{#1}}^{(qP)}_{kj}}

\def\chM{\vartheta}
\def\phM{\varphi}
\def\AM{A}

\def\nub{{\overline{\nu}}}

\def\ring#1{{\mathaccent'27 #1}}

\def\nrfctemplate#1#2{\nrtemplate{\ring{#1}}{#2}{}}
\def\anrfc#1{\nrfctemplate{a}{#1}}

\def\Vnrfc#1{\nrfctemplate{\mathcal{V}}{#1}}

\def\pd{\ol{p}\ol{d}}
\def \j{j}
\def\ii{a}
\def\AF{A}
\def\Ce#1#2{C^{#1}_{#2}}
\def\Ae#1#2{W^{#1}_{#2}}

\def\eiso{\de\ring{\ep}}
\def\LaF#1#2{\La_{#1}^{#2}}
\def\Mco#1#2{M_{#1}^{#2}}
\def\alg{\al'}
\def\Ht{\widetilde{H}}
\def\gt{\widetilde{g}}
\def\Va#1#2#3{V_{#1}^{(#2)#3}}
\def\Tg#1#2#3{T_{#1}^{(#2)#3}}
\def\dC{\la}
\def\caz{\rm Ca}
\def\Jj#1{{J_#1}}

\def\Bunit{\hat B}
\def\Bhat{\mbf\Bunit}

\def\TL{T_L}

\def\TqBnr#1{\nrtemplate{\T}{#1}{(qB)}}

\def\bVnrf#1#2{\nrtemplate{{\widebar{\V}_{#1}}}{#2}{}}

\begin{document}

\title{Lorentz and CPT Tests with Clock-Comparison Experiments}

\author{V.\ Alan Kosteleck\'y$^1$ and Arnaldo J.\ Vargas$^2$}

\affiliation{$^1$Physics Department, Indiana University,
Bloomington, Indiana 47405, USA\\
$^2$Physics Department, Loyola University, 
New Orleans, Louisiana 70118, USA}

\date{IUHET 628, May 2018}

\begin{abstract}
Clock-comparison experiments are among the sharpest existing tests
of Lorentz symmetry in matter.
We characterize signals in these experiments 
arising from modifications to electron or nucleon propagators
and involving Lorentz- and CPT-violating operators of arbitrary mass dimension.
The spectral frequencies of the atoms or ions used as clocks 
exhibit perturbative shifts 
that can depend on the constituent-particle properties 
and can display sidereal and annual variations in time.
Adopting an independent-particle model for the electronic structure
and the Schmidt model for the nucleus,
we determine observables for a variety of clock-comparison experiments
involving fountain clocks, comagnetometers, ion traps, lattice clocks,
entangled states, and antimatter.
The treatment demonstrates 
the complementarity of sensitivities to Lorentz and CPT violation
among these different experimental techniques.
It also permits the interpretation of some prior results 
in terms of bounds on nonminimal coefficients for Lorentz violation,
including first constraints on nonminimal coefficients in the neutron sector.
Estimates of attainable sensitivities in future analyses are provided.
Two technical appendices collect relationships 
between spherical and cartesian coefficients for Lorentz violation 
and provide explicit transformations converting cartesian coefficients 
in a laboratory frame to the canonical Sun-centered frame. 
\end{abstract}

\maketitle

\section{Introduction}
\label{Introduction}

Among the best laboratory tests of rotation invariance
are experiments measuring the ticking rate
of a clock as its orientation changes,
often as it rotates with the Earth.
A spatial anisotropy in the laws of nature would be revealed 
if the clock frequency varies in time
at harmonics of the rotation frequency.
Detecting any time variation requires a reference clock
that either is insensitive to the anisotropy
or responds differently to it.
Typically,
the two clock frequencies in these experiments 
are transition frequencies in atoms or ions,
and the spatial orientation of a clock
is the quantization axis established by an applied magnetic field.
These clock-comparison experiments
can attain impressive sensitivites to rotation violations,
as originally shown by Hughes \etal\ and Drever
\cite{hd}.
 
Rotation invariance is a key component of Lorentz symmetry,
the foundation of relativity.
Tests of this symmetry have experienced a revival in recent years,
stimulated by the possibility that minuscule violations 
could arise from a unification of quantum physics with gravity 
such as string theory
\cite{ksp}.
Using techniques from different subfields of physics,
numerous searches for Lorentz violation have now reached 
sensitivities to physical effects originating at the Planck scale
$M_P \simeq 10^{19}$~GeV
\cite{tables}.
Since the three boost generators of the Lorentz group
close under commutation into the three rotation generators,
any deviations from Lorentz symmetry in nature 
must necessarily come with violations of rotation invariance.
Searches for rotation violations 
therefore offer crucial tests of Lorentz symmetry.
In the present work,
we pursue this line of reasoning
by developing and applying a theoretical treatment
for the analysis of clock-comparison experiments
searching for Lorentz violation.

To date,
no compelling experimental evidence 
for Lorentz violation has been adduced.
Even if Lorentz violation does occur in nature,
identifying the correct realistic model among a plethora of options
in the absence of positive experimental guidance 
seems a daunting and improbable prospect.
An alternative is instead to adopt 
a general theoretical framework for Lorentz violation
that encompasses specific models
and permits a comprehensive study of possible effects.
Since any Lorentz violation is expected to be small,
it is reasonable to use effective field theory
\cite{sw}
for this purpose.
A realistic treatment then starts from well-established physics,
which can be taken as the action formed by coupling General Relativity 
to the Standard Model of particle physics,
and adds all possible Lorentz-violating operators
to yield the framework known as the Standard-Model Extension (SME)
\cite{ck,akgrav}.
Each Lorentz-violating operator in the SME 
is contracted with a coefficient
that determines the magnitude of its physical effects
while preserving coordinate independence of the theory.
The operators can be classified according to their mass dimension $d$,
with larger values of $d$ associated with greater suppression 
at low energies.
The limiting case with $d\leq 4$ is power-counting renormalizable
in Minkowski spacetime and is called the minimal SME.
Since CPT violation in effective field theory
is concurrent with Lorentz violation
\cite{ck,owg},
the SME also characterizes general effects from CPT violation.
Experimental constraints on the parameters 
of any Lorentz-violating model 
that is consistent with realistic effective field theory
can be found by identifying the model parameters 
with specific SME coefficients and their known constraints 
\cite{tables,reviews}.

Signals arising from Lorentz and CPT violation 
are predicted by the minimal SME
to appear in clock-comparison experiments
with atoms or ions
\cite{kl99}.
The signals include observable modifications of the spectra
that can exhibit time variations
and that depend on the electron and nucleon composition 
of the species used as clocks.
Null results from early clock-comparison experiments 
\cite{pr85,la89,ch89,be95}
can be reinterpreted as bounds 
on coefficients for Lorentz violation in the minimal SME 
\cite{kl99}.
Many minimal-SME coefficients have been directly constrained
in recent experiments,
including clock comparisons performed using 
a hydrogen maser
\cite{ph01,hu03},
$^{133}$Cs and $^{87}$Rb fountain clocks
\cite{cs06},
trapped ultracold neutrons and $^{199}$Hg atoms
\cite{al09},
$^3$He-K and $^{21}$Ne-Rb-K comagnetometers 
\cite{HeK,NeRb},
$^{133}$Cs and $^{199}$Hg magnetometers 
\cite{pe12},
transitions in $^{162}$Dy and $^{164}$Dy atoms 
\cite{ho13},
$^{129}$Xe and $^3$He atoms 
\cite{xema,xema2,Xe09,Xe14},
and entangled states of $^{40}$Ca$^{+}$ ions
\cite{Pr15}.
The results represent competitive tests of Lorentz and CPT symmetry
\cite{tables,ccreviews},
and additional constraints on minimal-SME coefficients
have been extracted by detailed theoretical analyses 
\cite{ba09,ro14,st15,cdl16,fl16,fl17,dz17}.

In this work,
we extend the existing theoretical treatment
of Lorentz and CPT violation in clock-comparison experiments
to include SME operators of nonminimal mass dimension $d>4$
that modify the Dirac propagators 
of the constituent electrons, protons, and neutrons in atoms and ions.
At an arbitrary given value of $d$,
all Lorentz- and CPT-violating operators affecting the propagation
have been identified and classified
\cite{km13},
which in the present context permits a perturbative analysis
of the effects of general Lorentz and CPT violation
on the spectra of the atoms or ions
used in clock-comparison experiments. 
Nonminimal SME operators are of direct interest
in various theoretical contexts 
associated with Lorentz-violating quantum field theories
including,
for instance,
formal studies of the underlying Riemann-Finsler geometry
\cite{finsler}
or of causality and stability 
\cite{causality}
and phenomenological investigations 
of supersymmetric models 
\cite{susy}
or noncommutative quantum field theories
\cite{ha00,chklo01}.
They are also of interest in experimental searches for geometric forces, 
such as torsion
\cite{torsion}
and nonmetricity
\cite{nonmetricity}.
Only a comparatively few constraints
on nonminimal SME coefficients for Lorentz violation 
in the electron and proton sectors
have been derived from laboratory experiments to date
\cite{km13,kv15,sc16,dk16,sm17,na17,dk18},
while the neutron sector is unexplored in the literature.
Here,
we seek to improve this situation
by developing techniques for analyzing clock-comparison experiments
and identifying potential signals from nonminimal Lorentz and CPT violation.
We use existing results to deduce numerous first constraints
on nonminimal SME coefficients in the neutron sector,
and we estimate sensitivities 
to electron, proton, and neutron nonminimal coefficients
that are attainable in future analyses. 

The organization of this work is as follows.
In Sec.\ \ref{Theory},
we present the theoretical techniques
that enable a perturbative treatment
of the effects of Lorentz and CPT violation
on the spectra of atoms and ions.
A description of the perturbation induced
by Lorentz- and CPT-violating operators of arbitrary mass dimension $d$
is provided in Sec.\ \ref{Description of the perturbation}.
The perturbative shifts in energy levels are discussed
in Sec.\ \ref{energy shift}
along with generic features of the resulting spectra,
and some useful formulae for subsequent calculations are derived.
In Sec.\ \ref{Independent-particle model for electrons},
we consider methods for determining expectation values of electronic states,
with emphasis on an independent-particle model.
The corresponding techniques for nucleon states
are presented in Sec.\ \ref{Nucleon expectation values},
primarily in the context of a comparatively simple nuclear model.
We then turn to evaluating the time variations in the spectrum
due to the noninertial nature of the laboratory frame,
first examining effects induced by the rotation of the Earth about its axis
in Sec.\ \ref{sidereal}
and next discussing ones induced by the revolution of the Earth about the Sun
in Sec.\ \ref{linearbe}.
The latter section also considers related issues 
associated with space-based missions.

Applications of these theoretical results
in the context of various clock-comparison experiments
are addressed in Sec.\ \ref{Applications}.
Searches for Lorentz and CPT violation using fountain clocks
are considered in Sec.\ \ref{cssec},
and estimates for attainable sensitivities are obtained.
Studies with comagnetometers 
are investigated in Sec.\ \ref{Comagnetometers},
and first sensitivities to many nonminimal coefficients for Lorentz violation
are deduced from existing data.
Optical transitions in ion-trap and lattice clocks
are discussed in Sec.\ \ref{Trapped ions and lattice clocks},
and potential sensitivities in available systems are considered.
Some comments about prospects for antimatter experiments 
are offered in Sec.\ \ref{Antimatter clocks},
where the first SME constraints from antihydrogen spectroscopy are presented.
A summary of the work is provided in Sec.\ \ref{Summary}.
Two appendices are also included.
Appendix \ref{sphtocar}
describes the general relationship between
spherical and cartesian coefficients for Lorentz violation
and tabulates explicit expressions for $3\leq d\leq 8$.
Appendix \ref{appSun}
presents techniques for transforming
cartesian coefficients for Lorentz violation
from the laboratory frame to the Sun-centered frame
and collects explicit results for $3\leq d\leq 8$.
In this work,
we use conventions and notation matching those of Ref.\ \cite{km13}
except as indicated.
Note that natural units with $c = \hbar = 1$ are adopted throughout.

\section{Theory}
\label{Theory}

This section discusses the theoretical techniques 
for the perturbative treatment of Lorentz and CPT violation
in atoms and ions.
The perturbation is described using a framework 
that encompasses all Lorentz-violating quantum operators 
affecting the motion of the component particles in the atom.
Generic restrictions on the induced energy shifts
arising from symmetries of the system are considered. 
The perturbative calculation of the energy shift is formulated,
and expressions useful for application to experiments are obtained.
Simple models are selected for the electronic and nuclear structure
so that derivations of the relevant expectation values
can be performed for a broad range of atomic species used in experiments.
The conversion from the laboratory frame 
to the Sun-centered frame is provided,
accounting both for the rotation of the Earth about its axis 
and for the revolution of the Earth about the Sun 
at first order in the boost parameter.

\subsection{Description of the perturbation}
\label{Description of the perturbation}

The experiments of interest here
involve comparisons of transitions in atoms or ions,
seeking shifts in energy levels due to Lorentz and CPT violation.
All possible shifts are controlled by SME coefficients,
which can be viewed as a set of background fields in the vacuum.
The energy-level shifts arise from the coupling of these background fields 
to the elementary particles and interactions comprising the atom or ion.
An exact theoretical treatment of the shifts is prohibitive.
However,
since any Lorentz and CPT violation is small,
a perturbative analysis is feasible 
and sufficient to establish the dominant effects.
 
From the perspective of perturbation theory,
the interaction between the electrons and the protons inside an atom or ion 
has some common features with the interaction 
between the nucleons inside the nucleus. 
In both cases,
the magnitude $|\pvec|$ of the momentum $\pvec$ 
of a fermion of flavor $\f$ in the zero-momentum frame 
is smaller than its rest mass $m_\f$.
One consequence is that the dominant contribution 
due to a perturbation added to the hamiltonian of the system 
can be obtained by expanding the perturbation 
in terms of the ratio $|\pvec|/m_\w$,
keeping only leading terms in the power series. 
In most cases,
it suffices to treat the system as effectively nonrelativistic.
Another feature of interest is that the energy per particle 
due to the interaction between the nucleons and the interaction 
between the electrically charged particles in the bound states 
is comparable to the nonrelativistic kinetic energy of the particles. 
The nonrelativistic kinetic energy 
is second order in the ratio $|\pvec|/m_\f$,
so the corrections to the propagation of the particles 
at order $(|\pvec|/m_\f)^0$ and $(|\pvec|/m_\f)^1$
and at leading order in Lorentz violation 
dominate any effects due to Lorentz-violating operators 
coupled to the interactions between the fermions. 

With these considerations in mind,
it is reasonable to proceed under the usual assumption
that the dominant Lorentz-violating shifts 
of the spectrum of the atom or ion arise 
from corrections to the propagation 
of the constituent particles.
For most purposes,
these particles can be taken as 
electrons $e$, protons $p$, and neutrons $n$,
so that $\f$ takes the values $e$, $p$, and $n$.
Applications to exotic atoms or ions can be accommodated
by extending appropriately the values of $\f$.
The relevant Lorentz-violating terms in the Lagrange density
are then quadratic in the fermion fields for the constituent particles.
All terms of this type have been classified and enumerated 
\cite{km13},
and applications to hydrogen and hydrogen-like systems have been established
\cite{kv15}. 
For convenience,
we reproduce in this subsection the key results relevant 
in the present context.

For a Dirac fermion $\ps_\f$ of flavor $\f$ and mass $m_{\f}$,
all quadratic terms in the Lagrange density $\cl$ 
can be expressed as
\cite{km13}
\beq
\cl \supset 
\half \psfb (\ga^\mu i\prt_\mu - m_\f + \Qhat_\f) \ps_\f 
+ {\rm h.c.}.
\label{lag}
\eeq
Here,
$\Qhat_\f$ is a spinor matrix
describing modifications of the standard fermion propagator,
including all Lorentz-invariant and Lorentz-violating contributions
obtained by contracting SME coefficients 
with operators formed from derivatives $i\prt_\mu$.
The matrix $\Qhat_\f$ 
can be decomposed in a basis of Dirac matrices
and can be converted to momentum space
with the identification $i\prt_\mu\to p_\mu$.
Individual operators with definite mass dimension $d$ in the Lagrange density 
incorporate $d-3$ momentum factors,
and the corresponding SME coefficients have dimension $4-d$.
The Lagrange density \rf{lag} has been extended 
to include operators at arbitrary $d$ in the photon sector
\cite{km08,km09,dk16}.
Analogous constructions exist for the neutrino
\cite{nonminnu}
and gravity
\cite{nonmingrav}
sectors.

For present purposes,
the SME coefficients can be assumed uniform and time independent 
within the solar system
\cite{ck}
and so can be taken as constants 
when specified in the canonical Sun-centered frame
\cite{sunframe}.
Using standard procedures,
an effective nonrelativistic one-particle hamiltonian
that includes the leading-order correction 
due to Lorentz- and CPT-violation 
to the propagation of a fermion of flavor $\f$ 
can be derived from the Lagrange density \rf{lag}. 
This hamiltonian can be separated into 
the conventional hamiltonian for a free nonrelativistic fermion 
and a perturbation term $\de h_\f^\nr$ 
containing the Lorentz- and CPT-violating contributions.
The perturbation $\de h^\nr_\f$ is thus a $2\times2$ matrix,
with each component being a function of the momentum operator 
and independent of the position. 
It can be expanded in terms of the identity matrix
and the vector $\sivec= (\si^1, \si^2, \si^3)$ 
of Pauli matrices. 
For convenience, 
this expansion can be performed 
using a helicity basis instead of a cartesian one.
The corresponding three basis vectors can be taken as 
$\ephat_r = \phat \equiv \pvec/\pmag$ 
and $\ephat_\pm = (\thhat \pm i\phhat)/\sqrt{2}$,
where $\thhat$ and $\phhat$ are the usual unit vectors 
for the polar angle $\th$ and azimuthal angle $\ph$ in momentum space,
with $\phat = (\sin\th\cos\ph,\sin\th\sin\ph,\cos\th)$. 
In this helicity basis,
the perturbation $\de h_\f^\nr$ takes the form
\cite{km13} 
\beq
\de h_\f^\nr
= h_{\f 0}+ h_{\f r} \sivec\cdot\ephat_r
+ h_{\f +} \sivec\cdot\ephat_-
+ h_{\f -} \sivec\cdot\ephat_+,
\label{nr}
\eeq
where $h_{\f 0}$ contains spin-independent effects
and the remaining terms describe spin-dependent ones.

Many experiments searching for Lorentz and CPT violation
focus on testing the rotation subgroup of the Lorentz group.
To facilitate the analysis of rotation properties,
it is useful to express the components 
$h_{\f 0}$, $h_{\f r}$, $h_{\f \pm}$ of the perturbation 
in spherical coordinates. 
It is opportune to express the spherical decomposition 
of the operators in the pertubation $\de h_\f^\nr$
in terms of spin-weighted spherical harmonics $\syjm{s}{\j m}(\phat)$ 
of spin weight $s$,
as these harmonics capture in a comparatively elegant form
the essential properties of the perturbation under rotations.
The usual spherical harmonics 
are spin-weighted harmonics with spin-weight $s=0$, 
$Y_{\j m}(\th,\ph) \equiv \syjm{0}{\j m}(\th, \ph)$. 
Definitions and some useful features of spin-weighted spherical harmonics 
are presented in Appendix A of Ref. \cite{km09}.

In terms of the spherical decomposition,
the components of the perturbation \rf{nr} can be expanded as
\beq
h_{\f 0} =
-\sum_{k \j m} \pmag^k 
\syjm{0}{\j m}(\phat) 
\Vnrf{\f}{\k \j m}
\label{FEHSI}
\eeq
for the spin-independent terms, 
and 
\bea
h_{\f r} &=&
-\sum_{k \j m} \pmag^k 
\syjm{0}{\j m}(\phat) 
\TzBnrf{\f}{k \j m},
\nonumber \\
h_{\f \pm} &=&
\sum_{kjm} \pmag^\k 
\syjm{\pm 1}{\j m}(\phat) 
\left(i\ToEnrf{\f}{k \j m} \pm \ToBnrf{\f}{k \j m}\right)
\qquad
\label{FEHSD}
\eea
for the spin-dependent ones.
The coefficients $\Vnrf{\f}{k \j m}$, ${\T_\f}^{\nr(qP)}_{k\j m}$,
where $qP$ takes values $0B$, $1B$, or $1E$, 
are nonrelativistic spherical coefficients for Lorentz violation. 
These effective coefficients
are linear combinations of SME coefficients for Lorentz violation 
that emerge naturally in the nonrelativistic limit 
of the one-particle hamiltonian obtained from the Lagrange density \rf{lag}. 

For applications,
it is useful to perform a further decomposition 
of the components of the perturbation hamiltonian 
according to their CPT handedness. 
In particular,
each nonrelativistic spherical coefficient can be separated into two pieces 
characterized by the CPT handedness of the corresponding operator.
This decomposition can be expressed as
\cite{km13}
\bea
\Vnrf{\f}{k\j m} &=&
\cnrf{\f}{k\j m} - \anrf{\f}{k\j m},
\nonumber \\
{\T_\f}^{\nr(qP)}_{k\j m} &=&
{g_\f}^{\nr(qP)}_{k\j m} - {H_\f}^{\nr(qP)}_{k\j m},
\label{cpt}
\eea 
where the $a$- and $g$-type coefficients 
are contracted with CPT-odd operators 
and the $c$- and $H$-type coefficients 
with CPT-even ones. 
The notation here parallels the standard assignments
in the minimal SME
\cite{ck}.
Each nonrelativistic coefficient
on the right-hand side of this equation
can be expressed as a sum of SME coefficients in the Lagrange density,
suitably weighted by powers of $m_\f$.
The explicit expressions for these sums are given in 
Eqs.\ (111) and (112) of Ref.\ \cite{km13}. 
The allowed ranges of values for the indices $k$, $\j$, $m$ 
and the numbers of independent components for each coefficient 
are listed in Table IV of Ref.\ \cite{km13}. 
Note that in the present work
we follow the convention of Ref.\ \cite{kv15}
and adopt the subscript index $k$ instead of $n$,
to avoid confusion with the principal quantum number of the atom or ion.

Given the perturbation $\de h_\f^\nr$
affecting the propagation of each fermion in an atom or ion,
we can formally express the perturbation $\de h^\atm$
of the system as a whole as
\beq
\de h^\atm=\sum_{\f}\sum_{\ii=1}^{N_\f}\left(\de h^\nr_{\f}\right)_{\ii},
\label{pert}
\eeq
where $\ii = 1,\ldots, N_\f$ labels the fermions of given flavor $\f$
in the atom or ion.
The Lorentz-violating operators considered in this work 
are functions of the momentum and the spin of the particle, 
so
$\left(\de h^\nr_{\f}\right)_{\ii} = \de h^\nr_{\f}(\pvec_\ii,\sivec_\ii)$ 
depends on the momentum operator $\pvec_\ii$ 
and the spin operator $\sivec_\ii$ 
for the $\ii$th fermion.
Each term
$\left(\de h^\nr_{\f}\right)_{\ii}$
is understood to be the tensor product 
of the perturbation \rf{nr} acting on the states 
of the $\ii$th fermion of flavor $\f$ 
with the identity operator acting on the Hilbert space 
of all other fermions.
Note that the index $\ii$ is tied to the momentum and spin,
whereas the index $\f$ controlling the flavor of the particle
is contained in the coefficient for Lorentz violation.
Note also that 
the perturbation \rf{pert} 
can be separated according to operator flavor as
\beq
\de h^\atm=\de h_{e}^\atm+ \de h_{p}^\atm+\de h_{n}^\atm ,
\eeq
where $\de h_{\f}^\atm$ 
is the sum of all operators of flavor $\f$ 
that contribute to $\de h^\atm$. 
For example, 
the expression for $\de h_{e}^\atm$ is given by
$\de h_{e}^\atm=\sum_{\ii=1}^{N_e}\left(\de h^\nr_{e}\right)_{\ii}$,
where $\ii$ ranges over the $N_e$ electrons in the atom.

\subsection{Energy shifts}
\label{energy shift}

The corrections to the spectrum of the atom or ion 
due to Lorentz and CPT violation 
can be obtained from the perturbation $\de h_\atm$ 
using Raleigh-Schr\"{o}dinger perturbation theory. 
At first order, 
the shift of an energy level 
is obtained from the matrix elements of the perturbation 
evaluated in the subspace spanned 
by the degenerate unperturbed energy eigenstates,
as usual.
In typical applications of relevance here,
the degeneracy in the energy levels 
is lifted by an external field 
such as an applied magnetic field.
In this scenario,
the first-order shift of an energy level
is obtained from the expectation value of the perturbation 
with respect to the unperturbed state.
Since the exact unperturbed energy states 
for multielectron atoms or ions are typically unknown,
approximations to these states must be used
to obtain the first-order Lorentz-violating shift of the spectrum.
However,
the symmetries of the unperturbed system
place restrictions on the expectation values of the perturbation. 
In this subsection,
we describe some of these constraints
and establish the general form of the perturbative energy shifts.

Assuming that the degeneracy of the energy levels 
is broken by an applied magnetic field,
parity is a symmetry of the system
and so the states of the atom or ion must be parity eigenstates.
As a result, 
the expectation values of parity-odd operators 
with respect to parity eigenstates must vanish,
so only parity-even operators can affect the spectrum.
This prevents some terms in the perturbation $\de h_\atm$ 
from contributing to the energy shift. 
Each operator in the spherical decomposition \rf{FEHSI} and \rf{FEHSD}
of the perturbation $\de h_\atm$ 
is either odd or even under parity,
with handedness determined by the indices $j$ and $k$ 
of the corresponding coefficient for Lorentz violation. 
The coefficients $\anrf{\f}{k\j m}$ and $\cnrf{\f}{k\j m}$
can contribute to energy shift at first-order in perturbation theory
only for even values of $j$ and $k$,
while
the coefficients ${g_\f}^{\nr(qB)}_{k\j m}$ and ${H_\f}^{\nr(qB)}_{k\j m}$ 
can contribute only if $\j$ is odd and $k$ even. 

Another constraint arises from time-reversal invariance
and the Wigner-Eckart theorem 
\cite{we},
and it concerns the expectation value in any angular-momentum eigenstate 
of Lorentz-violating operators 
controlled by spin-dependent coefficients with $P=E$.
It can be shown that
this expectation value must vanish when the Lorentz-violating operators
transform as spherical operators 
under rotations generated by the angular-momentum operator 
\cite{kv15},
which is the case for the perturbation $\de h^\atm$ of interest here. 
As a result,
none of the spin-dependent coefficients with $P=E$ 
contribute to the perturbative shift of the spectrum 
for any values of $k$ and $\j$. 
This result applies for all atoms and ions considered in the present work. 

In the absence of Lorentz violation,
the total angular momentum $\mbf F$ of the atom or ion
commutes with the hamiltonian of the system. 
When a magnetic field $\mbf B = B \Bhat$ is applied,
the rotational symmetry is broken. 
If the perturbative shift due to the magnetic field 
is smaller than the hyperfine structure,
both the quantum number $F$ corresponding to $\mbf F$ 
and the quantum number $m_F$ corresponding to ${\mbf F}\cdot\Bhat$
can be approximated as good quantum numbers.
Suppose the states $\ket{\alg m_F}$ 
represent a basis of eigenstates of the hamiltonian,
where $\alg$ is a set of quantum numbers including $F$ 
that together with $m_F$ forms a complete set of quantum numbers. 
Using the Wigner-Eckart theorem,
the energy shift due to the propagation of the fermions 
in the atom or ion can then be written as
\beq
\de \ep=\vev{\alg m_F|\de h^\atm|\alg m_F}=\sum_{j}\AF_{j0}\vev{Fm_Fj0|Fm_F},
\label{eatm}
\eeq
where $\vev{ j_2 m_2 j_3 m_3 | j_1 m_1 }$ denote Clebsch-Gordan coefficients. 
The factors $\AF_{j0}=\AF_{j0}(\alg)$ are independent of $m_F$. 
The sum over $j$ in Eq.\ \rf{eatm} 
involves the index $j$ labeling the coefficients for Lorentz violation
in $\de h^\atm$,
in parallel with the sums over $j$ in Eqs.\ \rf{FEHSI} and \rf{FEHSD}. 
Note that the Clebsch-Gordan coefficient $\vev{Fm_Fj0|Fm_F}$ 
vanishes when $j>2F$,
implying that no operator with $j>2F$ contributes to the energy shift. 

To find useful expressions for the factors $\AF_{j0}$,
we make some additional assumptions
that are broadly valid for the systems considered in this work. 
Except where stated otherwise,
we suppose that 
both the magnitude $J$ of the total angular momentum $\mbf J$ of the electrons 
and the magnitude $I$ of the nuclear spin $\mbf I$
are good quantum numbers for the system. 
We also assume that the states $\ket{\alg m_F}$ 
can be expressed as a tensor product 
$\ket{\Ps(\alg)}\otimes\ket{Fm_F}$, 
where
\beq
\ket{Fm_F}=\sum_{m_Jm_I}\vev{Im_IJm_J|Fm_F}\ket{I m_I}\otimes\ket{J m_J}.
\eeq
Here,
the kets $\ket{Fm_F}$, $\ket{Jm_J}$, and $\ket{Im_I}$ 
are associated with the angular momenta
 $\mbf F$, $\mbf J$, and $\mbf I$,
respectively. 
These states also depend on other quantum numbers
that are suppressed in the notation.
For example, 
the ket $\ket{Fm_F}$ depends on $J$, on $I$,
and also on other quantum numbers 
established by the couplings
of the orbital angular momenta and spins of the component particles
to form $\mbf F$.
For later use,
it is also convenient to introduce the notation
$\ket{\alg m_J}= \ket{\Ps{(\alg)}}\otimes\ket{Jm_J}$ 
and $\ket{\alg m_I}= \ket{\Ps{(\alg)}}\otimes\ket{I m_I}$.
Under these assumptions,
we can expand the factors $\AF_{j0}$ 
appearing in Eq.\ \rf{eatm} in the form 
\beq
\AF_{j0}=\Ce{e}{j}\Ae{e}{j0}+\Ce{p}{j}\Ae{p}{j0}+\Ce{n}{j}\Ae{n}{j0},
\label{AF}
\eeq
where $\Ce{\f}{j}(FJI)$ are weights for the quantities $\Ae{\f}{j0}(\alg)$
containing the expectation values of $\de h_{\f}^\atm$ 
with respect to the states $\ket{\alg m_J}$ and $\ket{\alg m_I}$.

The analytical expressions for the factors $\Ce{\f}{j}(FJI)$ 
in terms of Clebsch-Gordan coefficients are
\bea
\Ce{n}{j}&=&\Ce{p}{j}=\sum_{m_J m_I}
\dfrac{\vev{I m_I J m_J|FF}^2}{\vev{FFj0|FF}}\vev{Im_Ij0|Im_I}, 
\nonumber \\
\Ce{e}{j}&=&\sum_{m_J m_I}
\dfrac{\vev{I m_I J m_J|FF}^2}{\vev{FFj0|FF}}\vev{Jm_Jj0|Jm_J}.
\label{CJ}
\eea
Their numerical values can be obtained 
for any given allowed values of $F$, $J$, and $I$. 
Some of the properties of $\Ce{\f}{j}(FJI)$ are induced 
by features of the Clebsch-Gordan coefficients. 
For example,
$\Ce{\w}{0}=1$ for any values of $F$, $J$, and $I$
because
$\vev{Km_K00|Km_K}=1$ for $K$ equal to $F$, $J$, or $I$
and because
$\sum_{m_J m_I}\vev{I m_I J m_J|FF}^2=1$.
As another example,
$\Ce{e}{j}(FJI)=0$ whenever $j>2J$
and $\Ce{p}{j}(FJI)= \Ce{n}{j}(FJI) =0$ whenever $j> 2I$
because $\vev{Km_Kj0|Km_K}=0$ if $j>2K$. 

\renewcommand\arraystretch{1.5}
\begin{table*}
\caption{
Contributing nonrelativistic spherical coefficients.} 
\setlength{\tabcolsep}{6pt}
\begin{tabular}{cccccc}
\hline
\hline
${\K_\f}_{kj0}^\nr$	&  $\f$ & $K$ &$j$ values, integer $K$	        &$j$ values, half-integer$K$	                                 &$k$ values	\\
\hline													
$\Vnrf{\f}{kj0}$, $\anrf{\f}{kj0}$, $\cnrf{\f}{kj0}$         		&	$e$	            & $F$, $J$  &even, $2K\ge j\ge0$	  &	even, $2K-1\ge j\ge0$	        &	even, $k\ge j$\\
$\Vnrf{\f}{kj0}$, $\anrf{\f}{kj0}$, $\cnrf{\f}{kj0}$	        	&	$n$, $p$   & $F$, $I$  &even, $2K\ge j\ge0$	  &	even, $2K-1\ge j\ge0$	        &	even, $k\ge j$\\
$\TzBnrf{\f}{kj0}$, $\gzBnrf{\f}{kj0}$, $\HzBnrf{\f}{kj0}$	&	$e$	            & $F$, $J$  & odd, $2K-1\ge j\ge1$  &	odd, $2K\ge j\ge1$	                &	even, $k\ge j-1$\\
$\TzBnrf{\f}{kj0}$, $\gzBnrf{\f}{kj0}$, $\HzBnrf{\f}{kj0}$	&	$n$, $p$   & $F$, $I$  & odd, $2K-1\ge j\ge1$  &	odd, $2K\ge j\ge1$	                &	even, $k\ge j-1$\\
$\ToBnrf{\f}{kj0}$, $\goBnrf{\f}{kj0}$, $\HoBnrf{\f}{kj0}$	&	$e$	            & $F$, $J$  & odd, $2K-1\ge j\ge1$  &	odd, $2K\ge j\ge1$	                &	even, $k\ge j-1$\\
$\ToBnrf{\f}{kj0}$, $\goBnrf{\f}{kj0}$, $\HoBnrf{\f}{kj0}$	&	$n$, $p$   & $F$, $I$  & odd, $2K-1\ge j\ge1$  &	odd, $2K\ge j\ge1$	                &	even, $k\ge j-1$\\
\hline\hline
\end{tabular}
\label{cond}
\end{table*}

The explicit relationships between 
the expectation values of the perturbations $\de h^\atm_w$
and the $\Ae{\f}{j0}(\alg)$ in Eq.\ \rf{AF}
can be written as 
\bea
\vev{\alg m_J |\de h_{e}^\atm|\alg m_J}
&=&\sum_j \Ae{e}{j0}\vev{J m_J j0 |J m_J},
\nonumber \\
\vev{\alg m_I |\de h_{p}^\atm|\alg m_I}
&=&\sum_j \Ae{p}{j0}\vev{I m_I j0 |I m_I},
\nonumber \\
\vev{\alg m_I |\de h_{n}^\atm|\alg m_I}
&=&\sum_j \Ae{n}{j0}\vev{I m_I j0 |I m_I}.
\label{Aexp}
\eea
These expressions are based on using the Wigner-Eckart theorem,
which is valid because the single-particle operators 
in the spherical decomposition of $\de h^\atm$ 
transform as spherical operators 
with respect to rotations generated by $\mbf I$ and $\mbf J$. 
The $\Ae{\f}{j0}$ are combinations 
of coefficients for Lorentz violation 
with expectation values of the one-particle operators 
in Eqs.\ \rf{FEHSI} and \rf{FEHSD}. 
The combinations take the form 
\bea
\Ae{\f}{j0}&=& \sum_{k=j-1}^\infty
\Big(\TzBnrf{\f}{\k \j 0}\AzB{\f}{\k \j}
+\ToBnrf{\f}{\k \j 0}\AoB{\f}{\k \j}\Big)
\nonumber \\
&&
+\sum_{k=j}^\infty\Vnrf{\f}{\k \j 0}\AzE{\f}{\k\j},
\label{AJe}
\eea
where the indicated restrictions of the values of $k$ in the sums 
originate in the properties of the nonrelativistic coefficients
provided in Table IV of Ref.\ \cite{km13}. 
Generic expressions for the quantities $\Agen{\f}$ 
can be found in terms of expectation values 
of the states $\ket{\alg m_J}$ and $\ket{\alg m_I}$.
For the electron operators,
we have 
\bea
\AzB{e}{k\j}
&=&-\sum_{\ii=1}^{N_e}
\dfrac{\vev{\alg J||{\mbf p}_\ii|^k~\syjm{0} {\j 0} (\phat_\ii)
\sivec_\ii\cdot\phat_\ii|\alg J}}{\vev{JJ|j0JJ}},
\nonumber\\
\AoB{e}{k\j}
&=& 
\hskip -5pt
\sum_{\ii=1}^{N_e}
\dfrac{\vev{\alg J||{\mbf p}_\ii|^k\syjm{+1}{\j 0} (\phat_\ii)
\sivec_\ii\cdot(\ephat_{+\ii}+\ephat_{-\ii})|\alg J}}{\vev{JJ|j0JJ}},
\nonumber\\
\AzE{e}{k\j}
&=& -\sum_{\ii=1}^{N_e}
\dfrac{\vev{\alg J||{\mbf p}_\ii|^k~\syjm{0}{\j 0} (\phat_\ii)
|\alg J}}{\vev{JJ|j0JJ}},
\label{LaJ}
\eea
where the sum on $\ii$ 
ranges over the electrons in the atom. 
For the nucleon operators,
we find
\bea
\AzB{\f}{k\j}
&=&-\sum_{\ii=1}^{N_\f}
\dfrac{\vev{\alg I||{\mbf p}_\ii|^k~\syjm{0}{\j 0} (\phat_\ii)
\sivec_\ii\cdot\phat_\ii|\alg I}}{\vev{II|j0II}},
\nonumber\\
\AoB{\f}{k\j}
&=&
\hskip -5pt
\sum_{\ii=1}^{N_\f}
\dfrac{\vev{\alg I||{\mbf p}_\ii|^k\syjm{+1}{\j 0} (\phat_\ii)
\sivec_\ii\cdot(\ephat_{+\ii}+\ephat_{-\ii})|\alg I}}{\vev{II|j0II}},
\nonumber\\
\AzE{\f}{k\j}
&=&-\sum_{\ii=1}^{N_\f}
\dfrac{\vev{\alg I||{\mbf p}_\ii|^k~\syjm{0}{\j 0} (\phat_\ii)
|\alg I}}{\vev{II|j0II}},
\label{LaI}
\eea
where the sum on $\ii$ 
ranges over all particles with flavors $\f\in\{p,n\}$. 

Explicit determination of the nonvanishing expectation values 
in Eqs.\ \rf{LaJ} and \rf{LaI}
requires models for the electronic states and for the nuclear states,
as discussed below in 
Secs.\ \ref{Independent-particle model for electrons}
and \ref{Nucleon expectation values},
respectively.
However,
certain components of $\Agen{\f}$ vanish.
We saw above that only coefficients with even values of $k$ can contribute 
due to parity invariance.
This implies that $\Agen{\f}(\alg)$ vanishes 
if $k$ is even. 
For $\AzB{\f}{k\j}$ and $\AoB{\f}{k\j}$
it follows that $j$ must be odd, 
while for $\AzE{\f}{k\j}$ we find $j$ must be even. 
These results are a consequence of the relationships 
between the indices $k$ and $j$ 
of the nonrelativistic coefficients,
as listed in Table IV of Ref.\ \cite{km13}. 

Collecting the results discussed in this subsection
yields a set of constraints determining 
which coefficients for Lorentz violation
can affect the shift of an energy level in an atom or ion. 
Table \ref{cond} compiles information
about the nonrelativistic spherical coefficients 
that can contribute to spectral shifts.
The first column of the table lists the coefficients,
which we denote generically by ${\K_\f}_{k\j m}^\nr$.
The flavor of the operator associated to the coefficient 
is specified in the second column. 
The third column gives the angular momenta $K$ 
that restrict the values of the $j$ index on the coefficient
according to the constraint $2K\ge j$. 
For electron coefficients
these angular momenta are the total angular momentum $F$ of the system 
and the total angular momentum $J$ of the electronic shells, 
while for nucleon coefficients they are $F$ and the nuclear spin $I$. 
The next two columns provide conditions on the values of $j$
for the cases of integer $K$ and of half-integer $K$. 
The value of $j$ must be even
for coefficients in the first two rows 
and odd for other coefficients.
This can constrain the maximum allowed value of $j$.
For example,
for even $j$ and half-integer $K$ 
the equality in the condition $2K\ge j$ 
cannot be satisfied because $2K$ is odd,
so the maximum allowed value of $j$ is $2K-1$.
The final column in the table displays the allowed values of $k$.
Note that the appearance of a coefficient in the table
is necessary but not sufficient for it to contribute 
to a theoretical energy shift
because some $\Agen{\f}$ may vanish for other reasons 
when a particular model is used to compute the expectation values.

\subsection{Electron expectation values}
\label{Independent-particle model for electrons}

In this subsection,
the calculation of the electronic expectation values \rf{LaJ} 
is discussed. 
The situation where $F$ or $J$ vanish is considered first.
We then outline an approach to more complicated cases
that is general enough to cover systems of interest here 
while yielding a sufficient approximation
to the effects of Lorentz and CPT violation.
This involves modeling the electromagnetic interaction
between the electrons and the nucleus
via a central Coulomb potential
and treating the repulsion between the electrons
using a mean-field approximation.
The approach provides enough information about the states $\ket{\alg m_J}$
to permit reasonable estimation of the perturbative energy shift
due to Lorentz- and CPT-violating effects on the electron propagators.

\subsubsection{Case $F=0$ or $J=0$}
\label{CaseFJ}

The ground states of many atoms and ions considered in this work 
have quantum numbers $F=0$ or $J=0$. 
For example,
this holds for the ground state 
of any noble gas and any IIB transition metal such as Hg. 
It also holds for the ground states of many ions of interest,
including 
$^{27}\text{Al}^+$,
$^{113}\text{Cd}^+$ , 
$^{115}\text{In}^+$, 
$^{171}\text{Yb}^+$, 
and $^{199}\text{Hg}^+$. 
The excited states of some systems of interest 
also have these quantum numbers,
such as the $P_{0}$ state 
in $^{27}\text{Al}^+$ 
or $^{115}\text{In}^+$.

If either of the quantum numbers $F$ or $J$ vanishes,
then the electron coefficients for Lorentz violation 
that can contribute to the energy shift must have $j=0$.
These coefficients control isotropic Lorentz- and CPT-violating effects. 
The discussion in Sec.\ \ref{energy shift} 
reveals that the only relevant isotropic coefficients for electrons
are $\Vnrf{e}{k00}$.
These special coefficients are commonly denoted as $\Vnrfc{e,k}$,
where $\Vnrfc{e,k}\equiv\Vnrf{e}{k00}/\sqrt{4\pi}$. 

Since only $\Vnrfc{e,k}$ can affect the energy shift,
the quantities $\AzB{e}{k\j}$ and $\AoB{e}{k\j}$
cannot contribute to Eq.\ \rf{AJe} and so become irrelevant.
Moreoever,
when only the isotropic coefficients for electrons
can provide nonvanishing contributions,
we can simplify the expression \rf{LaJ} for $\AzE{e}{k0}$.
The values of the relevant Clebsch-Gordan coefficient 
and spherical harmonic 
are $\vev{JJ00|JJ}=1$ 
and $\syjm{0}{00}=1/\sqrt{4\pi}$. 
This yields
\beq
\AzE{e}{k0}=
-\sum_{\ii=1}^{N_e}\dfrac{1}{\sqrt{4\pi}}\vev{|{\mbf p}_\ii|^k},
\label{AJiso}
\eeq
where the sum on $\ii$ ranges over all the electrons in the atom. 
The quantities $\vev{{|\mbf p}_\ii|^{k}}$ are the expectation values 
of powers of the momentum magnitude. 

Denote the contribution to the energy shift 
due to the electron isotropic coefficients
by $\eiso_e$.
Recalling that $\Ce{\w}{0}=1$ for any value of $F$, $J$, and $I$,
it follows from Eqs.\ \rf{eatm} and \rf{AF}
that $\eiso_e$ takes the simple form
$\eiso_e=\Ae{e}{00}$.
Using equations $\rf{AJe}$ and $\rf{AJiso}$
then yields an expression for the energy shift 
due to the electron isotropic coefficients,
\beq
\eiso_e=-\sum_{q}\Vnrfc{e,2q}\sum_{\ii=1}^{N_e}\vev{{|\mbf p}_\ii|^{2q}}
\label{isoe}
\eeq
where the index $q$ is related to the index $k$ 
of the coefficients for Lorentz violation by $2q=k$. 
This enforces the condition 
that only even values of $k$ can contribute to the energy shift.

\subsubsection{One open subshell with one electron}
\label{one electron}

For atoms or ions with all electronic subshells closed 
except for a single subshell occupied by one electron,
we can find closed-form expressions for the 
expectation values $\Agen{e}$
under suitable simplifying approximations.
Treating the electrons in the closed subshells
as forming states with zero total angular momentum,
the value of $J$ for the whole system
can be identified with the total angular momentum 
of the electron in the open subshell. 
It follows that the only contribution to $\Agen{e}$ with $j\neq 0$ 
can arise from the valence electron.
Contributions from isotropic coefficients with $j=0$ 
are given by Eq.\ \rf{isoe}. 

The closed shells produce a spherically symmetric electronic distribution.
For present purposes,
the effective potential acting on the valence electron 
due to the repulsion from the closed-shell electrons 
can be approximated as central. 
One consequence of this is that 
the magnitude $L$ of the orbital angular momentum $\mbf L$ 
of the valence electron 
is a good quantum number for the system.
It then becomes feasible to obtain explicit expressions
for the quantities $\Agen{e}$ defined in Eq.\ \rf{LaJ}.
We find 
\bea
\AzE{e}{k\j}&=&
i^j
\dfrac{(j-1)!!}{j!!} \dfrac{\Mco{j}{J}}{\LaF{j}{J}} 
\vev{|\pvec|^k},
\nonumber\\
\AzB{e}{k\j}
&=&
i^{j-1}
\dfrac{j!!}{(j-1)!!} \Mco{j}{J}\LaF{j}{J}
\vev{|\pvec|^k},
\nonumber\\
\AoB{e}{k\j}
&=&
i^{j-1}\dfrac{2J+1}{L-J}\Mco{j}{J} \LaF{j}{J}
\sqrt{\dfrac{j!!\,(j-2)!!}{2\,(j+1)!!(j-1)!!}}
\vev{|\pvec|^k},
\nonumber\\
\label{Laval}
\eea
In this equation,
$\pvec$ is the momentum of the valence electron,
and we define 
\bea
\Mco{j}{J}&=&\sqrt{\dfrac{2j+1}{4\pi (2J+1)}},
\nonumber \\
\LaF{j}{J}&=&\sqrt{\dfrac{(2J+1)!!(2J-j-1)!!}{(2J-1)!!(2I+j+1)!!}}.
\eea

Note that the spin-independent operators in Eq.\ \rf{FEHSI} 
transform as spherical operators with respect to rotations 
generated by $\mbf L$, 
which suffices to exclude contributions from $\Vnrf{e}{kjm}$ 
to the energy shift when $j>2L$. 
However,
this requirement is already implied in the present context 
by the condition $2J-1\ge j$ presented in Table \ref{cond},
because the lowest value of $L$ for a given $J$ is $L=J-1/2$.

The results \rf{Laval} can be applied to alkali-metal atoms
and to singly ionized alkaline-earth ions.
In both cases,
the electrons in the closed subshells belong to closed shells,
so the approxiations made above are comparatively good.
This can be illustrated by comparing our results
with detailed calculations for specific states of particular systems. 
For example,
consider the numerical results 
presented in Table 1 of Ref.\ \cite{Dz16}
for the $D_{3/2}$ and $D_{5/2}$ states 
in ${\rm Ca}^{+}$, 
${\rm Ba}^{+}$,
and ${\rm Yb}^{+}$. 
The table provides the reduced matrix elements of the operator
\beq
T^{(2)}_{0}=-\sqrt{\dfrac{16\pi}{5}}\pmag^2\syjm{0}{20}(\phat),
\eeq
calculated using several many-body techniques
and defined in terms of Wigner 3-$j$ symbols 
instead of Clebsch-Gordan coefficients.
The ratio of the reduced matrix elements 
for the two states $D_{3/2}$ and $D_{5/2}$ 
is $ 0.77$ for ${\rm Ca}^{+}$ 
and $0.79$ for ${\rm Ba}^{+}$. 
Converting the notation appropriately,
we find that Eq.\ \rf{Laval} 
predicts a ratio of $0.76$ for both systems,
in reasonable agreement with the many-body calculations.
However,
for ${\rm Yb}^{+}$ 
the results obtained in Ref.\ \cite{Dz16} give a ratio of $0.82$,
revealing a greater deviation from our prediction.
This is unsurprising because in this ion 
some electrons in a closed subshell lie outside the closed shells,
so the accuracy of our approximation is expected to be reduced. 

The results \rf{isoe} and \rf{Laval} involve
expectation values of powers of the magnitude of the electron momentum.
An analytical evaluation of these expectation values is impractical,
even for comparatively simple cases such as
the expectation values $\vev{|\pvec|^k}$ for a valence electron.
Numerical methods can be adopted to resolve this issue,
in conjunction with techniques 
such as a self-consistent mean-field approximation.
However,
the principle goal of this work is to serve as a guide
to search for Lorentz and CPT violation.
In this context,
a precise determination of these expectation values is often inessential.
For example,
some transitions studied here are hyperfine or Zeeman transitions.
These involve two levels with similar momentum expectation values,
and the difference leaves unaffected the qualitative form
of experimental signals for Lorentz and CPT violation.
For these and many other transitions,
estimates of the expectation values of the electron momentum
suffice as a guide to the sensitivity of experiments
across a broad range of systems.
An accurate determination of the expectation values 
relevant to a given experimental setup 
may become useful once enough data are collected
and a detailed analysis is being performed to extract
the coefficients for Lorentz violation.
For a few transitions used in experiments,
estimates may be inadequate even as a guide to the sensitivity.
For example,
for optical transitions 
the difference between the expectation values in the two states
can be significant and must be included in the treatment,
as described in Sec.\ \ref{Trapped ions and lattice clocks} below.

For atoms or ions with more than one electron in an open subshell,
it is typically infeasible to obtain closed-form expressions
like Eqs.\ \rf{isoe} and \rf{Laval}.
These systems can have substantial many-body effects,
and their treatment requires a more sophisticated 
and individualized approach.
Investigations of such systems are likely also to be of interest 
in searches for Lorentz and CPT violation,
but a discussion along these lines lies beyond the scope of this work.

\subsection{Nucleon expectation values}
\label{Nucleon expectation values}

Next,
we turn to the evaluation of the nucleon expectation values 
\rf{LaI}.
The simplest situation arises when $F$ or $I$ vanishes,
for which a compact expression for the energy shift can be presented.
For more complicated scenarios,
a model accounting for the strong nuclear interactions is required.
The central effective potential and mean-field approximation
used above for the electronic structure
are inappropriate to describe the nucleon interactions.
Instead,
we adopt here a simple nuclear shell model
that permits analytical evaluation of the quantities $\Agen{\f}$.
This enables an evaluation of the effects
of Lorentz and CPT violation from nucleon propagators
on spectral shifts in a broad range of systems.

\subsubsection{Case $F=0$ or $I=0$}
\label{CaseFI}

A number of atoms or ions have either vanishing total angular momentum $F=0$ 
or vanishing nuclear spin $I=0$.
The latter situation arises in nuclei
with an even number of neutrons and an even number of protons.
In these cases,
independently of the nuclear model adopted,
the energy shift $\de \ep_\f$ 
due to a nucleon of flavor $\f = p$ or $n$ receives contributions 
only from isotropic coefficients for Lorentz violation.
The arguments here parallel those in Sec.\ \ref{CaseFJ}.

Introducing the special isotropic coefficients
$\Vnrfc{\f,k}\equiv\Vnrf{\f}{k00}/\sqrt{4\pi}$,
the expression for $\eiso_\f$ 
is found to be
\beq
\eiso_{\f} =
-\sum_{q=0}\Vnrfc{\f,2q}\sum_{\ii=1}^{N_\f}\vev{{|\mbf p}_\ii|^{2q}}
\label{isow}
\eeq
where the sum over $\ii$ spans the $N_\f$ nucleons of flavor $\f$ 
in the nucleus. 
Like the electron case,
this isotropic shift can also affect other energy levels 
having $F\neq 0$ or $I\neq 0$
through its contribution to Eq.\ \rf{AJe}.

\subsubsection{Schmidt model for one unpaired nucleon}
\label{Schmidt model for one unpaired nucleon}
 
The Schmidt model 
\cite{s37,bw52}
offers a comparatively simple description 
of a broad range of nuclei.
The model assumes a shell structure for the nucleus
in which any pair of nucleons of a given flavor 
combines to form states with total angular momentum equal to zero. 
If only one unpaired nucleon exists in the nucleus, 
then it is treated as a single-particle state 
with total angular momentum equal to the spin $I$ of the nucleus. 
The magnitude $L$ of the orbital angular momentum 
of the unpaired nucleon is treated as a good quantum number. 
The model can be expected to deviate significantly from observation
for nuclei lying away from a magic number. 
 
Mathematically, 
the Schmidt model represents a setup 
equivalent to the one described in Sec.\ \ref{one electron}
for a valence electron outside closed subshells.
The contribution to the perturbative energy shift
involving istropic coefficients is obtained from Eq.\ \rf{isow}.
When $j>0$,
the expressions for the quantities $\Agen{\f}$
can be calculated in closed form
and are given by
\bea
\AzE{\f}{k\j}
&=&
i^j
\dfrac{(j-1)!!}{j!!} \dfrac{\Mco{j}{I}}{\LaF{j}{I}}
\vev{|\pvec|^k},
\nonumber\\
\AzB{\f}{k\j}
&=&
i^{j-1}
\dfrac{j!!}{(j-1)!!} \Mco{j}{I} \LaF{j}{I}
\vev{|\pvec|^k},
\nonumber\\
\AoB{\f}{k\j}
&=&
i^{j-1}\dfrac{2I+1}{L-I}\Mco{j}{I} \LaF{j}{I}
\sqrt{\dfrac{j!!\,(j-2)!!}{2\,(j+1)!!(j-1)!!}}
\vev{|\pvec|^k},
\nonumber\\
\label{LavalI}
\eea
where $\pvec$ is the linear momentum 
of the unpaired nucleon of flavor $\f$.
The factors $\Mco{j}{I}$ and $\LaF{j}{I}$ are defined as
\bea
\Mco{j}{I}
&=&\sqrt{\dfrac{2j+1}{4\pi (2I+1)}},
\nonumber \\
\LaF{j}{I}
&=&\sqrt{\dfrac{(2I+1)!!(2I-j-1)!!}{(2I-1)!!(2I+j+1)!!}}.
\eea
 
The primary advantage of the Schmidt model
in the present context 
is its application to a broad range of systems 
for which the quantities $\Agen{\f}$ 
can be evaluated using Eq.\ \rf{LavalI}.
The model has previously been used
to determine signals 
arising from Lorentz- and CPT-violating operators in the minimal SME
for numerous experiments comparing atomic or ionic transitions
\cite{kl99}.
A significant limitation of the Schmidt model
in this respect
is that only one flavor of nucleon is assumed to contribute
to transitions in any given atom or ion,
which implies the corresponding experiment 
is sensitive only to coefficients for Lorentz violation 
in that flavor sector.
A more realistic treatment can be expected to reveal
dependence on coefficients for both values of $\f$.
This was illustrated in Ref.\ \cite{kl99} 
using more detailed wave functions 
for the nuclei of $^7$Li and $^9$Be.
Recently, 
calculations using semi-empirical models 
\cite{fl16} 
and many-body methods
\cite{Br} 
have obtained improved values for the coefficients $\Agen{\f}$,
particularly for the coefficient $\AzE{\f}{22}$. 
These improved values emphasize the disadvantages 
of using a single-valence model 
to study Lorentz- and CPT-violating effects involving the nucleus. 
Nonetheless,
to maintain generality in this work
and to permit the discussion of a broad range of atoms and ions, 
we adopt the Schmidt model throughout, 
commenting where appropriate on the likely consequences
of using improved nuclear modeling.
We remark also that
it suffices to estimate the expectation values 
of the magnitude of the linear nucleon momentum
for all experiments considered here
because no nuclear transitions are involved.

\subsection{Energy shift at zeroth boost order}
\label{sidereal}

In any cartesian inertial frame within the solar system,
the coefficients for Lorentz violation 
can reasonably be taken as constant in both time and space
\cite{ck,akgrav}.
However,
the energy shift \rf{eatm} is calculated in a laboratory frame.
Laboratories on the surface of the Earth or on orbiting satellites
generically correspond to noninertial frames,
so most coefficients appearing in Eq.\ \rf{eatm} 
vary with time due to the laboratory rotation and boost
\cite{ak98}.
Moreover,
the explicit forms of the coefficients for Lorentz violation
differ in distinct inertial frames.
To permit meaningful comparison of different experiments,
experimental coefficient values must therefore be reported
in a canonical inertial frame.
The standard frame adopted in the literature for this purpose 
is the Sun-centered celestial-equatorial frame 
\cite{sunframe},
with cartesian coordinates denoted $(T,X,Y,Z)$.
In this frame,
the origin of the time coordinate $T$
is defined as the vernal equinox 2000.
The $X$ axis points from the location of the Earth at this equinox
to the Sun,
the $Z$ axis is aligned with the Earth's rotation axis,
and the $X$, $Y$, $Z$ axes
form a right-handed coordinate system.
The Sun-centered frame is appropriate 
for reporting measurements of coefficients
because it is inertial to an excellent approximation
over the time scale of typical experiments.

The observer Lorentz transformation 
$ \La^\mu_{\ \nu}({\mbf\th},\bevec)$ 
between the laboratory frame and the Sun-centered frame 
can be expressed as the composition of 
an observer rotation 
$ \mathcal{R}^\mu_{\ \nu}({\mbf \th})$
with an observer boost
$ \mathcal{B}^\mu_{\ \nu}(\bevec)$,
\beq
\La^\mu_{\ \nu}({\mbf\th},\bevec)= 
\mathcal{R}^\mu_{\ \al}({\mbf \th})\mathcal{B}^\al_{\ \nu}(\bevec).
\label{lortr}
\eeq
The boost parameter $\bevec$ is the velocity of the laboratory frame 
with respect to the Sun-centered frame,
while the rotation parameter $\mbf \th$ fixes the relative orientation 
between the laboratory frame 
and the frame obtained via the boost.
The magnitude $\be$ of $\bevec$ is small compared to the speed of light.
For example,
the speed of the Earth in the Sun-centered frame 
in natural units is $\be\simeq10^{-4}$. 
At zeroth order in $\be$
the boost transformation is simply the identity map,
so the Lorentz transformation between the two frames becomes a pure rotation.  
In this subsection,
we consider the energy shift \rf{eatm} at zeroth boost order.
Effects at linear boost order are discussed in Sec.\ \ref{linearbe}.

In the laboratory frame,
only the nonrelativistic coefficients ${\K_\f}_{kjm}^{\rm NR,lab}$ 
for Lorentz and CPT violation 
with $m=0$ contribute to the energy shift. 
At zeroth boost order and for a laboratory on the Earth,
these coefficients can be converted to coefficients 
${\K_\f}_{kjm}^{\rm NR,Sun}$ in the Sun-centered frame via 
\beq
{\K_\f}_{kj0}^{\rm NR,lab} = 
\sum_{m} e^{i m\om_\oplus \TL}
d^{j}_{0m}(-\chM)
{\K_\f}^{\rm NR,Sun}_{kjm}.
\label{ltos}
\eeq
Here,
$\chM$ is the angle between the applied magnetic field 
and the Earth's rotation axis $Z$,
and the quantities $d^{j}_{mm'}$ are the little Wigner matrices 
given in Eq.\ (136) of Ref. \cite{km09}.
The conversion \rf{ltos} reveals the time variations
of the laboratory-frame coefficients,
which occur at harmonics of the Earth's sidereal frequency
$\om_\oplus\simeq 2\pi/(23{\rm ~h} ~56{\rm ~min})$.
The local sidereal time $\TL$
is a convenient local Earth sidereal time
with origin chosen as the time when the magnetic field
lies in the $XZ$ plane in the Sun-centered frame.
This choice yields the comparatively simple expression \rf{ltos}.
For some applications below it is preferable instead to adopt 
a different local sidereal time $T_\oplus$,
which is associated with the laboratory frame 
introduced in Ref.\ \cite{sunframe}
and has as origin the time at which
the tangential velocity of the laboratory frame points along the $Y$ axis.
The relationship between $\TL$ and $T_\oplus$ is
\beq
\om_\oplus T_\oplus = \om_\oplus \TL - \phM,
\eeq
where $\phM$ is the angle between the $X$ axis 
and the projection of the magnetic field 
on the $XY$ plane at $T_\oplus=0$.
Note that both $\TL$ and $T_\oplus$ are offset 
from the standard time $T$ in the Sun-centered frame
by an amount that depends on the longitude of the laboratory,
given explicitly for $T_\oplus$ in Eq.\ (43) of Ref.\ \cite{dk16}.

\begin{table}
\caption{Possible harmonics of $\om_\oplus$ at zeroth boost order.} 
\setlength{\tabcolsep}{10pt}
\begin{tabular}{cc}
\hline
\hline
Conditions on $F$, $J$, $I$ & Possible harmonics\\
\hline
$F\le J$ or $F\le I$ or both & $2F\ge |m|\ge 0$\\
$F\ge J$, $F\ge I$, $J\ge I$ & $2J\ge |m|\ge 0$\\
$F\ge J$, $F\ge I$, $I\ge J$ & $2I\ge |m|\ge 0$\\
\hline\hline
\end{tabular}
\label{side}
\end{table}

The factors $\AF_{j0}\equiv \AF_{j0}^{\rm lab}$
appearing in Eq.\ \rf{eatm} 
are defined in the laboratory frame.
They transform in the same way under rotations
as the nonrelativistic coefficients for Lorentz violation,
so we can convert them to factors $\AF_{j0}^{\rm Sun}$ 
defined in the Sun-centered frame
via the relation
\beq
\AF_{j0}^{\rm lab} = 
\sum_{m} e^{i m\om_\oplus T_\oplus}
d^{j}_{0m}(-\chM)
\AF^{\rm Sun}_{jm}.
\label{ltos2}
\eeq
The energy shift \rf{eatm} can therefore be expressed 
in the Sun-centered frame as
\bea
\de \ep &=&
\sum_{jm} 
d^{(j)}_{0|m|}(-\chM)
\vev{Fm_Fj0|Fm_F}
\nonumber\\ 
&&
\qquad
\times 
\big[ \Re{\AM_{j|m|}^{\rm Sun}} \cos{(|m| \om_\oplus T_\oplus )} 
\nonumber\\ 
&&
\qquad\qquad
- \Im{\AM_{j|m|}^{\rm Sun}} \sin{(|m| \om_\oplus T_\oplus )} \big],
\label{rotgen}
\eea
thereby explicitly demonstrating the time variation 
of the spectrum of the atom or ion
at harmonics of the sidereal frequency $\om_\oplus$. 

For any $m$,
a given factor $\AM_{j|m|}^{\rm Sun}$ contains
coefficients for Lorentz violation labeled with the same index $j$.
However,
as summarized in Table \ref{cond},
only restricted values of $j$ for nonrelativistic coefficients 
can contribute to a specific energy shift. 
Since the highest harmonic that can contribute 
to the sidereal variation is determined by the maximum value of $|m|$,
which in turn is fixed by the largest allowed value of the index $j$,
we can use the information in Table \ref{cond}
to deduce constraints on the possible harmonics
contributing to the time variation of any particular energy level.
Table \ref{side} summarizes these constraints
for various conditions on the quantum numbers $F$, $J$, and $I$. 
The first column of the table
lists the conditions,
while the second column displays the range of allowed values of $|m|$,
which corresponds to the possible harmonics of $\om_\oplus$ that can appear. 
For example,
the first row of the table shows that 
the time varation of an energy level 
with quantum numbers $F=3$, $I=7/2$, and $J=3/2$
can in principle contain up to the sixth harmonic of $\om_\oplus$.
Note,
however,
that special circumstances might contrive to lower the maximum harmonic
affecting a given transition frequency.
For example,
a factor $\AM_{j|m|}^{\rm Sun}$ might vanish identically,
or the two energy levels involved in the transition
might have identical contributions at a particular harmonic
so that the transition frequency is unaffected. 
Note also that 
time variations at higher harmonics than those displayed
in Table \ref{side} can become allowed 
when effects at linear or higher order in the boost are incorporated,
but any such variations are suppressed by powers of the boost.

\subsection{Energy shift at linear boost order}
\label{linearbe}

Since the magnitude of the boost between the laboratory frame
and the Sun-centered frame is small,
it is reasonable to expand the boost transformation 
$ \mathcal{B}^\mu_{\ \nu}(\bevec)$ 
of Eq.\ \rf{lortr}
in powers of the relative speed $\be$.
In this subsection,
we consider contributions to the energy shift 
that appear at linear order in the boost.
At this order,
the components of the observer Lorentz transformation 
$ \La^\mu_{\ \nu}({\mbf\th},\bevec)$ 
take the form
\beq
\La^{0}{}_{T}=1, 
\hskip 8pt
\La^{0}{}_{J}=-\bevec^J, 
\hskip 8pt
\La^{j}{}_{T}=-\mathcal{R}^{j}{}_{J}\bevec^{J},
\hskip 8pt
\La^{j}{}_{J}=\mathcal{R}^{j}{}_{J},
\label{LTlinear}
\eeq
where lower-case and upper-case indices 
represent spatial cartesian coordinates in the laboratory frame 
and in the Sun-centered frame,
respectively. 

Given an expression for the energy shift in the laboratory frame
in terms of spherical coefficients for Lorentz violation,
converting to the Sun-centered frame at linear boost order
can be performed in two steps.
First,
the spherical coefficients in the laboratory frame
can be rewritten in terms of cartesian coefficients in the same frame.
The transformation \rf{LTlinear}
can then be applied to express 
the cartesian coefficients in the laboratory frame
in terms of cartesian coefficients in the Sun-centered frame.

Explicit expressions relating spherical coefficients
to cartesian coefficients in any inertial frame
are given in Appendix \ref{sphtocar}.
To implement the conversion to the Sun-centered frame,
note that only spherical coefficients for Lorentz violation with $m=0$ 
contribute to the energy shift \rf{eatm} in the laboratory frame.
This implies that all uncontracted spatial indices
on the corresponding cartesian coefficients
are in the $x^3$ direction.
The relevant part of the rotation matrix $\mathcal{R}^{j}{}_{J}$
converting the cartesian components 
between the laboratory frame and the Sun-centered frame
therefore involves the row $\mathcal{R}^{3}{}_{J}$. 
This row can be viewed as the components of a unit vector along $x^3$.
Since by construction this is the quantization axis of the atom or ion,
which points in the direction $\Bhat$ of the applied magnetic field,
it follows that $\mathcal{R}^{3}{}_{J}=\hat{B}^J$.

To illustrate this idea with an example,
consider the spherical coefficient $g^{(4)(1B)}_{210}$ 
given in the laboratory frame.
From Appendix \ref{sphtocar} we see that
the spherical coefficient $g^{(4)(1B)}_{210}$ 
is proportional to the combination $\gt_{\eff}^{(4)j3j}$ 
of cartesian coefficients in the laboratory frame.
This combination can be converted to 
cartesian coefficients $\gt_{\eff, \rm rot}^{(4)\mu\nu\al}$
in the rotated frame as 
\bea
\gt_{\eff}^{(4)j3j}&=&
\mathcal{R}^{j}{}_K\mathcal{R}^{3}{}_L\mathcal{R}^{j}{}_M ~
\gt_{\eff, \rm rot}^{(4)KLM}
= \mathcal{R}^{3}{}_K ~ \gt_{\eff, \rm rot}^{(4)JKJ}
\nonumber\\
&=& \gt_{\eff, \rm rot}^{(4)JKJ} \hat B^K,
\eea
where the second equality follows from the identity
$\mathcal{R}^{j}{}_K\mathcal{R}^{j}{}_L=\de_{KL}$.

The above discussion shows that the number of factors of $\hat B^J$ 
appearing in a given term contributing to the energy shift
at linear boost order
is determined by the index structure 
of the corresponding coefficient for Lorentz violation. 
To keep the explicit tables appearing in Sec.\ \ref{Applications}
of reasonable size,
we limit attention below 
to Lorentz- and CPT-violating operators of mass dimension $d\leq 8$.
Expressions relating all the corresponding cartesian coefficients 
in the laboratory frame to those in the Sun-centered frame 
at linear boost order are given in Appendix \ref{appSun}. 
Inspection of these results reveals that
the form of the shift $\de\nu$ in a transition frequency for an atom or ion 
takes the generic form
\beq
\de\nu =\sum_{d=3}^8\sum_{s=0}^{5}
\Va{}{d}{J J_1 \ldots J_{s}} 
\hat{B}^{J_1} \ldots \hat{B}^{J_s}\bevec^{J}
\label{genboost}
\eeq
at linear boost order,
where the quantities $\Va{}{d}{J J_1 \ldots J_{s}}$ 
are linear combinations of cartesian coefficients for Lorentz violation
in the Sun-centered frame.
The explicit forms of $\bevec$ and $\Bhat$ in this equation
depend on the choice of laboratory frame. 
We consider here in turn two types of laboratory,
one located on the surface of the Earth
and another on a spacecraft orbiting the Earth.

For a laboratory on the surface of the Earth,
the boost velocity $\bevec$ in Eq.\ \rf{genboost}
can be taken as 
\beq
\bevec=\bevec_\oplus +\bevec_L,
\label{boostpar}
\eeq
where $\bevec_\oplus$ is the instantaneous Earth orbital velocity 
in the Sun-centered frame 
and $\bevec_L$ is the instantaneous tangential velocity 
of the laboratory relative to the Earth's rotation axis.
Approximating the Earth's orbit as circular,
the velocity $\bevec_\oplus$ can be written as 
\beq
\bevec_\oplus=
\be_\oplus \sin{\Om_\oplus T} ~\Xhat
-\be_\oplus \cos{\Om_\oplus T}
( \cos\et~\Yhat + \sin\et~\Zhat ) ,
\label{vorb}
\eeq
where $\be_\oplus\simeq 10^{-4}$ is the Earth's orbital speed,
$\Om_\oplus\simeq 2\pi/(365.26 \text{ d})$ 
is the Earth's orbital angular frequency,
and $\et\simeq23.4^\circ$ is the angle 
between the $XY$ plane and the Earth's orbital plane. 
Also,
treating the Earth as spherical,
the tangential velocity $\bevec_L$ takes the form 
\beq
\bevec_L=
- \be_L \sin{\om_\oplus T_\oplus}~\Xhat
+ \be_L \cos{\om_\oplus T_\oplus}~\Yhat ,
\label{vrot}
\eeq
where $\be_L \approx r_\oplus \om_\oplus \sin{\ch}$ 
is determined by the colatitude $\ch$ of the laboratory 
the radius $r_\oplus$ of the Earth,
and the sidereal frequency $\om_\oplus$.
The unit vector $\Bhat$ in Eq.\ \rf{genboost}
can conveniently be expressed 
in an instantaneous Earth-centered coordinate system 
with axes parallel to those of the Sun-centered frame,
\bea
\Bhat &=&
\sin{\chM}\cos{(\om_\oplus T_\oplus + \phM)}~\Xhat
\nonumber\\ 
&&
+ \sin{\chM}\sin{(\om_\oplus T_\oplus + \phM)}~\Yhat
+ \cos{\chM}~\Zhat,
\label{BhatE}
\eea
where $\chM$ and $\phM$ are the polar and azimuthal angles 
of the magnetic field at $T_\oplus =0$. 

Next,
consider a laboratory located on a space-based platform orbiting the Earth.
Examples include experiments on board the International Space Station (ISS)
such as the Atomic Clock Ensemble in Space (ACES) 
\cite{aces}
and the Quantum Test of the Equivalence Principle and Space Time (QTEST)
\cite{qtest},
or dedicated missions searching for Lorentz violation such as
the Space-Time Explorer and Quantum Equivalence Space Test (STE-QUEST)
\cite{ste-quest}
and the Boost Symmetry Test (BOOST)
\cite{boost}.
We adopt the coordinates depicted in Fig.\ 2 of Ref.\ \cite{space}.
Assuming for definiteness a trajectory with negligible eccentricity,
the parameters for the orbit are
the mean orbital radius $r_s$,
the mean orbital angular speed $\om_s$,
the angle $\ze$ between the satellite orbital axis 
and the Earth's rotation axis,
and the azimuthal angle $\al$ 
between the Earth and satellite orbital planes.
In this scenario,
the boost velocity $\bevec$ in Eq.\ \rf{genboost}
can be written as the vector sum 
\beq
\bevec=\bevec_\oplus+\bevec_s
\label{spacesum}
\eeq
of the Earth's orbital velocity $\bevec_\oplus$ 
in the Sun-centered frame and the satellite velocity $\bevec_s$
relative to an instantaneous Earth-centered frame.
Explicitly,
the components of the satellite velocity $\bevec_s$
in the Sun-centered frame
take the form
\beq
\bevec_s=
\left(
\begin{array}{c}
-\be_s\cos{\al}\sin{\om_s T_s}
-\be_s\cos{\ze}\sin{\al}\cos{\om_s T_s} \\
- \be_s\sin{\al}\sin{\om_s T_s}
+\be_s\cos{\al}\cos{\ze}\cos{\om_s T_s} \\
\be_s\sin{\ze}\cos{\om_s T_s}
\end{array}
\right),
\eeq
where $\be_s=r_s\om_s$
and the local satellite time $T_s$ 
has origin fixed as the satellite 
crosses the equatorial plane on an ascending orbit.
Obtaining also an explicit expression 
for the unit vector $\Bhat$ in Eq.\ \rf{genboost}
requires a further specification 
of the orientation of the space-based laboratory relative to the Earth.
For example,
when this orientation is fixed 
then an instantaneous satellite frame can be defined
with $x$ axis pointing radially towards the Earth 
and $z$ axis aligned along $\bevec_s$.
The components of the corresponding unit spatial vectors 
$\xhat_s$, $\yhat_s$, $\zhat_s$ 
can be expressed in the Sun-centered frame as
\bea
\xhat_s&=&
\left(
\begin{array}{c}
-\cos{\al}\cos{\om_s T_s}
+\cos{\ze}\sin{\al}\sin{\om_s T_s} \\
-\sin{\al}\cos{\om_s T_s}
-\cos{\al}\cos{\ze}\sin{\om_s T_s} \\
-\sin{\ze}\sin{\om_s T_s}
\end{array}
\right),
\nonumber\\
\yhat_s &\equiv&
\dfrac{\bevec_s\times\xhat_s}{\be_s}=
\left(
\begin{array}{c}
\sin{\al}\sin{\ze} \\
-\cos{\al}\sin{\ze}\\
\cos{\ze}
\end{array}
\right) ,
\quad
\zhat_s =
\fr{\bevec_s}{\be_s}.
\qquad
\eea
Using this basis,
the direction $\Bhat$ of the magnetic field 
in the space-based experiment can be expressed as
\beq
\Bhat=
\sin{\th_s}\cos{\ph_s}~\xhat_s
+\sin{\th_s}\sin{\ph_s}~\yhat_s
+\cos{\th_s}~\zhat_s,
\label{spacemag}
\eeq
where $\cos{\th_s}=\bevec_s\cdot\Bhat/\be_s$ 
and $\cos{\ph_s}=\xhat_s\cdot\hat{B}$.

\begin{table*}
\caption{Frequency shift $\de \nu_{c,0}$ at zeroth boost order.} 
\setlength{\tabcolsep}{10pt}
\renewcommand{\arraystretch}{1.8}
\begin{tabular}{cccc}
\hline\hline
$\om_\oplus \TL$ & $\chM$ & $\vev{\pmag^k}$ & Coefficient\\
\hline
1	 & 	1	 & 	 $\vev{\pmag^2}$	 & 	 $-\tfrac{3}{56}\sqrt{\tfrac{5}{\pi}}\sVnrf{p}{220}$\\
1	 & 	 $\cos{2\chM}$ 	 & 	 $\vev{\pmag^2}$	 & 	 $-\tfrac{9}{56}\sqrt{\tfrac{5}{\pi}}\sVnrf{p}{220}$\\
1	 & 	1	 & 	 $\vev{\pmag^4}$	 & 	 $-\tfrac{3}{56}\sqrt{\tfrac{5}{\pi}}\sVnrf{p}{420}+\tfrac{405}{4928\sqrt{\pi}}\sVnrf{p}{440}$\\
1	 & 	 $\cos{2\chM}$ 	 & 	 $\vev{\pmag^4}$	 & 	 $-\tfrac{9}{56}\sqrt{\tfrac{5}{\pi}}\sVnrf{p}{420}+\tfrac{225}{1232\sqrt{\pi}}\sVnrf{p}{440}$\\
1	 & 	 $\cos{4\chM}$ 	 & 	 $\vev{\pmag^4}$	 & 	 $\tfrac{225}{704\sqrt{\pi}}\sVnrf{p}{440}$\\
$\sin{\om_\oplus \TL}$ 	 & 	 $\sin{2\chM}$ 	 & 	 $\vev{\pmag^2}$	 & 	 $\tfrac{3}{14}\sqrt{\tfrac{15}{2\pi}}\,\Im{\sVnrf{p}{221}}$\\
$\sin{\om_\oplus \TL}$ 	 & 	 $\sin{2\chM}$ 	 & 	 $\vev{\pmag^4}$	 & 	 $\tfrac{3}{14}\sqrt{\tfrac{15}{2\pi}}\,\Im{\sVnrf{p}{421}}-\tfrac{45}{616}\sqrt{\tfrac{5}{\pi}}\,\Im{\sVnrf{p}{441}}$\\
$\sin{\om_\oplus \TL}$ 	 & 	 $\sin{4\chM}$ 	 & 	 $\vev{\pmag^4}$	 & 	 $-\tfrac{45}{176}\sqrt{\tfrac{5}{\pi}}\,\Im{\sVnrf{p}{441}}$\\
$\cos{\om_\oplus \TL}$ 	 & 	 $\sin{2\chM}$ 	 & 	 $\vev{\pmag^2}$	 & 	 $-\tfrac{3}{14}\sqrt{\tfrac{15}{2\pi}}\,\Re{\sVnrf{p}{221}}$\\
$\cos{\om_\oplus \TL}$ 	 & 	 $\sin{2\chM}$ 	 & 	 $\vev{\pmag^4}$	 & 	 $-\tfrac{3}{14}\sqrt{\tfrac{15}{2\pi}}\,\Re{\sVnrf{p}{421}}+\tfrac{45}{616}\sqrt{\tfrac{5}{\pi}}\,\Re{\sVnrf{p}{441}}$\\
$\cos{\om_\oplus \TL}$ 	 & 	 $\sin{4\chM}$ 	 & 	 $\vev{\pmag^4}$	 & 	 $\tfrac{45}{176}\sqrt{\tfrac{5}{\pi}}\,\Re{\sVnrf{p}{441}}$\\
$\sin{2\om_\oplus \TL}$ 	 & 	 $1$ 	 & 	 $\vev{\pmag^2}$	 & 	 $\tfrac{3}{28}\sqrt{\tfrac{15}{2\pi}}\,\Im{\sVnrf{p}{222}}$\\
$\sin{2\om_\oplus \TL}$ 	 & 	 $\cos{2\chM}$ 	 & 	 $\vev{\pmag^2}$	 & 	 $-\tfrac{3}{28}\sqrt{\tfrac{15}{2\pi}}\,\Im{\sVnrf{p}{222}}$\\
$\sin{2\om_\oplus \TL}$ 	 & 	 $1$ 	 & 	 $\vev{\pmag^4}$	 & 	 $\tfrac{3}{28}\sqrt{\tfrac{15}{2\pi}}\,\Im{\sVnrf{p}{422}} -\tfrac{135}{1232}\sqrt{\tfrac{5}{2\pi}}\,\Im{\sVnrf{p}{442}}$\\
$\sin{2\om_\oplus \TL}$ 	 & 	 $\cos{2\chM}$ 	 & 	 $\vev{\pmag^4}$	 & 	 $-\tfrac{3}{28}\sqrt{\tfrac{15}{2\pi}}\,\Im{\sVnrf{p}{422}} -\tfrac{45}{308}\sqrt{\tfrac{5}{2\pi}}\,\Im{\sVnrf{p}{442}}$\\
$\sin{2\om_\oplus \TL}$ 	 & 	 $\cos{4\chM}$ 	 & 	 $\vev{\pmag^4}$	 & 	 $\tfrac{45}{176}\sqrt{\tfrac{5}{2\pi}}\,\Im{\sVnrf{p}{442}}$\\
$\cos{2\om_\oplus \TL}$ 	 & 	 $1$ 	 & 	 $\vev{\pmag^2}$	 & 	 $-\tfrac{3}{28}\sqrt{\tfrac{15}{2\pi}}\,\Re{\sVnrf{p}{222}}$\\
$\cos{2\om_\oplus \TL}$ 	 & 	 $\cos{2\chM}$ 	 & 	 $\vev{\pmag^2}$	 & 	 $\tfrac{3}{28}\sqrt{\tfrac{15}{2\pi}}\,\Re{\sVnrf{p}{222}}$\\
$\cos{2\om_\oplus \TL}$ 	 & 	 $1$ 	 & 	 $\vev{\pmag^4}$	 & 	 $-\tfrac{3}{28}\sqrt{\tfrac{15}{2\pi}}\,\Re{\sVnrf{p}{422}} +\tfrac{135}{1232}\sqrt{\tfrac{5}{2\pi}}\,\Re{\sVnrf{p}{442}}$\\
$\cos{2\om_\oplus \TL}$ 	 & 	 $\cos{2\chM}$ 	 & 	 $\vev{\pmag^4}$	 & 	 $\tfrac{3}{28}\sqrt{\tfrac{15}{2\pi}}\,\Re{\sVnrf{p}{422}} +\tfrac{45}{308}\sqrt{\tfrac{5}{2\pi}}\,\Re{\sVnrf{p}{442}}$\\
$\cos{2\om_\oplus \TL}$ 	 & 	 $\cos{4\chM}$ 	 & 	 $\vev{\pmag^4}$	 & 	 $-\tfrac{45}{176}\sqrt{\tfrac{5}{2\pi}}\,\Re{\sVnrf{p}{442}}$\\
$\sin{3\om_\oplus \TL}$ 	 & 	 $\sin{2\chM}$ 	 & 	 $\vev{\pmag^4}$	 & 	 $-\tfrac{45}{88}\sqrt{\tfrac{5}{7\pi}}\,\Im{\sVnrf{p}{443}}$\\
$\sin{3\om_\oplus \TL}$ 	 & 	 $\sin{4\chM}$ 	 & 	 $\vev{\pmag^4}$	 & 	 $\tfrac{45}{176}\sqrt{\tfrac{5}{7\pi}}\,\Im{\sVnrf{p}{443}}$\\
$\cos{3\om_\oplus \TL}$ 	 & 	 $\sin{2\chM}$ 	 & 	 $\vev{\pmag^4}$	 & 	 $\tfrac{45}{88}\sqrt{\tfrac{5}{7\pi}}\,\Re{\sVnrf{p}{443}}$\\
$\cos{3 \om_\oplus \TL}$ 	 & 	 $\sin{4\chM}$ 	 & 	 $\vev{\pmag^4}$	 & 	 $-\tfrac{45}{176}\sqrt{\tfrac{5}{7\pi}}\,\Re{\sVnrf{p}{443}}$\\
$\sin{4\om_\oplus \TL}$ 	 & 	 $1$ 	 & 	 $\vev{\pmag^4}$	 & 	 $-\tfrac{135}{352}\sqrt{\tfrac{5}{14\pi}}\,\Im{\sVnrf{p}{444}}$\\
$\sin{4\om_\oplus \TL}$ 	 & 	 $\cos{2\chM}$ 	 & 	 $\vev{\pmag^4}$	 & 	 $\tfrac{45}{88}\sqrt{\tfrac{5}{14\pi}}\,\Im{\sVnrf{p}{444}}$\\
$\sin{4\om_\oplus \TL}$ 	 & 	 $\cos{4\chM}$ 	 & 	 $\vev{\pmag^4}$	 & 	 $-\tfrac{45}{352}\sqrt{\tfrac{5}{14\pi}}\,\Im{\sVnrf{p}{444}}$\\
$\cos{4\om_\oplus \TL}$ 	 & 	 $1$ 	 & 	 $\vev{\pmag^4}$	 & 	 $ \tfrac{135}{352}\sqrt{\tfrac{5}{14\pi}}\,\Re{\sVnrf{p}{444}}$\\
$\cos{4\om_\oplus \TL}$ 	 & 	 $\cos{2\chM}$ 	 & 	 $\vev{\pmag^4}$	 & 	 $-\tfrac{45}{88}\sqrt{\tfrac{5}{14\pi}}\,\Re{\sVnrf{p}{444}}$\\
$\cos{4\om_\oplus \TL}$ 	 & 	 $\cos{4\chM}$ 	 & 	 $\vev{\pmag^4}$	 & 	 $\tfrac{45}{352}\sqrt{\tfrac{5}{14\pi}}\,\Re{\sVnrf{p}{442}}$\\
\hline\hline
\end{tabular}
\label{csrot}
\end{table*}

\section{Applications}
\label{Applications}

In this section,
we comment on some applications of the formulae derived above.
Many existing searches for Lorentz and CPT violation
are based on the study of transitions
in fountain clocks,
in comagnetometers,
and in trapped ions or lattice clocks.
Each of these experimental approaches is considered in turn.
We present expressions relevant 
to the analysis of data from a variety of experiments,
and we estimate the attainable sensitivities
to coefficients for Lorentz violation 
along with actual constraints from existing data
where possible.

\subsection{Fountain Clocks}
\label{cssec}

Fountain clocks using $^{133}$Cs atoms 
have been widely adopted as primary time and frequency standards.
The standard transition in these clocks,
$\ket{F=3, m_F=0}\leftrightarrow \ket{F=4,m_F=0}$,
is insensitive to the Lorentz- and CPT-violating spectral shifts
discussed above.
This implies that the $^{133}$Cs frequency standard
can be used as a reference 
in experimental studies searching for Lorentz violation,
in parallel with the hydrogen-maser standard
\cite{kv15}.
Searches for violations of Lorentz and CPT symmetry
using a $^{133}$Cs fountain clock can instead be performed 
by studying the frequencies $\nu_{m_F}$ for transitions 
$\ket{F=3, m_F}\leftrightarrow \ket{F=4,m_F}$ with $m_F\ne0$.
These transitions are individually sensitive to the linear Zeeman shift 
and hence their precision is limited by systematic effects.
However,
the systematics can be significantly reduced
by measuring the observable $\nu_c=\nu_{+3}+\nu_{-3}-2\nu_0$ 
\cite{cs06,cs17}. 

The total electronic angular momentum for the states with $m_F=\pm1$ 
is $J=1/2$.
Consulting Table \ref{cond} reveals that 
only electron operators with $j\leq1$ 
can in principle shift the frequencies $\nu_{\pm 3}$. 
However,
the observable $\nu_c$ remains unaffected by these shifts.
To evaluate the nucleon contributions to $\nu_c$
we adopt the Schmidt model as discussed above,
in which the nuclear spin 
$I=7/2$ of $^{133}$Cs is assigned to the unpaired proton. 
With this assumption,
the Lorentz-violating shift $\de\nu_c$ of the observable $\nu_c$ 
is given by
\bea
2\pi \de \nu_c&=&
- \dfrac{3}{14}\sqrt{\dfrac{5}{\pi}}
\Big(\vev{\pmag^2}\Vnrf{p}{220}+\vev{\pmag^4}\Vnrf{p}{420}\Big)
\nonumber\\ 
&&
+ \dfrac{45}{77\sqrt{\pi}}\vev{\pmag^4}\Vnrf{p}{440},
\label{csvc}
\eea
where $\pvec$ is the momentum of the valence proton.
Note that the results presented in Ref.\ \cite{cs06,cs17} 
based on the minimal SME analysis in Ref.\ \cite{kl99} 
can be recovered from the above expression
by excluding contributions from nonminimal Lorentz-violating operators. 
In practice,
this correspond to making the replacements 
\bea
\Vnrf{p}{220}&\rightarrow& 
\dfrac{1}{3m_p}\sqrt{\dfrac{4\pi}{5}}
\Big(3{c_p}_{zz}^{(4)}-{c_p}_{jj}^{(4)}\Big),
\nonumber\\
\Vnrf{p}{420}&\rightarrow& 0, 
\quad
\Vnrf{p}{440}\rightarrow 0 
\eea
in Eq. \rf{csvc}. 

To convert the above expression to the Sun-centered frame,
consider first the frequency shift $\de \nu_{c,0}$
at zeroth boost order.
Applying the transformation rule \rf{ltos} 
for nonrelativistic coefficients to the result \rf{csvc}
yields a somewhat lengthy form for $\de \nu_{c,0}$
in the Sun-centered frame.
The result is presented in tabular form in Table \ref{csrot}.
In each row of the table,
the first entry contains the harmonic dependence
on the sidereal frequency $\om_\oplus$
and the local sidereal time $\TL$.
The second entry describes the dependence 
on the orientation of the magnetic field in the laboratory frame.
The third entry provides the relevant expectation value
of the proton momentum magnitude $|\pvec|$,
while the fourth entry contains the numerical factor
and the coefficient for Lorentz violation. 
To obtain the frequency shift $\de \nu_{c,0}$,
it suffices to multiply the columns and add the rows
in the table.
For example,
the contributions to $\de \nu_{c,0}$ from the first and second rows are
\beq
-\dfrac{3}{56}\sqrt{\dfrac{5}{\pi}}
\vev{\pmag^2}
\sVnrf{p}{220}
- \dfrac{9}{56}\sqrt{\dfrac{5}{\pi}}
\cos{2\chM} 
\vev{\pmag^2}
\sVnrf{p}{220}.
\eeq
Note that 
the corresponding expression for $\de\nu_{c,0}$ in the minimal SME
can be obtained by making the replacements
\bea
\sVnrf{p}{4jm}&\rightarrow& 0, 
\nonumber\\
\sVnrf{p}{220}&\rightarrow& 
-\dfrac{1}{3m_p^2}\sqrt{\dfrac{4\pi}{5}}\widetilde{c}_p{}_Q^{(4)},
\nonumber\\
\Re{\sVnrf{p}{221}}&\rightarrow& 
-\dfrac{1}{m_p^2}\sqrt{\dfrac{2\pi}{15}}\widetilde{c}_p{}^{(4)}_Y, 
\nonumber\\
\Im{\sVnrf{p}{221}}&\rightarrow& 
\dfrac{1}{m_p^2}\sqrt{\dfrac{2\pi}{15}}\widetilde{c}_p{}^{(4)}_X,
\nonumber\\
\Re{\sVnrf{p}{222}}&\rightarrow& 
\dfrac{1}{m_p^2}\sqrt{\dfrac{2\pi}{15}}\widetilde{c}_p{}^{(4)}_-, 
\nonumber\\
\Im{\sVnrf{p}{222}}&\rightarrow&
- \dfrac{1}{m_p^2}\sqrt{\dfrac{2\pi}{15}}\widetilde{c}_p{}^{(4)}_Z
\label{ctosph}
\eea
in the entries in the fourth column of Table \ref{csrot}.

\begin{table}
\caption{The quantities 
$\Va{{\cs133},k}{d}{JJ_1\dots J_s}$ for $5\leq d\leq 8$.} 
\renewcommand\arraystretch{1.7}
\setlength{\tabcolsep}{3pt}
\begin{tabular}{lc}
\hline
\hline
$\Va{{\cs133},k}{d}{JJ_1\dots J_s}$ & Combination\\ 
\hline
$\Va{\cs133,2}{5}{J}$ 
&$\frac{3}{7} \big({a_p}_\eff^{(5)JKK}+2{a_p}_\eff^{(5)JTT}\big)$ \\ 
$\Va{\cs133,2}{5}{J\Jj{1}\Jj{2}}$ 
&$-\frac{9}{7} \big({a_p}_\eff^{(5)J\Jj{1}\Jj{2}}+2\de^{J\Jj{1}}{a_p}_\eff^{(5)TT\Jj{2}}\big)$ \\ 
$\Va{\cs133,2}{6}{J}$ 
&$-\frac{12}{7} \big({c_p}_\eff^{(6)JTKK}+{c_p}_\eff^{(6)JTTT}\big)$ \\ 
$\Va{\cs133,2}{6}{J\Jj{1}\Jj{2}}$ 
&$\frac{36}{7} \big({c_p}_\eff^{(6)JT\Jj{1}\Jj{2}}+\de^{J\Jj{1}}{c_p}_\eff^{(6)TTT\Jj{2}}\big)$ \\ 
$\Va{\cs133,2}{7}{J}$ 
&$\frac{10}{7} \big(3 {a_p}_\eff^{(7)JTTKK}+2{a_p}_\eff^{(7)JTTTT}\big)$ \\   
$\Va{\cs133,2}{7}{J\Jj{1}\Jj{2}}$ 
&$-\frac{30}{7} \big(3{a_p}_\eff^{(7)JTT\Jj{1}\Jj{2}}
+2\de^{J\Jj{1}}{a_p}_\eff^{(7)TTTT\Jj{2}}\big)$ \\ 
$\Va{\cs133,4}{7}{J}$ 
&$\frac{60}{77} \big({a_p}_\eff^{(7)JKKLL}+4{a_p}_\eff^{(7)JTTKK}\big)$ \\ 
$\Va{\cs133,4}{7}{J\Jj{1}\Jj{2}}$ 
&$-\frac{540}{77} \big({a_p}_\eff^{(7)JTT\Jj{1}\Jj{2}}
+\de^{J\Jj{1}}{a_p}_\eff^{(7)TT\Jj{1}KK} \big)$ \\
[-4pt]
&$-\frac{270}{77} {a_p}_\eff^{(7)J\Jj{1}\Jj{2}KK}$\\
$\Va{\cs133,4}{7}{J\Jj{1}\Jj{2}\Jj{3}\Jj{4}}$ 
&$\frac{150}{77}\big({a_p}_\eff^{(7)J\Jj{1}\Jj{2}\Jj{3}\Jj{4}}
+4\de^{J \Jj{1}}{a_p}_\eff^{(7)TT\Jj{2}\Jj{3}\Jj{4}}\big)$ \\
$\Va{\cs133,2}{8}{J}$ 
&$-\frac{30}{7} \big(2 {c_p}_\eff^{(8)JTTTKK}
+{c_p}_\eff^{(8)JTTTTT}\big)$ \\   
$\Va{\cs133,2}{8}{J\Jj{1}\Jj{2}}$ 
&$\frac{90}{7} \big(2{c_p}_\eff^{(8)JTTT\Jj{1}\Jj{2}}
+\de^{J\Jj{1}}{c_p}_\eff^{(8)TTTTT\Jj{2}} \big)$ \\ 
$\Va{\cs133,4}{8}{J}$ 
&$-\frac{360}{77} \big({c_p}_\eff^{(8)JTKKLL}
+2{c_p}_\eff^{(8)JTTTKK}\big)$ \\ 
$\Va{\cs133,4}{8}{J\Jj{1}\Jj{2}}$ 
&$\frac{1620}{77} \big({c_p}_\eff^{(8)JTTT\Jj{1}\Jj{2}}
+\de^{J\Jj{1}}{c_p}_\eff^{(8)TTT\Jj{2}KK}\big)$ \\
[-4pt]
&$+\frac{1620}{77} {c_p}_\eff^{(8)JT\Jj{1}\Jj{2}KK}$\\
$\Va{\cs133,4}{8}{J\Jj{1}\Jj{2}\Jj{3}\Jj{4}}$ 
&$-\frac{900}{77}{c_p}_\eff^{(8)JT\Jj{1}\Jj{2}\Jj{3}\Jj{4}}$\\
[-4pt]
&$-\frac{1800}{77}\de^{J\Jj{1}}{c_p}_\eff^{(7)TTT\Jj{2}\Jj{3}\Jj{4}}$ 
\\\hline\hline
\end{tabular}
\label{csboost}
\end{table}

Next,
consider the contribution to the frequency shift $\de\nu_{c,1}$ 
at linear order in the boost.
Applying the transformation \rf{LTlinear} to Eq. \rf{csvc}
and writing the result in the form \rf{genboost}
yields 
\bea
2\pi \de \nu_{c,1} &=&
\sum_{d}\dfrac{\vev{\pmag^2}}{m_p^{5-d}}
\big(\Va{\cs133,2}{d}{J}\bevec^J
+\Va{\cs133,2}{d}{JKL} \hat{B}^{K} \hat{B}^{L} \bevec^J \big)
\nonumber\\
&&
+\sum_d\dfrac{\vev{\pmag^4}}{m_p^{7-d}}
\big(\Va{\cs133,4}{d}{J}\bevec^J
+\Va{\cs133,4}{d}{JKL} \hat{B}^{K} \hat{B}^{L} \bevec^J\big)
\nonumber\\
&&
+\sum_{d}\dfrac{\vev{\pmag^4}}{m_p^{7-d}}
\Va{\cs133,4}{d}{JKLMN}
\hat{B}^{K} \hat{B}^{L} \hat{B}^{M} \hat{B}^{N} \bevec^J.
\nonumber\\
\label{csboostf}
\eea
Expressions for the quantities 
$\Va{{\cs133},k}{d}{JJ_1\dots J_s}$ 
in terms of the effective cartesian coefficients 
can be deduced from the results presented in Appendix \ref{appSun}
and are displayed in Table \ref{csboost}
for mass dimensions $5\leq d \leq 8$.
In each row of this table,
the first entry lists a specific quantitiy
$\Va{{\cs133},k}{d}{JJ_1\dots J_s}$,
while the second entry gives its expression
as a combination of effective cartesian coefficients
in the Sun-centered frame. 

Note that the minimal-SME limit of the result \rf{csboostf} 
can be obtained by setting to zero 
all the quantities $\Va{{\cs133},k}{d}{JJ_1\dots J_s}$
except for
\beq
\Va{\cs133,2}{4}{J} = -\frac{2}{7} {c_p}_\eff^{(4)TJ},
\quad
\Va{\cs133,2}{4}{JKL}= \frac{6}{7} {c_p}_\eff^{(4)TK}\de^{JL},
\eeq
where the coefficients $c^{(4)\mu \nu}_{\eff}$ 
are defined as the symmetric combination
$\tfrac{1}{2}(c^{(4)\mu\nu}+c^{(4)\nu \mu})$,
as in Ref.\ \cite{km13}.
In contrast to the minimal-SME case,
the nonminimal terms introduce sidereal variations 
incorporating the third, fourth, and fifth harmonics 
of the sidereal frequency.
For example,
the contribution to $\de \nu_{c,1}$
from the fifth harmonic is given by
\begin{widetext}
\bea
2\pi \de \nu_{c,1,5\om_\oplus}&=&
\dfrac{1}{16}\be_L \sin^4{\chM}
\sum_{d}\dfrac{\vev{\pmag^4}}{m_p^{7-d}}
\Big[
\sin{5\om_\oplus T}
\big(10\Va{\cs133,4}{d}{XXXYY}
-\Va{\cs133,4}{d}{XXXXX}
-5\Va{\cs133,4}{d}{XYYYY}\big)
\nonumber\\
&&
\hskip 100pt
+\cos{5\om_\oplus T}
\big(5\Va{\cs133,4}{d}{XXXXY}
-10\Va{\cs133,4}{d}{XXYYY}
+\Va{\cs133,4}{d}{YYYYY}\big)
\Big]
\label{csboos5}
\eea
\end{widetext}
and is suppressed by the boost factor $\be_L$, 
in agreement with the discussion following Eq.\ \rf{rotgen}.

\begin{table}
\caption{Potential sensitivities to coefficients in the Sun-centered frame 
from sidereal and annual variations in a $^{133}$Cs fountain clock.} 
\renewcommand\arraystretch{1.5}
\setlength{\tabcolsep}{6pt}
\begin{tabular}{ll}
\hline
\hline
Coefficient & Sensitivity\\\hline
$\big|\anrf{p}{221}\big|$, $\big|\cnrf{p}{221}\big|$ & $ 10^{-24}$ GeV$^{-1}$\\
$\big|\anrf{p}{421}\big|$, $\big|\cnrf{p}{421}\big|$ & $ 10^{-22}$ GeV$^{-3}$\\
$\big|\anrf{p}{222}\big|$, $\big|\cnrf{p}{222}\big|$ & $ 10^{-24}$ GeV$^{-1}$\\
$\big|\anrf{p}{422}\big|$, $\big|\cnrf{p}{422}\big|$ & $ 10^{-22}$ GeV$^{-3}$ \\
$\big|{c_p}_\eff^{(4)TJ}\big|$ & $10^{-20}$\\
$\big|{a_p}_\eff^{(5)TTJ}\big|$, $\big|{a_p}_\eff^{(5)KKJ}\big|$ & $10^{-20}$ GeV$^{-1}$\\
$\big|{c_p}_\eff^{(6)TTTJ}\big|$, $\big|{c_p}_\eff^{(6)TKKJ}\big|$ & $10^{-21}$ GeV$^{-2}$\\
$\big|{a_p}_\eff^{(7)TTTTJ}\big|$, $\big|{a_p}_\eff^{(7)TTKKJ}\big|$ & $10^{-21}$ GeV$^{-3}$\\
$\big|{a_p}_\eff^{(7)KKLLJ}\big|$ & $10^{-18}$ GeV$^{-3}$\\
$\big|{c_p}_\eff^{(8)TTTTTJ}\big|$, $\big|{c_p}_\eff^{(8)TTTKKJ}\big|$ & $10^{-21}$ GeV$^{-4}$\\
$\big|{c_p}_\eff^{(8)TKKLLJ}\big|$ & $10^{-19}$ GeV$^{-4}$\\\hline\hline
\end{tabular}
\label{csestimates}
\end{table}

Taken together,
the above results permit estimates 
of the potential sensitivity to Lorentz and CPT violation
that is attainable in experiments with $^{133}$Cs fountain clocks
via studies of sidereal and annual variations.
Adopting as a benchmark
the measurements of minimal-SME coefficients
reported in Ref.\ \cite{cs17},
it is reasonable to expect sensitivities
in the Sun-centered frame
of the orders of magnitude listed in Table \ref{csestimates}. 
The first four lines of this table
provides estimated sensitivities 
to the nonrelativistic coefficients $\Vnrf{p}{k2m}$,
while the remainder of the table
concerns the effective cartesian coefficients $\Va{{\cs133},k}{d}{J}$.
For the entries involving the latter,
the uncontracted cartesian spatial index $J$ 
represents any of the possible values $X$, $Y$, $Z$. 
These estimated attainable sensitivities are competitive,
so extracting real constraints from data
would be of definite interest.
 
Further developments of these results are also possible.
Corrections at second boost order 
that are sensitive to isotropic coefficients
in the minimal SME are analyzed in Ref.\ \cite{cs17}.
Generalizing this analysis to the nonminimal sector
is a worthwhile open project.
Another line of reasoning extending the above results
would involve replacing the Schmidt model 
with a more realistic nuclear model
for the $^{133}$Cs nucleus.
In the minimal SME,
this replacement reveals
that neutron Lorentz-violating operators 
with $j=2$ also contribute to the frequency shift,
thereby leading to constraints 
on coefficients for Lorentz violation in the neutron sector
\cite{cs17}.
For the nonrelativistic coefficients,
the neutron-sector corrections
can be incorporated into the expressions given above
via the replacement
\beq
\Vnrf{p}{22m}\rightarrow\Vnrf{p}{22m}+0.021 \Vnrf{n}{22m}.
\eeq
We can therefore estimate the attainable sensitivities 
to these neutron-sector coefficients 
by reducing by about two orders of magnitude
the corresponding proton-sector estimates 
given in Table \ref{csestimates}. 
Note that neutron Lorentz-violating operators 
with $j=2$, $k=4$ 
or with $j=4$ 
may also affect the energy shift,
but this possibility remains unexplored in the literature to date.

Atomic clocks placed on orbiting satellites or other spacecraft
offer qualitatively different experimental opportunities
for studying Lorentz and CPT symmetry.
Since typical space missions involve different clock trajectories
than those relevant to Earth-based laboratories,
they provide access to different combinations 
of coefficients for Lorentz violation 
\cite{space}. 
For example,
the orbital plane of space-based laboratories 
like the International Space Station (ISS)
is inclined relative to the equatorial plane and precesses,
thereby sampling orientation-dependent effects in a unique way.
The satellite orbital velocity 
can also exceed the rotational velocity of the Earth,
which can enhance some signals for Lorentz violation
and can permit faster data accumulation.
For instance,
the period of the ISS is approximately 92 minutes,
so over 15 orbits are completed during a sidereal day.

A $^{133}$Cs cold-atom clock is a component
of the ACES platform on the ISS
\cite{aces}.
The proposed STE-QUEST mission
\cite{ste-quest}
may also involve a $^{133}$Cs clock.
Frequency data obtained from operating a $^{133}$Cs clock 
in the spin-polarized mode
can be converted to the Sun-centered frame
using Eqs.\ \rf{spacesum}-\rf{spacemag} or similar expressions,
leading to bounds on combinations of coefficients for Lorentz violation
that are inaccessible to Earth-based experiments. 
For example,
Table \ref{csrot} shows that the coefficients $\Vnrf{p}{kj0}$ 
produce no sidereal effects in an Earth-based laboratory,
but they can be measured on a space platform.

Fountain clocks using $^{87}$Rb atoms
have been considered as interesting alternatives
for a primary frequency standard
\cite{Rb12,Rb17}
and for studying Lorentz symmetry
via the proposed space-based mission QTEST 
\cite{qtest}. 
A double $^{133}$Cs and $^{87}$Rb fountain clock
has been used to search for Lorentz and CPT violation 
\cite{cs06,cs17}.
Like its $^{133}$Cs and H analogues,
the $^{87}$Rb clock transition 
$\ket{F=1, m_F=0}\leftrightarrow \ket{F=2,m_F=0}$
is insensitive to Lorentz and CPT violation
and can thus be used as a reference in experiments.
However,
the frequencies $\nu^{\rm Rb}_{m_F}$
associated with the transitions 
$\ket{F=1, m_F}\leftrightarrow \ket{F=2,m_F}$ with $m_F\neq 0$
do experience Lorentz- and CPT-violating shifts.
The systematics associated with the linear Zeeman shift
can in this case be reduced by considering
the observable 
$\nu_c^{\rm Rb}=\nu^{\rm Rb}_{+1}+\nu^{\rm Rb}_{-1}-2\nu^{\rm Rb}_0$.
Coefficients in the electron sector leave $\nu_c^{\rm Rb}$ unaffected.
In the context of the Schmidt model
the valence nucleon is a proton with spin $I=3/2$,
so the shift $\de \nu_c^{\rm Rb}$ in the observable $\nu_c^{\rm Rb}$
depends on coefficients for Lorentz violation in the proton sector.
We find
\beq
2\pi \de \nu_c^{\rm Rb}
= -\dfrac{1}{\sqrt{5\pi}} 
\Big(\vev{\pmag^2}\Vnrf{p}{220}+\vev{\pmag^4}\Vnrf{p}{420}\Big).
\label{rbvc}
\eeq
Since the nuclear spin of $^{87}$Rb is smaller than that of $^{133}$Cs,
fewer coefficients appear in Eq.\ \rf{rbvc} than in Eq.\ \rf{csvc}.
All results for $^{133}$Cs fountains 
discussed in the present subsection
can be transcribed to results for $^{87}$Rb fountains
by matching the changes between Eqs.\ \rf{csvc} and \rf{rbvc}.

\subsection{Comagnetometers}
\label{Comagnetometers}

Comagnetometers form another category of sensitive tools 
used for studies of Lorentz and CPT symmetry.
High-sensitivity searches for Lorentz and CPT violation 
in both sidereal and annual variations
have been achieved using $^{129}$Xe-$^3$He comagnetometers 
\cite{xema,xema2,Xe09,Xe14}. 
The experiments compared the angular frequencies
$\om_{\rm Xe}$ and $\om_{\rm He}$ 
of Larmor transitions in the ground states 
of $^{129}$Xe and $^3$He atoms
by measuring the observable
\beq
\om = \om_{\rm He}-\dfrac{\ga_{\rm He}}{\ga_{\rm Xe}} \om_{\rm Xe},
\eeq
which is insensitive to the linear Zeeman shift. 
Here,
$\ga_{\rm Xe}$ is the gyromagnetic ratio 
for the ground state of $^{129}$Xe 
and $\ga_{\rm He}$ is that for the ground state of $^3$He. 

Since the total electronic angular momentum of the noble gases 
in the ground state is $J=0$,
the Larmor transitions are unaffected by 
electron coefficients for Lorentz violation.
The contributions from the nucleon coefficients
can be estimated using the Schmidt model,
in which the nuclear spin $I=1/2$ of each species
is assigned to the unpaired neutron.
The analysis in Sec.\ \ref{Theory}
then yields the Lorentz-violating shift $\de\om$
of the observable $\om$ as 
\bea
\de \om &=&
-\dfrac{1}{\sqrt{3\pi}} \sum_{q=0}^2
\Big(\vev{\pmag^{2q}}_{\rm He}
-\dfrac{\ga_{\rm He}}{\ga_{\rm Xe}} \vev{\pmag^{2q}}_{\rm Xe}\Big)
\nonumber\\
&&
\hspace{50pt}
\times\Big(\TzBnrf{n}{(2q)10}+2\ToBnrf{n}{(2q)10}\Big),
\eea
evaluated in the laboratory frame.
In this expression,
$\vev{\pmag^k}_{\rm He}$ and $\vev{\pmag^k}_{\rm Xe}$ 
are the expectation values of the Schmidt neutron 
in $^3$He and $^{129}$Xe.
These quantities can reasonably be taken 
as roughly the same order of magnitude,
$\vev{\pmag^k}_{\rm He}\sim\vev{\pmag^k}_{\rm Xe}$,
so the shift $\de\om$ can be written as 
\bea
\de \om &=&
\sum_{q=0}^2
\left(\dfrac{\ga_{\rm He}}{\ga_{\rm Xe}} -1\right)
\dfrac{\vev{\pmag^{2q}}}{\sqrt{3\pi}}
\Big(\TzBnrf{n}{(2q)10}+2\ToBnrf{n}{(2q)10}\Big).
\nonumber\\
\label{omXe}
\eea
This result reduces to the minimal-SME expressions 
presented in Refs.\ \cite{xema,xema2,Xe09,Xe14} 
based on the theoretical treatment of Ref.\ \cite{kl99},
by taking the limit
\bea
\TzBnrf{n}{010}+2\ToBnrf{n}{010}
&\rightarrow& 
2\sqrt{3 \pi} ~\widetilde{b}_3^n,
\nonumber\\
\TzBnrf{n}{210}+2\ToBnrf{n}{210}
&\rightarrow &
0,
\nonumber\\
\TzBnrf{n}{410}+2\ToBnrf{n}{410}
&\rightarrow &
0,
\eea
as expected.

\begin{table}
\caption{
Constraints on the moduli of the real and imaginary parts
of neutron nonrelativistic coefficients 
determined from $^{129}$Xe-$^3$He comparisons 
using Eq.\ \rf{hexecons}.}
\renewcommand\arraystretch{1.6}
\setlength{\tabcolsep}{10pt}
\begin{tabular}{cl}
\hline\hline
Coefficient &  Constraint on\\[-6pt]
$\K$ & $|\Re{\K}|, |\Im{\K}|$
\\\hline
$\sHzBnrf{n}{011},\,\sgzBnrf{n}{011}$ & $<4\times 10^{-33}$ GeV\\
$\sHoBnrf{n}{011},\,\sgoBnrf{n}{011}$ & $<2\times 10^{-33}$ GeV\\
$\sHzBnrf{n}{211},\,\sgzBnrf{n}{211}$ & $<4\times 10^{-31}$ GeV$^{-1}$\\
$\sHoBnrf{n}{211},\,\sgoBnrf{n}{211}$ & $<2\times 10^{-31}$ GeV$^{-1}$\\
$\sHzBnrf{n}{411},\,\sgzBnrf{n}{411}$ & $<4\times 10^{-29}$ GeV$^{-3}$\\
$\sHoBnrf{n}{411},\,\sgoBnrf{n}{411}$ & $<2\times 10^{-29}$ GeV$^{-3}$\\
\hline\hline
\end{tabular}
\label{Xelimits}
\end{table}

Conversion of Eq.\ \rf{omXe} to the Sun-centered frame
reveals the time variations in the observable $\om$. 
At zeroth boost order,
the nonminimal terms produce time variations 
at the first harmonic of the sidereal frequency,
which can be explicitly obtained using Eq. \rf{ltos}. 
We can then translate existing bounds 
on the minimal SME coefficients $\widetilde{b}^n_X$ and $\widetilde{b}^n_Y$
obtained from studies of this harmonic 
to constraints on nonminimal coefficients for Lorentz violation.
For this purpose,
it suffices to implement the identifications
\bea
\widetilde{b}_{X}^n
&\rightarrow& 
-\dfrac{1}{\sqrt{6\pi}}\sum_{q=0}^2
\vev{\pmag^{2q}}
\Re\Big[\TzBnrf{n}{(2q)11}+2\ToBnrf{n}{(2q)11}\Big],
\nonumber\\
\widetilde{b}_{Y}^n
&\rightarrow& 
\dfrac{1}{\sqrt{6\pi}}\sum_{q=0}^2
\vev{\pmag^{2q}}
\Im\Big[\TzBnrf{n}{(2q)11}+2\ToBnrf{n}{(2q)11}\Big]
\qquad
\eea
on existing minimal-SME limits.
For example,
the bound on the coefficient $\widetilde{b}^n_\bot$ 
reported in Ref.\ \cite{Xe14}
then yields the constraint 
\bea
\Big|\sum_{q=0}^2\vev{\pmag^{2q}}
\big( \sTzBnrf{n}{(2q)11}+2\sToBnrf{n}{(2q)11}\big)\Big|
&&
\nonumber\\
&& 
\hspace{-80pt}
<3.7\times 10^{-33} ~ {\rm GeV}
\qquad
\label{hexecons}
\eea
at the one sigma level. 
Following the standard procedure in the literature
of taking one coefficient to be nonzero at a time
\cite{tables},
we find the maximal sensitivities to nonrelativistic coefficients
shown in Table \ref{Xelimits}. 
These are the first constraints 
on neutron nonrelativistic coefficients in the literature.
They correspond to substantially greater sensitivities 
to Lorentz and CPT violation than those attained to date 
on electron or proton nonrelativistic coefficients,
and they exceed even the comparatively tight constraints 
on muon nonminimal coefficients 
obtained from laboratory measurements of the muon anomalous magnetic moment
\cite{muon,gkv14}.

\begin{table}
\caption{The quantities $\Tg{{\rm HeXe}, k}{d}{JK}$ for $3\leq d\leq8$.} 
\renewcommand\arraystretch{1.7}
\setlength{\tabcolsep}{8pt}
\begin{tabular}{lc}
\hline
\hline
$\Tg{{\rm HeXe}, k}{d}{JK}$ & Combination \\
\hline
$\Tg{{\rm HeXe}, 0}{3}{JK}$ & $2 \Htf{n}{3}{JK}$ \\ 
$\Tg{{\rm HeXe}, 0}{4}{JK}$ & $4 \gtf{n}{4}{J(TK)}$ \\ 
$\Tg{{\rm HeXe}, 0}{5}{JK}$ & $6 \Htf{n}{5}{J(TTK)}$ \\ 
$\Tg{{\rm HeXe}, 0}{6}{JK}$ & $8 \gtf{n}{6}{J(TTTK)}$ \\ 
$\Tg{{\rm HeXe}, 0}{7}{JK}$ & $10 \Htf{n}{7}{J(TTTTK)}$ \\ 
$\Tg{{\rm HeXe}, 0}{8}{JK}$ & $12 \gtf{n}{8}{J(TTTTTK)}$ \\ 
$\Tg{{\rm HeXe}, 2}{5}{JK}$ 
& $\frac{4}{3} \Htf{n}{5}{TLTL}\de^{JK}+4 \Htf{n}{5}{J(TTK)}$\\
[-4pt] 
&$+2 \Htf{n}{5}{J(LLK)}$ \\
$\Tg{{\rm HeXe}, 2}{6}{JK}$ 
& $2 \gtf{n}{6}{TLTTL}\de^{JK}+8 \gtf{n}{6}{J(TTTK)}$\\
[-4pt] 
&$+8 \gtf{n}{6}{J(TLLK)}$ \\ 
$\Tg{{\rm HeXe}, 2}{7}{JK}$ 
& $\frac{8}{3} \Htf{n}{7}{TLTTTL}\de^{JK}+\frac{40}{3} \Htf{n}{7}{J(TTTTK)}$\\
[-4pt] 
&$+20 \Htf{n}{7}{J(TTLLK)}$ \\
$\Tg{{\rm HeXe}, 2}{8}{JK}$ 
& $\frac{10}{3} \gtf{n}{8}{TLTTTTL}\de^{JK}+20 \gtf{n}{8}{J(TTTTTK)}$\\
[-4pt] 
&$+40 \gtf{n}{8}{J(TTTLLK)}$ \\
$\Tg{{\rm HeXe}, 4}{7}{JK}$ 
& $\frac{8}{5} \Htf{n}{7}{TLTMML}\de^{JK}+8 \Htf{n}{7}{J(TTLLK)}$\\
[-4pt] 
&$+2 \Htf{n}{7}{J(LLMMK)}$ \\
$\Tg{{\rm HeXe}, 4}{8}{JK}$ 
& $4 \gtf{n}{8}{TMTTLLM}\de^{JK}+24 \gtf{n}{8}{J(TTTLLK)}$\\
[-4pt] 
&$+12 \gtf{n}{8}{J(TLLMMK)}$ \\
\hline\hline
\end{tabular}
\label{Xeboost}
\end{table}

\begin{table}
\caption{The quantities $\dC^{(d)k}_*$ 
in terms of $\Tg{{\rm HeXe}, k}{d}{JK}$.}
\renewcommand\arraystretch{1.6}
\setlength{\tabcolsep}{4pt}
\begin{tabular}{lc}
\hline
\hline
$\dC^{(d)k}_*$ & Combination\\
\hline
$\dC_{\oplus}^{(d)k}$ & $\cos{\ph}\,\Tg{{\rm HeXe}, (2q)}{d}{[XY]}+\frac{1}{2}\sin{\ph}(\Tg{{\rm HeXe}, k}{d}{XX}+\Tg{{\rm HeXe}, k}{d}{YY})$\\ 
$\dC_{L}^{(d)k}$ & $\Tg{{\rm HeXe}, k}{d}{ZX}$\\
$\dC_{c\Om}^{(d)k}$ &$-(\cos{\et}\,\Tg{{\rm HeXe}, k}{d}{ZY}+\sin{\et}\,\Tg{{\rm HeXe}, k}{d}{ZZ})$\\
$\dC_{s\Om}^{(d)k}$ & $\Tg{{\rm HeXe}, k}{d}{ZX}$\\
$\dC_{c\om}^{(d)k}$ & $\Tg{{\rm HeXe}, k}{d}{ZY}$ \\
$\dC_{s\om}^{(d)k}$ & $-\Tg{{\rm HeXe}, k}{d}{ZX}$\\
$\dC_{c2\om}^{(d)k}$ & $\cos{\ph}\,\Tg{{\rm HeXe}, k}{d}{(XY)}+\sin{\ph}\,\frac{1}{2}(\Tg{{\rm HeXe}, k}{d}{YY}-\Tg{{\rm HeXe}, k}{d}{XX})$\\
$\dC_{s2\om}^{(d)k}$ & $-\sin{\ph}\,\Tg{{\rm HeXe}, k}{d}{(XY)}-\cos{\ph}\,\frac{1}{2}(\Tg{{\rm HeXe}, k}{d}{YY}-\Tg{{\rm HeXe}, k}{d}{XX})$\\
$\dC_{c\Om s\om}^{(d)k}$ & $\cos{\et}(\sin{\ph}\Tg{{\rm HeXe}, k}{d}{XY}-\cos{\ph}\Tg{{\rm HeXe}, k}{d}{YY})$\\
& $+\sin{\et}(\sin{\ph}\Tg{{\rm HeXe}, k}{d}{XZ}-\cos{\ph}\Tg{{\rm HeXe}, k}{d}{YZ})$ \\
$\dC_{c\Om c\om}^{(d)k}$ & $-\sin{\et}(\cos{\ph}\Tg{{\rm HeXe}, k}{d}{XZ}+\sin{\ph}\Tg{{\rm HeXe}, k}{d}{YZ})$\\
& $-\cos{\et}(\cos{\ph}\Tg{{\rm HeXe}, k}{d}{XY}+\sin{\ph}\Tg{{\rm HeXe}, k}{d}{YY})$\\
$\dC_{s\Om s\om}^{(d)k}$ & $\cos{\ph}\,\Tg{{\rm HeXe}, k}{d}{YX}-\sin{\ph}\,\Tg{{\rm HeXe}, k}{d}{XX}$\\
$\dC_{s\Om c\om}^{(d)k}$ & $\sin{\ph}\,\Tg{{\rm HeXe}, k}{d}{YX}+\cos{\ph}\,\Tg{{\rm HeXe}, k}{d}{XX}$\\\hline\hline
\end{tabular}
\label{Xeboost2}
\end{table}

At linear boost order in the Sun-centered frame,
the Lorentz-violating shift $\de\om_1$ in $\om$
follows the generic structure \rf{genboost}
and can be written as 
\beq
\de\om_1 =
\sum_{d=3}^8\sum_{q=0}^2
\left(\dfrac{\ga_{\rm He}}{\ga_{\rm Xe}} -1\right)
\dfrac{\vev{\pmag^{2q}}}{m_n^{3+2q-d}}
\Tg{{\rm HeXe},(2q)}{d}{JK}
\hat{B}^J \bevec^K .
\label{sunXe}
\eeq
The quantities $\Tg{{\rm HeXe},k}{d}{JK}$ 
are the linear combinations of effective cartesian coefficients
displayed in Table \ref{Xeboost}. 
In this table,
parentheses around indices are understood to represent symmetrization
with a suitable factor,
e.g., 
$\gtf{n}{4}{J(TK)}=
(\gtf{n}{4}{JTK}+\gtf{n}{4}{JKT})/2!$.
Also, 
repeated dummy indices are understood to be summed,
e.g., 
$\Htf{n}{5}{TJTJ}=
\Htf{n}{5}{TXTX}+\Htf{n}{5}{TYTY}+\Htf{n}{5}{TZTZ}$.
The explicit form of the result \rf{sunXe} can be displayed 
by substituting Eqs.\ \rf{boostpar}-\rf{BhatE}
given in Sec.\ \ref{linearbe} 
for the boost velocity of the laboratory 
and for the direction of the magnetic field.
This ensuing expression takes the form 
\bea
\de \om_1 &=&
\be_\oplus\sin{\chM}\dC_{\oplus}+\be_L\cos{\chM}\dC_{L}
\nonumber\\
&&
+\be_\oplus \cos{\chM}\cos{(\Om_\oplus T)}\dC_{c\Om}
\nonumber\\
&&
+\be_\oplus \cos{\chM}\sin{(\Om_\oplus T)}\dC_{s\Om}
\nonumber\\
&&
+\be_L \cos{\chM}\cos{(\om_\oplus T_\oplus)}\dC_{c\om}
\nonumber\\
&&
+\be_L \cos{\chM}\sin{(\om_\oplus T_\oplus)}\dC_{s\om}
\nonumber\\
&&
+\be_\oplus\sin{\chM}\cos{(\om_\oplus T_\oplus)}
\cos{(\Om_\oplus T)}\dC_{c\om c\Om}
\nonumber\\
&&
+\be_\oplus\sin{\chM}\sin{(\om_\oplus T_\oplus)}
\cos{(\Om_\oplus T)}\dC_{s\om c\Om}
\nonumber\\
&&
+\be_\oplus\sin{\chM}\cos{(\om_\oplus T_\oplus)}
\sin{(\Om_\oplus T)}\dC_{c\om s\Om}
\nonumber\\
&&
+\be_\oplus\sin{\chM}\sin{(\om_\oplus T_\oplus)}
\sin{(\Om_\oplus T)}\dC_{s\om s\Om}
\nonumber\\
&&
+\be_L\sin{\chM} \cos{(2\om_\oplus T_\oplus)} \dC_{c2\om}
\nonumber\\
&&
+\be_L\sin{\chM} \sin{(2\om_\oplus T_\oplus)} \dC_{s2\om},
\label{sunXe2}
\eea
where the twelve quantities $\dC_*$ with subscripts $*$ ranging over 
the values $\oplus$, $L$, $c\Om$, $s\Om$, $c\om$, $s\om$, 
$c\om c\Om$, $s\om c\Om$, $c\om s\Om$, $s\om s\Om$, $c2\om, s2\om$
can be decomposed in terms of quantities $\dC^{(d)k}_*$
with fixed values of $k$ and $d$ via the relation 
\bea
\dC_* &=& 
\sum_{d=3}^8\sum_{q=0}^2
\left(\dfrac{\ga_{\rm He}}{\ga_{\rm Xe}} -1\right)
\dfrac{\vev{\pmag^{2q}}}{m_n^{3+2q-d}}
\dC^{(d)(2q)}_*.
\eea
Expressions for the quantities $\dC^{(d)k)}_*$ 
in terms of $\Tg{{\rm HeXe}, 2}{8}{JK}$ 
are given in Table \ref{Xeboost2}.

\begin{table*}
\caption{
Constraints on the moduli of neutron effective cartesian coefficients 
determined from $^{129}$Xe-$^3$He comparisons 
using Eq.\ \rf{hexelim2}.}
\renewcommand\arraystretch{1.5}
\setlength{\tabcolsep}{10pt}
\begin{tabular}{ll|ll}
\hline
\hline
Coefficient & Constraint & Coefficient & Constraint\\ 
\hline
$\Htf{n}{5}{X(TXT)}$ & $< 1\times 10^{-27}$ GeV$^{-1}$  &  $\gtf{n}{6}{X(TXTT)}$ & $< 9\times 10^{-28}$ GeV$^{-2}$ \\
$\Htf{n}{5}{X(TYT)}$ & $< 8\times 10^{-28}$ GeV$^{-1}$  & $\gtf{n}{6}{X(TYTT)}$ & $< 7\times 10^{-28}$ GeV$^{-2}$ \\
$\Htf{n}{5}{X(TZT)}$ & $< 2\times 10^{-27}$ GeV$^{-1}$  & $\gtf{n}{6}{X(TZTT)}$ & $< 2\times 10^{-27}$ GeV$^{-2}$ \\
$\Htf{n}{5}{Y(TXT)}$ & $< 8\times 10^{-28}$ GeV$^{-1}$  & $\gtf{n}{6}{Y(TXTT)}$ & $< 6\times 10^{-28}$ GeV$^{-2}$ \\
$\Htf{n}{5}{Y(TYT)}$ & $< 8\times 10^{-28}$ GeV$^{-1}$  & $\gtf{n}{6}{Y(TYTT)}$ & $< 7\times 10^{-28}$ GeV$^{-2}$ \\
$\Htf{n}{5}{Y(TZT)}$ & $< 2\times 10^{-27}$ GeV$^{-1}$  & $\gtf{n}{6}{Y(TZTT)}$ & $< 2\times 10^{-27}$ GeV$^{-2}$ \\
$\Htf{n}{5}{X(JXJ)}$ & $< 4\times 10^{-25}$ GeV$^{-1}$  & $\gtf{n}{6}{X(JXJT)}$ & $< 9\times 10^{-26}$ GeV$^{-2}$ \\
$\Htf{n}{5}{X(JYJ)}$ & $< 3\times 10^{-25}$ GeV$^{-1}$  & $\gtf{n}{6}{X(JYJT)}$ & $< 7\times 10^{-26}$ GeV$^{-2}$ \\
$\Htf{n}{5}{X(JZJ)}$ & $< 6\times 10^{-25}$ GeV$^{-1}$  & $\gtf{n}{6}{X(JZJT)}$ & $< 2\times 10^{-25}$ GeV$^{-2}$ \\
$\Htf{n}{5}{Y(JXJ)}$ & $< 2\times 10^{-25}$ GeV$^{-1}$  & $\gtf{n}{6}{Y(JXJT)}$ & $< 2\times 10^{-25}$ GeV$^{-2}$ \\
$\Htf{n}{5}{Y(JYJ)}$ & $< 3\times 10^{-25}$ GeV$^{-1}$  & $\gtf{n}{6}{Y(JYJT)}$ & $< 7\times 10^{-26}$ GeV$^{-2}$ \\
$\Htf{n}{5}{Y(JZJ)}$ & $< 6\times 10^{-25}$ GeV$^{-1}$  & $\gtf{n}{6}{Y(JZJT)}$ & $< 2\times 10^{-25}$ GeV$^{-2}$ \\
$\Htf{n}{5}{TJTJ}$   & $< 6\times 10^{-25}$ GeV$^{-1}$ & $\gtf{n}{6}{TJTJT}$    & $< 5\times 10^{-25}$ GeV$^{-2}$ \\
$\Htf{n}{7}{X(TXTTT)}$ & $< 8\times 10^{-28}$ GeV$^{-3}$  &$\gtf{n}{8}{X(TXTTTT)}$ & $< 7\times 10^{-28}$ GeV$^{-4}$ \\
$\Htf{n}{7}{X(TYTTT)}$ & $< 6\times 10^{-28}$ GeV$^{-3}$ &$\gtf{n}{8}{X(TYTTTT)}$  & $< 5\times 10^{-28}$ GeV$^{-4}$\\
$\Htf{n}{7}{X(TZTTT)}$ & $< 2\times 10^{-27}$ GeV$^{-3}$ &$\gtf{n}{8}{X(TZTTTT)}$  & $< 1\times 10^{-27}$ GeV$^{-4}$\\
$\Htf{n}{7}{Y(TXTTT)}$ & $< 6\times 10^{-28}$ GeV$^{-3}$ &$\gtf{n}{8}{Y(TXTTTT)}$  & $< 5\times 10^{-28}$ GeV$^{-4}$\\
$\Htf{n}{7}{Y(TYTTT)}$ & $< 6\times 10^{-28}$ GeV$^{-3}$ &$\gtf{n}{8}{Y(TYTTTT)}$  & $< 5\times 10^{-28}$ GeV$^{-4}$\\
$\Htf{n}{7}{Y(TZTTT)}$ & $< 2\times 10^{-27}$ GeV$^{-3}$ &$\gtf{n}{8}{Y(TZTTTT)}$  & $< 1\times 10^{-27}$ GeV$^{-4}$\\
$\Htf{n}{7}{X(JXJTT)}$ & $< 4\times 10^{-26}$ GeV$^{-3}$ &$\gtf{n}{8}{X(JXJTTT)}$  & $< 2\times 10^{-26}$ GeV$^{-4}$\\
$\Htf{n}{7}{X(JYJTT)}$ & $< 3\times 10^{-26}$ GeV$^{-3}$ &$\gtf{n}{8}{X(JYJTTT)}$  & $< 1\times 10^{-26}$ GeV$^{-4}$\\
$\Htf{n}{7}{X(JZJTT)}$ & $< 7\times 10^{-26}$ GeV$^{-3}$ &$\gtf{n}{8}{X(JZJTTT)}$  & $< 4\times 10^{-26}$ GeV$^{-4}$\\
$\Htf{n}{7}{Y(JXJTT)}$ & $< 3\times 10^{-26}$ GeV$^{-3}$ &$\gtf{n}{8}{Y(JXJTTT)}$  & $< 1\times 10^{-26}$ GeV$^{-4}$\\
$\Htf{n}{7}{Y(JYJTT)}$ & $< 3\times 10^{-26}$ GeV$^{-3}$ &$\gtf{n}{8}{Y(JYJTTT)}$  & $< 1\times 10^{-26}$ GeV$^{-4}$\\
$\Htf{n}{7}{Y(JZJTT)}$ & $< 7\times 10^{-26}$ GeV$^{-3}$ &$\gtf{n}{8}{Y(JZJTTT)}$  & $< 3\times 10^{-26}$ GeV$^{-4}$\\
$\Htf{n}{7}{TJTJTT}$   & $< 2\times 10^{-25}$ GeV$^{-3}$  &$\gtf{n}{8}{TJTJTTT}$   & $< 4\times 10^{-25}$ GeV$^{-4}$ \\
$\Htf{n}{7}{X(JXJKK)}$ & $< 4\times 10^{-23}$ GeV$^{-3}$ &$\gtf{n}{8}{X(JXJTKK)}$  & $< 7\times 10^{-24}$ GeV$^{-4}$\\
$\Htf{n}{7}{X(JYJKK)}$ & $< 3\times 10^{-23}$ GeV$^{-3}$ &$\gtf{n}{8}{X(JYJTKK)}$  & $< 5\times 10^{-24}$ GeV$^{-4}$\\
$\Htf{n}{7}{X(JZJKK)}$ & $< 7\times 10^{-23}$ GeV$^{-3}$ &$\gtf{n}{8}{X(JZJTKK)}$  & $< 1\times 10^{-23}$ GeV$^{-4}$\\
$\Htf{n}{7}{Y(JXJKK)}$ & $< 3\times 10^{-23}$ GeV$^{-3}$ &$\gtf{n}{8}{Y(JXJTKKK)}$ & $< 5\times 10^{-24}$ GeV$^{-4}$\\
$\Htf{n}{7}{Y(JYJKK)}$ & $< 3\times 10^{-23}$ GeV$^{-3}$ &$\gtf{n}{8}{Y(JYJTKK)}$  & $< 5\times 10^{-24}$ GeV$^{-4}$\\
$\Htf{n}{7}{Y(JZJKK)}$ & $< 7\times 10^{-23}$ GeV$^{-3}$ &$\gtf{n}{8}{Y(JZJTKK)}$  & $< 1\times 10^{-23}$ GeV$^{-4}$\\
$\Htf{n}{7}{TJTJKK}$   & $< 6\times 10^{-23}$ GeV$^{-3}$  &$\gtf{n}{8}{TJTJTKK}$   & $< 2\times 10^{-23}$ GeV$^{-4}$ \\\hline\hline
\end{tabular}
\label{Xelimbe}
\end{table*}

An experiment using a dual $^{129}$Xe-$^3$He maser
to study sidereal variations in the observable $\om$
at different times of the year
was performed at the Harvard-Smithsonian Center for Astrophysics
\cite{xema2}. 
In this experiment,
the magnetic field was oriented west to east,
corresponding to $\chM=0^\circ$ and $\ph=90^\circ$.
We can use the bounds on $\de\om_1$ 
reported in Ref.\ \cite{xema2}
to determine limits on nonminimal effective cartesian coefficients
for the neutron.
The published analysis neglected contributions proportional 
to the laboratory velocity $\be_L$ in the Sun-centered frame,
so we can deduce the four bounds
\bea
\dC_{c\om c\Om}&=&
(-3.9\pm3.5)\times10^{-27} ~{\rm GeV},
\nonumber\\
\dC_{c\om s\Om}&=&
(0.7\pm6.3)\times10^{-27} ~{\rm GeV},
\nonumber\\
\dC_{s\om s\Om}&=&
(-6.3\pm6.7)\times10^{-27} ~{\rm GeV},
\nonumber\\
\dC_{s\om c\Om}&=&
(-3.9\pm2.8)\times10^{-27} ~{\rm GeV}.
\label{hexelim2}
\eea
Note that the dependence of the quantities $\dC_*$ on the angle $\ph$
means that these bounds hold only at $\ph=90^\circ$. 
Using the results in Tables \ref{Xeboost} and \ref{Xeboost2}
and the bounds \rf{hexelim2},
we can extract maximal sensitivities 
to many nonminimal effective cartesian coefficients for the neutron.
These constraints are listed in Table \ref{Xelimbe}.
They are the first of their kind reported in the literature for neutrons.

Improvements over the results in Table \ref{Xelimbe}
are within reach of existing experiments.
The sensitivity recently attained in the Heidelberg apparatus 
described in Ref.\ \cite{Xe14} 
represents a gain of about two orders of magnitude,
so sufficient sidereal data accumulated 
at the annual frequency with this apparatus
could in principle better the constraints in Table \ref{Xelimbe}
by a similar factor.
Moreover,
with the sidereal data already in hand,
the time variations at the second harmonic of the sidereal frequency 
appearing in Eq. \rf{sunXe2} 
could in principle be studied 
and would be expected to yield additional measurements of interest.
Although this signal is suppressed by about two orders of magnitude
compared to annual-variation effects,
the greater sensitivity of the Heidelberg apparatus
suggests constraints of the same order of magnitude
as those in Table \ref{Xelimbe} could be obtained.
Note also that
direct measurements of the annual modulation 
would lead to new constraints on SME coefficients,
as sidereal variations are insensitive
to the combinations $\la_{c\Om}$ and $\la_{s\Om}$
even when monitored throughout the year.

Another avenue offering potential improvements
is the adoption of better nuclear models beyond the Schmidt model.
These techniques have already been used 
to show that contributions from proton coefficients to Eq.\ \rf{omXe} 
are significant,
being suppressed only by a factor of about five 
for coefficients with $k=0$ 
\cite{st15}. 
If a similar relationship for coefficients with $k=2,4$
can be demonstrated,
then the constraints on the neutron coefficients 
listed in Tables \ref{Xelimits} and \ref{Xelimbe} 
could be extended to bounds on the corresponding proton coefficients 
by multiplying by a factor of five. 
This would represent a striking gain in sensitivity 
to the proton nonrelativistic coefficients
compared to the existing results obtained
using data from a hydrogen maser 
\cite{kv15}. 

Other comagnetometers can also be used to test Lorentz and CPT symmetry
and may offer sensitivities to additional coefficients.
One potential example is the $^{21}$Ne-Rb-K comagnetometer 
described in Ref.\ \cite{NeRb},
which is designed to extend the reach 
achieved earlier by a $^3$He-K self-compensating comagnetometer 
\cite{HeK}.
The addition of $^{21}$Ne to the system is of particular interest here 
because the nuclear spin of $^{21}$Ne is $I=3/2$ 
and so this comagnetometer 
can access more coefficients for Lorentz violation.
A glance at Table \ref{cond} reveals that there are prospects 
for measuring the coefficients with $j=2$ and $j=3$. 
The underlying physics of this comagnetometer system 
differs significantly from that of the other systems 
discussed in this work,
so the results obtained in Sec.\ \ref{Theory}
cannot be directly applied to estimate sensitivities.
However,
some of the bounds presented in Ref.\ \cite{NeRb} 
can be converted to constraints on nonrelativistic coefficients
for the neutron
by applying the relationship \rf{ctosph}
between the nonrelativistic coefficients and 
the coefficients $c_{\mu\nu}^{(4)}$.
Table \ref{NeRblimits} lists the corresponding 
maximal sensitivites achieved,
which are the first of this kind in the literature.
As before, 
these results can be expected to extend 
to constraints on nonrelativistic coefficients for the proton
because nuclear models beyond the Schmidt model 
are known to allow contributions from proton operators 
to Lorentz-violating expectation values with $j=2$ 
\cite{fl17}.
It is also plausible that a similar situation holds
for coefficients with $j=3$. 
All these interesting issues are open for future investigation. 

\begin{table}
\caption{
Constraints on neutron nonrelativistic coefficients 
determined using data from the $^{21}$Ne-Rb-K comagnetometer.} 
\setlength{\tabcolsep}{10pt}
\begin{tabular}{lc}
\hline
\hline
Coefficient & Constraint \\
\hline
$\Re\anrf{n}{221}$, $\Re\cnrf{n}{221}$ 
& $-(3.3\pm 3.0)\times10^{-29}$ GeV$^{-1}$\\ 
$\Im\anrf{n}{221}$, $\Im\cnrf{n}{221}$ 
& $-(1.9\pm 2.3)\times10^{-29}$ GeV$^{-1}$\\ 
$\Re\anrf{n}{222}$, $\Re\cnrf{n}{222}$ 
& $(1.0\pm 1.2)\times10^{-29}$ GeV$^{-1}$\\ 
$\Im\anrf{n}{222}$, $\Im\cnrf{n}{222}$ 
& $(0.83\pm 0.96)\times10^{-29}$ GeV$^{-1}$\\ 
$\Re\anrf{n}{421}$, $\Re\cnrf{n}{421}$ 
& $-(3.7\pm 3.4)\times10^{-27}$ GeV$^{-3}$\\ 
$\Im\anrf{n}{421}$, $\Im\cnrf{n}{421}$ 
& $-(2.2\pm 2.6)\times10^{-27}$ GeV$^{-3}$\\ 
$\Re\anrf{n}{422}$, $\Re\cnrf{n}{422}$ 
& $(1.1\pm 1.3)\times10^{-27}$ GeV$^{-3}$\\ 
$\Im\anrf{n}{422}$, $\Im\cnrf{n}{422}$ 
& $(0.9\pm 1.1)\times10^{-27}$ GeV$^{-3}$\\
\hline\hline
\end{tabular}
\label{NeRblimits}
\end{table}

\subsection{Trapped ions and lattice clocks}
\label{Trapped ions and lattice clocks}

The stability and accuracy of optical frequency standards 
currently exceeds the performance of fountain clocks. 
It is thus natural to consider the prospects
for testing Lorentz and CPT symmetry using optical transitions.
However, 
sensitivities to many coefficients for Lorentz violation
depend on the absolute uncertainty of the frequency measurement
rather than on its relative precision.
As the absolute uncertainties of fountain clocks
still surpass those of optical clocks,
the advantages of the latter lie primarily in their ability 
to access distinct Lorentz- and CPT-violating effects.
In particular,
optical clocks offer sensitivities 
to coefficients for Lorentz violation in the electron sector
that are unattainable in other clock-comparison experiments.
In this subsection,
we study sensitivities to electron coefficients
in trapped-ion and lattice optical clocks.
Since any signals from proton and neutron coefficients 
are better accessed via other techniques,
we disregard nucleon contributions in what follows.

The transition $^1S_0$-$^3P_0$ is commonly used
in optical frequency standards.
It has been studied with trapped ions,
including $^{27}$Al$^+$ 
\cite{Al27a,AlHgion,Al27b,Al27c} 
and $^{115}$In$^+$ 
\cite{In115a,In115b,In115c},
and also in the context of optical lattice clocks 
based on $^{87}$Sr 
\cite{Sr87a,Sr87b,Sr87c,Sr87d,Sr87e,Sr87f}, 
$^{171}$Yb 
\cite{Ybla,YbSr87,YbSr87Hg,Yblb}, 
and $^{199}$Hg 
\cite{HgSr87,Hgla,Hglb}. 
For this transition,
the total electronic angular momentum of the two states involved is $J=0$,
so only isotropic electron Lorentz-violating operators can contribute. 
The Lorentz-violating shift $\de \nu$ in the transition frequency $\nu$
therefore involves only coefficients with $jm=00$.
In the independent-particle model 
discussed in Sec.\ \ref{Independent-particle model for electrons}, 
the shift in the laboratory frame is given by
\beq
2\pi \de \nu=
-\dfrac{1}{\sqrt{4\pi}}
\left(\De p^2 \Vnrf{e}{200}+\De p^4 \Vnrf{e}{400}\right),
\label{fopt}
\eeq
where $\De p^k$ is the difference 
in the expectation values $\vev{\pmag^k}$ 
of the energy levels involved in the transition. 

Some optical frequency standards involve transitions 
between energy levels with $J\ne0$.
For example, 
the transition $^2S_{1/2}$-$^2D_{5/2}$
is used as a frequency standard in ion-trap clocks 
based on $^{40}$Ca$^+$ 
\cite{Ca40a,CaSr87,Ca40c,Ca40d} 
and $^{88}$Sr$^+$
\cite{Sr88a,Sr88b,Sr88c}.
For these systems,
certain systematic effects can be minimized 
by measuring transitions involving different Zeeman sublevels. 
These techniques typically also eliminate
sensitivity to some Lorentz-violating effects,
as is to be expected given that the coefficients for Lorentz violation
behave in many ways as effective external fields.

One common technique to remove 
the linear Zeeman shift of the clock transition 
is averaging over the Zeeman pair 
$^2S_{1/2,1/2}$-$^2D_{5/2,m_J}$ 
and $^2S_{1/2,-1/2}$-$^2D_{5/2,-m_J}$.
Similarly, 
the electric quadrupole shift can be removed 
by averaging over three different Zeeman pairs. 
Implementing this process eliminates any contributions 
from Lorentz-violating operators with $j\ne0$ 
at linear order in perturbation theory,
due to the identity
\beq
\sum_{m_j=-5/2}^{5/2} \vev{\tfrac52 m_J j0| \tfrac52 m_J}=6 \de_{j0}.
\eeq
As a result,
the Lorentz- and CPT-violating frequency shift $\de\nu$
measurable in these systems is still given by Eq.\ \rf{fopt},
despite the nonzero value of $J$.

Another technique to remove the quadrupole shift 
uses instead two Zeeman pairs to interpolate 
the value of the frequency at $m_J^2={35}/{12}$,
which corresponds to zero quadrupole shift 
because the shift is proportional to $35/12-m^2_J$. 
This method eliminates contributions involving 
coefficients for Lorentz violation with $j=2$,
but it retains contributions with $j=4$. 
In this scenario,
the Lorentz- and CPT-violating shift \rf{fopt}
is replaced by the expression
\bea
2\pi \de \nu&=&
-\dfrac{1}{\sqrt{4\pi}}
\left(\De p^2 \Vnrf{e}{200}+\De p^4 \Vnrf{e}{400}\right)
\nonumber\\
&&
+\dfrac{7}{27\sqrt{\pi}}\vev{\pmag^4}\Vnrf{e}{440},
\label{fcasr}
\eea
where the expectation value $\vev{\pmag^4}$ 
is evaluated in the state $^2$D$_{5/2}$.

The clock transition 
$^2S_{1/2}(F=0)$-$^2D_{3/2}(F=2)$ with $\De m_F=0$ 
in $^{171}$Yb$^+$ 
has also been used as a frequency standard 
\cite{Ybq,Ybqo}. 
In the context of the independent-particle model
described in Sec.\ \ref{Independent-particle model for electrons}, 
the Lorentz- and CPT-violating frequency shift $\de\nu$
for this system is given in the laboratory frame by
\bea
2\pi \de \nu&=&
-\dfrac{1}{\sqrt{4\pi}}
\left(\De p^2 \Vnrf{e}{200}+\De p^4 \Vnrf{e}{400}\right)
\nonumber\\
&&
+\dfrac{1}{2\sqrt{5\pi}}
\left(\vev{\pmag^2}\Vnrf{e}{420}
+\vev{\pmag^4}\Vnrf{e}{420}\right),
\label{fyb}
\eea
where the expectation value $\vev{\pmag^4}$ 
is evaluated in the state $^2D_{3/2}$. 
However,
to suppress the contribution from the electric quadrupole shift,
an averaging of the frequency over three orthogonal directions 
of the magnetic field is performed.
This procedure suppresses the contribution from coefficients with $j=2$.
As a result,
in the limit that the three directions are exactly orthogonal,
the shift \rf{fyb} reduces to the expression \rf{fopt}. 

Other frequency standards are provided by the electric octopole transitions 
in $^{171}$Yb$^+$
\cite{Ybo,Ybqo} 
and $^{199}$Hg$^+$
\cite{Hg199a,Hg199b}. 
The clock transition used in these systems 
is the transition $\De m_F=0$,
which is insensitive to $B$-type coefficients for Lorentz violation. 
As before,
the contribution to the Lorentz- and CPT-violating frequency shift $\de\nu$
arising from coefficients with $j=0$ is given by Eq.\ \rf{fopt}.
The contribution from coefficients with $j=2$ 
is again eliminated by the averaging procedure
over three different directions of the magnetic field,
which is designed to cancel the electric quadrupole shift.
It is conceivable that coefficients with $j=4$ contribute
to the frequency shift,
but establishing this lies outside our present scope.

The coefficients in the above expressions are in the laboratory frame
and hence may vary with time.
In converting to the Sun-centered frame,
the isotropic frequency shift \rf{fopt} receives contributions 
that depend on the boost velocity of the laboratory frame. 
At linear boost order,
we find that the shift $\de\nu_1$ is given by 
\bea
2\pi \de \nu_1 &=&
-\sum_{d,k}\dfrac{\De p^k}{\sqrt{4\pi}} 
\Big[
\be_\oplus \sin\Om_\oplus T~\Va{e,k}{d}{X}
\nonumber\\
&&
\hskip 10pt
-\be_\oplus \cos\Om_\oplus T 
(\cos\et ~\Va{e,k}{d}{Y} + \sin\et ~\Va{e,k}{d}{Z} )
\nonumber\\
&&
\hskip 10pt
+\be_L
(\cos\om_\oplus T_\oplus ~\Va{e,k}{d}{Y}
- \sin\om_\oplus T _\oplus ~\Va{e,k}{d}{X})
\Big],
\nn\\
\label{optsun}
\eea 
where expressions for the quantities $\Va{e,k}{d}{J}$ 
in terms of effective cartesian coefficients 
are given in Table \ref{Vopt}. 

\begin{table}
\caption{The quantities $\Va{e,k}{d}{J}$ for $5\leq d\leq8$.}
\setlength{\tabcolsep}{10pt}
\begin{tabular}{lc}
\hline\hline
$\Va{e,k}{d}{J}$ & Combination\\
\hline
$\Va{e,2}{5}{J}$ 	& 	$-2a^{(5)JTT}_\eff-a^{(5)JKK}_\eff$	\\
$\Va{e,2}{6}{J}$ 	& 	$4c^{(6)JTTT}_\eff+4c^{(6)JTKK}_\eff$	\\
$\Va{e,2}{7}{J}$ 	& 	$-\frac{10}{3}(2a^{(7)JTTTT}_\eff+3a^{(7)JTTKK}_\eff)$	\\
$\Va{e,2}{8}{J}$ 	& 	$10c^{(8)JTTTTT}_\eff+20c^{(8)JTTTKK}_\eff$	\\
$\Va{e,4}{7}{J}$ 	& 	$-a^{(7)JKKLL}_\eff-4 a^{(7)JTTKK}_\eff$	\\
$\Va{e,4}{8}{J}$ 	& 	$6 c^{(8)JTKKLL}_\eff+12 c^{(8)JTTTKK}_\eff$	\\
\hline\hline
\end{tabular}
\label{Vopt}
\end{table}

The result \rf{optsun} predicts annual and sidereal variations 
of the transition frequency,
which can in principle be detected by comparision to a reference. 
Since optical clocks can outperform other frequency standards,
an effective way to search for the effects predicted by Eq.\ \rf{optsun} 
is to compare two optical clocks
and search for a sidereal or annual modulation 
of their frequency difference.
For systems with long-term stability,
studying annual variations is preferable 
because the speed $\be_\oplus$ 
is typically about two orders of magnitude bigger than $\be_L$. 
Note also that the two clocks can be located in different laboratories:
Using Eq.\ \rf{optsun},
we see that the annual and sidereal modulations 
of the frequency difference between clocks A and B
are given by 
\bea
2\pi \de \nu_{\rm AB} &=&
\sum_{d,k}\dfrac{\De p^k_{\rm B}-\De p^k_{\rm A}}{\sqrt{4\pi}} 
\Big[
\be_\oplus \sin\Om_\oplus T~\Va{e,k}{d}{X}
\nonumber\\
&&
\hskip 10pt
-\be_\oplus \cos\Om_\oplus T 
(\cos\et ~\Va{e,k}{d}{Y} + \sin\et ~\Va{e,k}{d}{Z} )
\Big]
\nonumber\\
&&
\hskip -40pt
+ \sum_{d,k}
\dfrac{\De p^k_{\rm B}\be_{L,{\rm B}}
-\De p^k_{\rm A}\be_{L,{\rm A}}}{\sqrt{4\pi}} 
( \cos\om_\oplus T_\oplus ~\Va{e,k}{d}{Y}
\nonumber\\
&&
\hskip 80pt
- \sin\om_\oplus T _\oplus ~\Va{e,k}{d}{X}),
\label{compopt}
\eea 
where $\De p^k_{\rm A}$ is the expectation value $\vev{\pmag^k}$ 
for the transitions in clock A,
$\be_{L,{\rm A}}$ is the speed of the laboratory containing clock A, 
and $\De p^k_{\rm B}$, $\be_{L,{\rm B}}$ 
are defined similarly for clock B.

Several laboratories have the potential to compare two clocks
at the same location,
searching for the effects predicted in Eq.\ \rf{compopt}
in a scenario with $\be_{L,{\rm A}}=\be_{L,{\rm B}}$.
For example,
scanning the literature cited above 
suggests that comparisons of any two lattice clocks based 
on $^{87}$Sr, $^{171}$Yb, or $^{199}$Hg 
could in principle be performed at Rikagaku Kenky\= usho (RIKEN) in Japan.
Similarly,
$^{87}$Sr and $^{199}$Hg lattice clocks can be compared
at the Syst\`eme de R\'ef\'erence Temps-Espace (SYRTE) in France,
ones based on $^{87}$Sr and $^{171}$Yb 
can be compared at the National Metrology Institute of Japan (NMIJ),
and the $^{27}$Al$^+$ ion clock could be compared 
to the $^{171}$Yb lattice clock at 
the National Institute of Standards and Technology (NIST) in the United States.
Many individual comparisons between clocks located at 
different institutions are also possible in principle,
by using Eq.\ \rf{compopt} with $\be_{L,{\rm A}}\neq \be_{L,{\rm B}}$.
Moreover,
some Lorentz- and CPT-violating effects that are absent in Eq.\ \rf{compopt}
and hence cannot be studied with any of these clock combinations
might become accessible 
given suitable care for the treatment of systematics
and its implication for cancellations of signals.
Some examples of such experiments with clocks at a single location
might include comparison of the 
the $^{88}$Sr$^+$ and $^{171}$Yb$^+$ ion clocks
at the National Physical Laboratory (NPL) in England,
the $^{87}$Sr lattice clock and the $^{171}$Yb$^+$ ion clock
at the Physikalisch-Technische Bundesanstalt (PTB) in Germany,
or the $^{27}$Al$^+$ and $^{199}$Hg$^+$ ion clocks at NIST.
 
A qualitatively different approach to testing Lorentz and CPT symmetry
is to create an entangled state and monitor its time evolution.
In Ref.\ \cite{Pr15},
the entangled state combines the states
$(\ket{\pm 5/2}\ket{\mp 5/2}+\ket{\pm 1/2}\ket{\mp 1/2})/\sqrt{2}$
of two $^{40}$Ca$^{+}$ ions, 
where the kets $\ket{m_F}$ represent the $m_F$ Zeeman level 
of the energy state $^2D_{5/2}$. 
The experimental observable $\ol f$ is obtained 
by averaging the energy difference between the product states 
$\ket{\pm 5/2}\ket{\mp 5/2}$ and $\ket{\pm 1/2}\ket{\mp 1/2}$.

Following the approach in Sec.\ \ref{Independent-particle model for electrons}, 
we assign angular momenta $J=5/2$ and $L=2$ to the valence electron. 
The Lorentz- and CPT-violating shift $\de \ol{f}$ 
of the observable $\ol{f}$ in the laboratory frame is found to be 
\bea
2\pi \de \ol{f}&=&
\dfrac{18}{7\sqrt{5\pi}}
\left(\vev{\pmag^2}\Vnrf{e}{220}+\vev{\pmag^4}\Vnrf{e}{420}\right)
\nonumber\\ 
&&
+\dfrac{1}{7\sqrt{\pi}}\vev{\pmag^4}\Vnrf{p}{440}.
\label{cafbar}
\eea
This expression has a structure similar to that 
of the frequency shift \rf{csvc} in fountain clocks,
so we can adapt the results presented in Sec.\ \ref{cssec} 
to convert the expression \rf{cafbar} to the Sun-centered frame. 
The expression for the shift $\de \ol{f}_0$ 
at zeroth boost order is therefore given by Table \ref{csrot} 
with the replacements
\bea
\Vnrf{p}{k2m}\rightarrow -\dfrac{12}{5}\Vnrf{e}{k2m},
\quad
\Vnrf{p}{k4m}\rightarrow \dfrac{11}{45}\Vnrf{e}{k4m}.
\label{Carepla}
\eea
At linear boost order, 
the contribution $\de \ol{f}_1$ is
\bea
2\pi \de \ol{f}_1 &=&
\sum_{d}\dfrac{\vev{\pmag^2}}{m_p^{5-d}}
\big(\Va{\caz,2}{d}{J}\bevec^J
+\Va{\caz,2}{d}{JKL} \hat{B}^{K} \hat{B}^{L} \bevec^J \big) 
\nonumber\\
&&
\hskip -2pt
+\sum_d\dfrac{\vev{\pmag^4}}{m_p^{7-d}}
\big(\Va{\caz,4}{d}{J}\bevec^J
+\Va{\caz,4}{d}{JKL} \hat{B}^{K} \hat{B}^{L} \bevec^J\big)
\nonumber\\
&&
\hskip -2pt
+\sum_{d}\dfrac{\vev{\pmag^4}}{m_p^{7-d}}\Va{\caz,4}{d}{JKLMN}
\hat{B}^{K} \hat{B}^{L} \hat{B}^{M} \hat{B}^{N} \bevec^J,
\qquad
\label{cazboostf}
\eea
where expressions for the quantities $\Va{{\caz},k}{d}{JJ_1\dots J_s}$ 
in terms of effective cartesian coefficients 
are displayed in Table \ref{cazboost}.

\begin{table}
\caption{The quantities 
$\Va{{\caz},k}{d}{JJ_1\dots J_s}$ for $5\leq d\leq 8$.} 
\renewcommand\arraystretch{1.6}
\setlength{\tabcolsep}{0pt}
\begin{tabular}{lc}
\hline\hline
$\Va{{\caz},k}{d}{JJ_1\dots J_s}$ & Combination\\
\hline
$\Va{\caz,2}{5}{J}$ & $-\frac{36}{35} \big({a_e}_\eff^{(5)JKK}+2{a_e}_\eff^{(5)JTT}\big)$ \\ 
$\Va{\caz,2}{5}{J\Jj{1}\Jj{2}}$ & $\frac{108}{35} \big({a_e}_\eff^{(5)J\Jj{1}\Jj{2}}+2\de^{J\Jj{1}}{a_e}_\eff^{(5)TT\Jj{2}}\big)$ \\ 
$\Va{\caz,2}{6}{J}$ & $\frac{144}{35} \big({c_e}_\eff^{(6)JTKK}+{c_e}_\eff^{(6)JTTT}\big)$ \\ 
$\Va{\caz,2}{6}{J\Jj{1}\Jj{2}}$ & $-\frac{432}{35} \big({c_e}_\eff^{(6)JT\Jj{1}\Jj{2}}+ \de^{J\Jj{1}}{c_e}_\eff^{(6)TTT\Jj{2}}\big)$ \\ 
$\Va{\caz,2}{7}{J}$ & $-\frac{24}{7} \big(3 {a_e}_\eff^{(7)JTTKK}+2{a_e}_\eff^{(7)JTTTT}\big)$ \\   
$\Va{\caz,2}{7}{JKL}$ & $\frac{72}{7} \big(3{a_p}_\eff^{(7)JTT\Jj{1}\Jj{2}}+2\de^{J\Jj{1}}{a_p}_\eff^{(7)TTTT\Jj{2}}\big)$ \\ 
$\Va{\caz,4}{7}{J}$ & $-\frac{10}{7} \big({a_e}_\eff^{(7)JKKLL}+4{a_e}_\eff^{(7)JTTKK}\big)$ \\ 
$\Va{\caz,4}{7}{J\Jj{1}\Jj{2}}$ & $8\big({a_e}_\eff^{(7)JTT\Jj{1}\Jj{2}}+\de^{J\Jj{1}}{a_e}_\eff^{(7)TT\Jj{2}KK}\big)$ \\[-3pt]
& $+4 {a_e}_\eff^{(7)J\Jj{1}\Jj{2}KK}$\\
$\Va{\caz,4}{7}{J\Jj{1}\Jj{2}\Jj{3}\Jj{4}}$ & $\frac{10}{21}\big({a_e}_\eff^{(7)J\Jj{1}\Jj{2}\Jj{3}\Jj{4}}+4\de^{J\Jj{1}}{a_e}_\eff^{(7)TT\Jj{2}\Jj{3}\Jj{4}}\big)$ \\
$\Va{\caz,2}{8}{J}$ & $\frac{72}{7} \big(2 {c_e}_\eff^{(8)JTTTKK}+{c_e}_\eff^{(8)JTTTTT}\big)$ \\   
$\Va{\caz,2}{8}{J\Jj{1}\Jj{2}}$ & $-\frac{216}{7} \big(2{c_e}_\eff^{(8)JTTT\Jj{1}\Jj{2}}+\de^{J\Jj{1}}{c_e}_\eff^{(8)TTTTT\Jj{2}}\big)$ \\ 
$\Va{\caz,4}{8}{J}$ & $\frac{60}{7} \big({c_e}_\eff^{(8)JTKKLL}+2{c_e}_\eff^{(8)JTTTKK}\big)$ \\ 
$\Va{\caz,4}{8}{J\Jj{1}\Jj{2}}$ & $-24 \big({c_e}_\eff^{(8)JTTT\Jj{1}\Jj{2}}+\de^{J\Jj{1}}{c_e}_\eff^{(8)TTT\Jj{2}KK}\big)$ \\[-3pt]
& $-24\, {c_e}_\eff^{(8)JT\Jj{1}\Jj{2}KK}$\\
\hskip-4pt
$\Va{\caz,4}{8}{J\Jj{1}\Jj{2}\Jj{3}\Jj{4}}$ & $-\frac{20}{7}
\big({c_e}_\eff^{(8)JT\Jj{1}\Jj{2}\Jj{3}\Jj{4}}
-2\de^{J\Jj{1}}{c_e}_\eff^{(7)TTT\Jj{2}\Jj{3}\Jj{4}}\big)$\\
\hline\hline
\end{tabular}
\label{cazboost}
\end{table}

For the nonminimal terms considered in this work,
the result \rf{cazboostf}
incorporates time variation at the first five harmonics 
of the sidereal frequency
along with annual variations. 
At the sidereal frequency,
the dominant contributions to the variations in the first four harmonics 
are given by Table \ref{csrot} with the substitutions \rf{Carepla}. 
The variation at the fifth harmonic is suppressed by $\be_L$,
and it is given by Eq.\ \rf{csboos5} with the replacement 
$\Va{\cs133,k}{d}{J\dots K}\rightarrow \Va{\caz,k}{d}{J\dots K}$.
Using these results,
we can estimate the sensitivities of the $^{40}$Ca$^{+}$ experiment 
\cite{Pr15}
to the nonrelativistic coefficients.
Table \ref{Calimits} displays these sensitivities. 
In deriving them,
we take $\vev{\pmag^2}\sim 10^{-11}$ GeV$^2$ 
and $\vev{\pmag^4}\sim 10^{-22}$~GeV$^4$. 
We also suppose the experimental reach is 0.03 Hz. 
With sufficient stability and data collection over a long time period,
constraints could also be placed on coefficients for Lorentz violation
associated with the annual variation signal predicted by Eq.\ \rf{cazboostf}. 

\begin{table}
\caption{Potential sensitivities to coefficients in the Sun-centered frame 
from sidereal variations in entangled $^{40}$Ca$^+$ ions.} 
\setlength{\tabcolsep}{14pt}
\begin{tabular}{ll}
\hline\hline
Coefficient & Sensitivity \\
\hline
$|\anrf{e}{22m}|$, $|\cnrf{e}{22m}|$ & $10^{-14}$ GeV$^{-1}$\\
$|\anrf{e}{42m}|$, $|\cnrf{e}{42m}|$ & $10^{-3}$ GeV$^{-3}$\\
$|\anrf{e}{44m}|$, $|\cnrf{e}{44m}|$ & $10^{-3}$ GeV$^{-3}$\\
\hline\hline
\end{tabular}
\label{Calimits}
\end{table}

Related experiments have been proposed 
using Zeeman transitions of the $F_{7/2}$ state in Yb$^+$
\cite{Dz16}.
Experiments using a dynamical decoupling technique
have also been proposed for a broad class
of trapped ions and lattice clocks
\cite{sh17}.
All are expected to permit significant improvements
over existing constraints on the coefficients $c_{\mu\nu}^{(4)}$.
The energy shift produced by these minimal-SME coefficients
can be generalized to incorporate also
the contributions from the nonminimal coefficients $\Vnrf{e}{22m}$
in the nonrelativistic limit.
The proposed experiments can therefore be expected 
to yield substantial improvements over the estimated sensitivities 
to the coefficients $\anrf{e}{22m}$ and $\cnrf{e}{22m}$ 
given in Table \ref{Calimits}.

\subsection{Antimatter clocks}
\label{Antimatter clocks}

In this final subsection on applications,
we offer some comments about the prospects
for spectroscopic experiments using antimatter.
Comparisons of the properties of matter and antimatter
are of particular interest 
for testing the CPT symmetry of quantum field theory.
The line of reasoning outlined in the introduction reveals that
effective field theory also provides the general model-independent framework 
for analysing antimatter systems,
so the results of experiments testing CPT symmetry
can be expressed in a model-independent way
as constraints on SME coefficients.

A diverse set of such constraints has already been obtained
via precision spectroscopy of positrons and antiprotons 
confined in a Penning trap
\cite{bkr97,de99,ga99,dk16,sm17,na17}. 
Studying antimatter instead of antiparticles
offers advantages in searches for CPT violation
\cite{bkr99,kv15},
and several collaborations are developing techniques
for the precision spectroscopy of antihydrogen.
Recently, 
the Antihydrogen Laser Physics Apparatus 
(ALPHA)
collaboration has measured 
the antihydrogen ground-state hyperfine transitions 
\cite{ah17} 
and the 1$S$-2$S$ transition 
\cite{ah18},
heralding an era of precision antimatter spectroscopy.
Other collaborations pursuing this goal
include
the Atomic Spectroscopy and Collisions Using Slow Antiprotons 
(ASACUSA) 
collaboration
\cite{asacusa1,asacusa2},
and the Antihydrogen Trap 
(ATRAP)
collaboration
\cite{ga02}.
Experiments investigating the gravitational response of antihydrogen
are also being developed,
including
the Antihydrogen Experiment: Gravity, Interferometry, Spectroscopy
(AEGIS)
collaboration
\cite{aegis},
the ALPHA collaboration
\cite{alphagrav},
and the Gravitational Behavior of Antihydrogen at Rest
(GBAR)
collaboration
\cite{gbar},
and the corresponding techniques may also enhance 
future spectroscopic studies of antihydrogen.

One signal for nonzero CPT violation would be a measured difference 
$\De \nub_{1S2S} \equiv \nu_{1S2S} - \nub_{1S2S}$
between the resonance frequency $\nu_{1S2S}$ 
of the 1$S$-2$S$ transition in hydrogen 
and the analogous resonance frequency $\nub_{1S2S}$ in antihydrogen. 
Performing a general analysis in the context of effective field theory
\cite{kv15}
reveals that CPT-violating effects
contributing to a nonzero value of $\De \nub_{1S2S}$
can be classified as spin independent or spin dependent
and as isotropic or anisotropic, 
and they can exhibit time variations induced
by the noninertial nature of the experimental laboratory.
It turns out that the spin-dependent effects are more readily studied
using ground-state hyperfine transitions,
while the time variations are better explored
by directly studying modulations of $\nub_{1S2S}$.
However,
the difference $\De \nub_{1S2S}$
is particularly sensitive to isotropic, spin-independent, 
and time-constant CPT violation
controlled by the coefficients
$\anrfc{e,2}$, $\anrfc{e,4}$, $\anrfc{p,2}$, $\anrfc{p,4}$.
An explicit expression for $\De \nub_{1S2S}$ 
in terms of these coefficients
is given by Eq.\ (86) of Ref.\ \cite{kv15},
with the correction $8\to 16$ in the denominator.
Note that these nonrelativistic coefficients
incorporate effects from CPT-violating operators of arbitrary mass dimension
\cite{km13}.

Based on an analysis that assumes
no spin-, geometry-, or time-dependent CPT violation,
the ALPHA collaboration reported agreement 
between the 1$S$-2$S$ resonance frequencies of hydrogen and antihydrogen 
at a precision of $2\times10^{-12}$ 
\cite{ah18}.
We can therefore deduce the constraint 
\beq
\Big| \anrfc{e,2}+ \anrfc{p,2}
+\tfrac{67}{12} (\al m_r)^2 (\anrfc{e,4} + \anrfc{p,4}) \Big|
< 1\times10^{-9} ~{\rm GeV}^{-1},
\label{bhbar}
\eeq
where $\al$ is the fine-structure constant
and $m_r$ is the reduced mass of hydrogen.
The result \rf{bhbar} represents the first constraint
on SME coefficients extracted from antihydrogen spectroscopy.
Table \ref{Hbarlimits} lists the corresponding maximal sensitivities 
obtained by taking each coefficient to be nonzero in turn,
following the standard procedure in the literature
\cite{tables}.
Note that several factors currently limit
the precision of the measurement of $\nub_{1S2S}$,
including the comparatively smaller number and higher temperature
of atoms in antihydrogen experiments relative to hydrogen ones.
However,
there is every reason to expect improvements in the future.
One proposal along these lines is to trap the ultracold antiatoms 
from the GBAR antihydrogen beam in an optical lattice 
\cite{pa17},
which could enable measurements of the 1$S$-2$S$ transition in antihydrogen 
at a level approaching the precision of $4.2\times 10^{-15}$ 
already obtained with hydrogen 
\cite{pa11}. 

\begin{table}
\caption{Constraints on electron and proton nonrelativistic coefficients
determined from 1$S$-2$S$ hydrogen and antihydrogen spectroscopy.}
\setlength{\tabcolsep}{14pt}
\begin{tabular}{ll}
\hline\hline
Coefficient & Constraint\\
\hline
$|\anrfc{e,2}|$, $|\anrfc{p,2}|$ & $1\times10^{-9}$ GeV$^{-1}$\\
$|\anrfc{e,4}|$, $|\anrfc{p,4}|$ & $14$ GeV$^{-3}$\\
\hline\hline
\end{tabular}
\label{Hbarlimits}
\end{table}

Other signals for CPT violation can appear
in comparisons of the hyperfine structure
of hydrogen and antihydrogen
\cite{bkr99,kv15}.
High-precision measurements of the hyperfine transition of hydrogen 
can be obtained using a hydrogen maser 
\cite{ra90}, 
but these methods are impractical for antihydrogen due, 
for example, 
to collisions with the walls in the maser bulb.
One different approach already realized by the ALPHA collaboration
is to perform hyperfine spectroscopy on trapped antihydrogen
\cite{ah17}. 
An alternative option being pursued by the ASACUSA collaboration 
involves using instead an antihydrogen beam 
\cite{asacusa2}.
Testing the latter method with hydrogen has demonstrated 
a precision only three orders of magnitude below 
that achieved via the hydrogen maser. 
The prospects are excellent for further substantial improvements 
in hyperfine spectroscopy using advanced tools 
such as ultracold antihydrogen beams,
and perhaps ultimately adapting techniques 
similar to those used for atomic fountain clocks. 

In the longer term,
antiatom spectroscopy could conceivably evolve 
to include also experiments with heavier antiatoms.
The simplest candidate system is antideuterium,
which has the antideuteron as its nucleus.
Unlike antiprotonic deuterium,
antideuterium is expected to be stable
and is therefore in principle a candidate for precision spectroscopy
and hence for CPT tests.
Deuterium spectroscopy is known 
to be many orders of magnitude more sensitive than hydrogen spectroscopy 
to certain kinds of Lorentz and CPT violation
\cite{kv15},
and the same arguments hold for
the comparative sensitivities of antideuterium and antihydrogen spectroscopy.
The production of a single heavier antiion is also of real interest,
as it could in principle be confined in an ion trap
and repeatedly interrogated to perform high-precision spectroscopy.

Whatever the future of antimatter experiments
with heavier systems than antihydrogen,
the theoretical treatments presented in Section \ref{Theory}
and in Ref.\ \cite{kv15}
can readily be adapted to antiatoms and antiions.
In particular,
the expression for the shift in an energy level of an antiatom or antiion
can be obtained from the corresponding expression for an atom or ion
by implementing the substitutions
\bea
\anrf{w}{jkm}\rightarrow -\anrf{w}{jkm},
&& 
\cnrf{w}{jkm}\rightarrow \cnrf{w}{jkm},
\nonumber\\
{H_{w}}^{\nr(sB)}_{jkm}\rightarrow - {H_{w}}^{\nr(sB)}_{jkm},
&& 
{g_{w}}^{\nr(sB)}_{jkm}\rightarrow {g_{w}}^{\nr(sB)}_{jkm}
\qquad
\eea 
for the SME coefficients.
For example,
an expression for the frequency shift
$\de\nu_{\ol{\rm D}}$ of the $nL$-$n'L'$ transition in antideuterium 
due to isotropic Lorentz and CPT violation 
can be obtained from the corresponding expression for the shift 
$\de\nu_{\rm D}$ in deuterium given as Eq.\ (103) of Ref.\ \cite{kv15},
yielding the result
\bea
2\pi \de \nu_{\ol{\rm D}} &=&
\fr \mr {\sqrt{\pi}}
(\ve_{n'}-\ve_{n}) 
\bigg[
\bVnrf{e}{200} 
+ \frac 14 
\left(
\bVnrf{p}{200} + \bVnrf{n}{200}
\right)
\nonumber\\
&&
\hskip 60pt
+ \vev{\mbf p_{\pd}^2}
\left(
\bVnrf{p}{400} + \bVnrf{n}{400}
\right)
\bigg]
\nonumber \\
&&
\hskip -40pt
- \fr {2 \mr^2} {\sqrt{\pi}} 
\Bigg[
\ve_{n'}^2
\left(\fr{8n'}{2L'+1} - 3 \right)
-\ve_{n}^2 
\left(\fr{8n}{2L+1} - 3 \right)
\Bigg] 
\nonumber\\
&&
\hskip 20pt
\times
\left(\bVnrf{e}{400}
+\frac 1 {16} (\bVnrf{p}{400}
+\bVnrf{n}{400})
\right) ,
\label{degenf}
\eea
where $\bVnrf{\f}{kjm}=\cnrf{\f}{kjm}+\anrf{\f}{kjm}$,
$\mr$ is the reduced mass of antideuterium,
$\ve_{n}\equiv -\al^2 \mr/2n^2$,
and $\vev{\mbf p_{\pd}^2}\simeq 10^4$ MeV$^2$.

\section{Summary}
\label{Summary}

This work studies Lorentz and CPT violation in clock-comparison experiments
by incorporating effects on electron and nucleon propagators 
arising from SME operators of arbitrary mass dimension $d$.
It begins with a discussion of theoretical issues in Sec.\ \ref{Theory}.
The general Lagrange density \rf{lag} for a fermion
propagating in the presence of arbitrary Lorentz and CPT violation
implies the perturbative result \rf{nr}
for the corresponding nonrelativistic one-particle hamiltonian.
Combining the expressions for the constituent particles 
yields the hamiltonian \rf{pert} for an atom or ion,
which is the basis for our analysis of clock-comparison experiments. 

The experimental observables are transition frequencies
in atoms or ions.
The Lorentz- and CPT-violating signals in these frequencies
can be calculated from the perturbative shifts \rf{eatm} in energy levels. 
These shifts involve products of Clebsch-Gordan coefficients
with expectation values of the perturbative hamiltonian.
The symmetries of the system imply that
contributions to the energy shifts can arise 
only from specific nonrelativistic spherical coefficients 
for Lorentz violation,
as listed in Table \ref{cond}.
Explicit computation of the expectation values
requires modeling the electronic and nuclear states.
Our approach for electrons adopts the independent-particle model
described in Sec.\ \ref{Independent-particle model for electrons},
while for the nucleus we use the Schmidt model
as discussed in Sec.\ \ref{Nucleon expectation values}.

A laboratory on the surface of the Earth
or on an orbiting satellite typically represents a noninertial frame.
As a result,
most SME coefficients measurable in the laboratory
acquire a dependence on time due to the laboratory rotation and boost
relative to the canonical Sun-centered frame,
which is an approximately inertial frame over the experimental timescale.
Determining the time dependence induced by the rotation of the Earth
is the subject of Sec.\ \ref{sidereal}.
This treatment is extended in Sec.\ \ref{linearbe}
to include effects at linear order in the Earth's boost
as it orbits the Sun.
The time dependence in a space-based laboratory
arising from the orbital motion of a satellite
is also discussed.
 
The application of our results to the analysis of clock-comparison experiments
is described in Sec.\ \ref{Applications}.
We first consider fountain clocks,
deriving expressions for the frequency shift 
in terms of coefficients expressed in the Sun-centered frame.
At zeroth boost order
the frequency shift is given in Table \ref{csrot},
while at linear boost order
it is given by Eq.\ \rf{csboostf}
and the entries in Table \ref{csboost}.
Estimates for attainable sensitivities to SME coefficients 
using existing devices 
are provided in Table \ref{csestimates}.
The discussion covers both $^{133}$Cs and $^{87}$Rb fountain clocks,
and it is also applicable to clocks located on a space-based platform. 
The primary sensitivity in these systems
is to coefficients for Lorentz violation in the proton sector.

We next consider the prospects for using comagnetometers 
to search for nonminimal violations of Lorentz and CPT symmetry.
The methodology developed in Sec.\ \ref{Theory} 
is well suited for application to investigations 
using $^{129}$Xe and $^3$He atoms as comagnetometers.
Within the nuclear model adopted here,
the Lorentz- and CPT-violating signals
are affected predominantly by SME coefficients in the neutron sector.
In Sec.\ \ref{Comagnetometers},
we determine the shift in the experimental frequency observable
at zeroth boost order
and extract the bound \rf{hexecons}
by extending to arbitrary $d$ the known results for the minimal SME.
This leads to the constraints on neutron nonrelativistic coefficients
listed in Table \ref{Xelimits}.
We also establish the Lorentz- and CPT-violating shift
at linear boost order,
using existing data to place constraints
on neutron effective cartesian coefficients
in Table \ref{Xelimbe}.
Other comagnetometers can also place competitive limits
on the neutron sector of the SME.
We derive a partial map from known minimal-SME bounds
to nonminimal coefficients,
which permits using data from a $^{21}$Ne-Rb-K comagnetometer
to place the additional constraints on the neutron sector 
given in Table \ref{NeRblimits}.
All these constraints on neutron coefficients for Lorentz violation 
are the first of their kind reported in the literature.

As another application,
we consider the attainable reach in clock-comparison experiments
using trapped ions and lattice clocks.
In this case,
interesting sensitivities are in principle attainable 
to coefficients for Lorentz violation
in the electron sector.
Various transitions are considered for a range of atoms and ions.
The expression \rf{compopt} is found to describe 
the annual and sidereal modulations 
of the frequency difference between two clocks,
including ones located in distinct laboratories. 
In this section,
we also consider tests of Lorentz and CPT symmetry
based on studying the time evolution of an entangled state.
The shift in the experimental frequency observable
is determined at both zeroth and first boost order
and is used to estimate attainable sensitivities
to electron nonrelativistic coefficients,
as listed in Table \ref{Calimits}.

Our final application considers 
the prospects for experiments using antimatter.
Signals for Lorentz and CPT violation in antihydrogen 
have previously been investigated theoretically
both in the minimal SME
\cite{bkr99}
and allowing for nonminimal terms of arbitrary mass dimension
\cite{kv15}.
These treatments are combined with recent spectroscopic measurements
of the 1$S$-2$S$ transition in antihydrogen
to extract first constraints on SME coefficients from this system,
summarized in Table \ref{Hbarlimits}.
We also propose that in the long term it may become feasible to perform
experiments with heavier antiatoms and antiions,
with options possibly including the precision spectroscopy 
of antideuterium or of trapped antiions.
A technique is presented
to convert theoretical results for frequency shifts in atoms or ions 
to the corresponding ones in antiatoms or antiions.

The two appendices following the present summary 
collect some results that are useful 
in handling coefficients for Lorentz violation.
Appendix \ref{sphtocar}
includes relations connecting spherical and cartesian coefficients
and provides explicit expressions
between them for the cases $3\leq d \leq 8$.
Appendix \ref{appSun}
discusses the transformation 
between the laboratory frame and the Sun-centered frame
and tabulates explicit results
connecting cartesian coefficients in the two frames
for the cases $3\leq d \leq 8$.
The results in these appendices are generally applicable
and so have implications extending outside
the analysis of clock-comparison experiments.

Throughout this work,
we have noted possibilities for pursuing investigations
that go beyond our present scope
while remaining within the context
of Lorentz- and CPT-violating corrections
to the propagators of the constituents of atoms and ions.
In principle,
our scope could also be extended by incorporating 
effects arising from other SME sectors.
For instance,
the Maxwell equations acquire modifications
due to Lorentz and CPT violation in the pure-photon sector.
Including these might further enhance the reach 
of clock-comparison experiments,
though in practice most relevant photon-sector coefficients
are already tightly bounded from analyses of other systems
\cite{tables,photonexpt,photonastro}.
Effects involving U(1)-covariant Lorentz- and CPT-violating couplings 
between photons and fermions are of interest as well,
with only a few SME coefficients currently constrained by experiment
\cite{tables,dk16,Fcoeffexpt}.
One could also envisage the inclusion of SME effects
arising in the strong, electroweak, or gravitational sectors,
although some of these are expected either to be suppressed
or to be more readily studied by other means.
An exception might be countershaded Lorentz and CPT violation 
\cite{kt09},
for which unexpectedly large effects can appear
in the context of special measurements.
For example,
sensitivity to countershaded coefficients
has been demonstrated using atom interferometry,
which can be interpreted in terms of clock comparisons
\cite{ho11}.

Overall, 
the content of this paper provides a broad methodology
for exploring Lorentz and CPT symmetry
using clock-comparison experiments.
While our treatment has yielded many first constraints,
numerous coefficients for Lorentz violation are unmeasured to date.
The striking potential sensitivities attainable
either from reanalysis of existing data or in future searches 
suggests that further work with clock-comparison experiments 
remains one of the most interesting prospects 
for uncovering these novel physical effects in nature.

\section*{Acknowledgments}

This work was supported in part 
by the United States Department of Energy
under grant number {DE}-SC0010120
and by the Indiana University Center for Spacetime Symmetries.

\appendix

\section{Relation between spherical and cartesian coefficients}
\label{sphtocar}

\renewcommand\arraystretch{1.5}
\begin{table*}
\caption{Relations between spherical coefficients $\V_{kj0}^{(d)}$ 
and effective cartesian coefficients for $3\leq d \leq 8$.}
\setlength{\tabcolsep}{0pt}
\begin{tabular}{clccl}
\hline\hline
Spherical & \quad Cartesian & & Spherical & \quad Cartesian \\
\hline
$a^{(3)}_{000}$ 	 & 	 $\sqrt{4\pi}\,a^{(3)t}_\eff$ 	&&	$ c^{(6)}_{400}$ 	 & 	 $\frac{1}{5}\sqrt{4\pi}\,c^{(6)llmm}_\eff$ 	\\
$a^{(5)}_{000}$ 	 & 	 $\sqrt{4\pi}\,a^{(5)ttt}_\eff$ 	&&	$c^{(6)}_{220}$ 	 & 	 $4 \sqrt{\frac{ \pi}{5}}(3 c^{(6)ttzz}_\eff-c^{(6)ttmm}_\eff)$ 	\\
$a^{(5)}_{200}$ 	 & 	 $\sqrt{4\pi}\,a^{(5)tll}_\eff$ 	&&	$c^{(6)}_{420}$ 	 & 	 $\frac{4}{7} \sqrt{\frac{ \pi}{5}}(3 c^{(6)llzz}_\eff-c^{(6)llmm}_\eff)$ 	\\
$a^{(5)}_{220}$ 	 & 	 $\sqrt{\frac{4 \pi}{5}}(3 a^{(5)tzz}_\eff-a^{(5)tll}_\eff)$ 	&&	$c^{(6)}_{440}$ 	 & 	 $\frac{\sqrt{4\pi}}{21} (7 c^{(6)zzzz}_\eff-6 c^{(6)zzll}_\eff)+\frac{\sqrt{4\pi}}{35} c^{(6)mmll}_\eff$ 	\\
$a^{(7)}_{000}$ 	 & 	 $\sqrt{4\pi}a^{(7)ttttt}_\eff$ 	&&	$ c^{(8)}_{000}$ 	 & 	 $\sqrt{4\pi}c^{(8)tttttt}_\eff$ 	\\
$a^{(7)}_{200}$ 	 & 	 $\frac{10}{3}\sqrt{4\pi}a^{(7)tttll}_\eff$ 	&&	$ c^{(8)}_{200}$ 	 & 	 $5\sqrt{4\pi}c^{(8)ttttll}_\eff$ 	\\
$a^{(7)}_{400}$ 	 & 	 $\sqrt{4\pi}a^{(7)tmmll}_\eff$ 	&&	$ c^{(8)}_{400}$ 	 & 	 $3\sqrt{4\pi}c^{(8)ttmmll}_\eff$ 	\\
$a^{(7)}_{220}$ 	 & 	 $\frac{4}{3} \sqrt{5\pi}(3 a^{(7)tttzz}_\eff-a^{(7)tttmm}_\eff)$ 	&&	$ c^{(8)}_{600}$ 	 & 	 $\frac{1}{7}\sqrt{4\pi}c^{(8)nnllmm}_\eff$ 	\\
$a^{(7)}_{420}$ 	 & 	 $\frac{4}{7} \sqrt{5\pi}(3 a^{(7)tllzz}_\eff-a^{(7)tllmm}_\eff)$ 	&&	$c^{(8)}_{220}$ 	 & 	 $2 \sqrt{5\pi}(3 c^{(8)ttttzz}_\eff-c^{(8)ttttmm}_\eff)$ 	\\
$ a^{(7)}_{440}$ 	 & 	 $\frac{\sqrt{100\pi}}{21} (7 a^{(7)tzzzz}_\eff-6 a^{(7)tzzll}_\eff) + \frac{\sqrt{4\pi}}{7}  a^{(7)tmmll}_\eff$ 	&&	$c^{(8)}_{420}$ 	 & 	 $\frac{12}{7} \sqrt{5\pi}(3 c^{(8)ttllzz}_\eff-c^{(8)ttllmm}_\eff)$ 	\\
$c^{(4)}_{000}$ 	 & 	 $\sqrt{4\pi}\,c^{(4)tt}_\eff$ 	&&	$ c^{(8)}_{620}$ 	 & 	 $\frac{2}{21} \sqrt{5\pi}(3 c^{(8)nnllzz}_\eff-c^{(8)nnllmm}_\eff)$ 	\\
$c^{(4)}_{200}$ 	 & 	 $\frac{1}{3}\sqrt{4\pi}\,c^{(4)ll}_\eff$ 	&&	$ c^{(8)}_{440}$ 	 & 	 $\frac{\sqrt{100\pi}}{7} (7 c^{(8)ttzzzz}_\eff-6 c^{(8)ttzzll}_\eff) +3\frac{\sqrt{4\pi}}{7} c^{(8)ttmmll}_\eff$ 	\\
$c^{(4)}_{220}$ 	 & 	 $\frac{2}{3}\sqrt{\frac{\pi}{5}}(3 c^{(4)zz}_\eff-c^{(4)ll}_\eff)$ 	&&	$ c^{(8)}_{640}$ 	 & 	 $\frac{\sqrt{100\pi}}{77} (7 c^{(8)mmzzzz}_\eff-6 c^{(8)mmzzll}_\eff) +\frac{\sqrt{4\pi}}{77} 3 c^{(8)nnmmll}_\eff$ 	\\
$ c^{(6)}_{000}$ 	 & 	 $\sqrt{4\pi}\,c^{(6)tttt}_\eff$ 	&&	$c^{(8)}_{660}$ 	 & 	 $\frac{\sqrt{4\pi}}{231} (231 c^{(8)zzzzzz}_\eff-5 c^{(8)nnmmll}_\eff)$ 	\\
$ c^{(6)}_{200}$ 	 & 	 $2\sqrt{4\pi}\,c^{(6)ttll}_\eff$ 	&&		 & 	$\hskip 30pt +\frac{\sqrt{100\pi}}{11}( c^{(8)zzmmll}_\eff-3 c^{(8)zzzzmm}_\eff)$ 	\\
\hline\hline
\end{tabular}
\label{appa-1}
\end{table*}

This appendix presents relationships between spherical coefficients 
and effective cartesian coefficients 
and tabulates explicit results for $d\leq 8$.
We focus on spherical coefficients for Lorentz violation 
with even $k$ and $m=0$,
which are centrally relevant to analyses of clock-comparison experiments.
The coefficients $\T_{kjm}^{(d)(1E)}$
controlling spin-dependent operators of the $E$ type 
are disregarded here as they leave unaffected the energy shifts.
As elsewhere in this work,
we follow Ref.\ \cite{km13} in using the symbol $\V$ 
with appropriate subscripts and superscripts
to indicate the difference of $c$- and $a$-type coefficients
and $\T$ to indicate the difference of $g$- and $H$-type coefficients.
For instance,
$\V^{(d)}_{kjm}$ represents the difference
$\V^{(d)}_{kjm} = c^{(d)}_{kjm} - a^{(d)}_{kjm}$.
The spherical coefficients are assigned indices $kj0$,
while $t$, $x$, $y$, $z$ are used 
for specific index values on the effective cartesian coefficients 
in a chosen frame.
Dummy spatial cartesian indices are represented by $l$, $m$, $n$,
and repeated cartesian indices are summed.
For example,
$c^{(4)ll}_\eff$ represents the sum
$c^{(4)ll}_\eff=c^{(4)xx}_\eff+c^{(4)yy}_\eff+c^{(4)zz}_\eff$.

The single-particle hamiltonian can be decomposed 
in either the spherical or the cartesian bases.
The connection between these decompositions is presented 
in Sec.\ IV of Ref.\ \cite{km13}. 
Consider first the spin-independent component of the hamiltonian.
The corresponding match between the cartesian and spherical bases 
is fixed by 
\bea
\widehat{\V}^{(d)\mu}_\eff p_\mu &=&
\V_{\eff}^{(d)\mu\al_1\al_2\dots\al_{d-3}}
p_\mu p_{\al_1}p_{\al_2}\ldots p_{\al_{d-3}}
\nonumber\\
&=& \sum_{kjm} E_0^{d-2-k}\pmag^k Y_{jm}(\phat) 
\V^{(d)}_{kjm},
\eea
where $p_\mu=(E_0,-\pvec)$.
Using the orthogonality of the spherical harmonics,
the connection between the cartesian and spherical terms 
can be written as 
\beq
\int d\Om ~ Y^*_{jm}(\phat)\widehat{\V}^{(d)\mu}_\eff p_\mu 
=\sum_{k=j}^{d-2}E_0^{d-2-k}\pmag^k\V^{(d)}_{kjm},
\label{app3}
\eeq
where $d\Om$ is the differential element of solid angle in momentum space. 
The upper and lower bounds for the summation index $k$ 
are determined by the spherical-index relations 
listed in Table III of Ref.\ \cite{km13}.

Using Eq. \rf{app3},
we can extract explicit expressions 
for the spin-independent spherical coefficients 
in terms of effective cartesian coefficients. 
Table \ref{appa-1} contains the results
for spherical coefficients with
$3\leq d\leq 8$, 
$m=0$, 
even values of $j$ in the range $0\leq j\leq k$, 
and even values of $k$ in the range $0\leq k\leq d-2$.
The table consists of two pairs of columns.
In each pair,
the first entry in a given row lists a spherical coefficient,
while the second entry provides its equivalent
as a linear combination of effective cartesian coefficients. 

\renewcommand\arraystretch{1.8}
\begin{table*}
\caption{Relations between spherical coefficients $\T_{kj0}^{(d)(0B)}$ 
and effective cartesian coefficients for $3\leq d \leq 8$.}
\setlength{\tabcolsep}{2pt}
\begin{tabular}{clccl}
\hline\hline
Spherical & \quad Cartesian & \hskip 10pt & Spherical & \quad Cartesian \\
\hline
$H^{(3)(0B)}_{010}$ 	 & 	 $\sqrt{\frac{4\pi}{3}}\Ht^{(3)tz}_\eff$ 	&&	$g^{(4)(0B)}_{010}$ 	 & 	 $\sqrt{\frac{4\pi}{3}}\gt^{(4)tzt}_\eff$ 	\\
$H^{(5)(0B)}_{010}$ 	 & 	 $\sqrt{\frac{4\pi}{3}}\Ht^{(5)tztt}_\eff$ 	&&	$g^{(6)(0B)}_{010}$ 	 & 	 $\sqrt{\frac{4\pi}{3}}\gt^{(6)tzttt}_\eff$ 	\\
$H^{(5)(0B)}_{210}$ 	 & 	 $\frac{1}{15}\sqrt{\frac{4\pi}{3}}(2\Ht^{(5)tllz}_\eff+\Ht^{(5)tzll}_\eff)$ 	&&	$g^{(6)(0B)}_{210}$ 	 & 	 $\frac{1}{5}\sqrt{\frac{4\pi}{3}}(2\gt^{(6)tltlz}_\eff+\gt^{(6)tztll}_\eff)$ 	\\
$H^{(5)(0B)}_{230}$ 	 & 	 $\frac{2}{15}\sqrt{\frac{\pi}{7}}(5\Ht^{(5)tzzz}_\eff-\Ht^{(5)tzll}_\eff)-\frac{4}{15}\sqrt{\frac{\pi}{7}}\Ht^{(5)tllz}_\eff$ 	&&	$g^{(6)(0B)}_{230}$ 	 & 	 $\frac{2}{5}\sqrt{\frac{\pi}{7}}(5\gt^{(6)tztzz}_\eff-\gt^{(6)tztll}_\eff)-\frac{4}{5}\sqrt{\frac{\pi}{7}}\gt^{(6)tltlz}_\eff$ 	\\
$H^{(7)(0B)}_{010}$ 	 & 	 $\sqrt{\frac{4\pi}{3}}\Ht^{(7)tztttt}_\eff$ 	&&	$g^{(8)(0B)}_{010}$ 	 & 	 $\sqrt{\frac{4\pi}{3}}\gt^{(8)tzttttt}_\eff$ 	\\
$H^{(7)(0B)}_{210}$ 	 & 	 $\frac{2}{5}\sqrt{\frac{4\pi}{3}}(2\Ht^{(7)tmttmz}_\eff+\Ht^{(7)tzttmm}_\eff)$ 	&&	$g^{(8)(0B)}_{210}$ 	 & 	 $\frac{2}{3}\sqrt{\frac{4\pi}{3}}(2\gt^{(8)tmtttmz}_\eff+\gt^{(8)tztttmm}_\eff)$ 	\\
$H^{(7)(0B)}_{410}$ 	 & 	 $\frac{2}{175}\sqrt{3\pi}(4\Ht^{(7)tmmllz}_\eff+\Ht^{(7)tzllmm}_\eff)$ 	&&	$g^{(8)(0B)}_{410}$ 	 & 	 $\frac{2}{35}\sqrt{3\pi}(4\gt^{(8)tmtmllz}_\eff+\gt^{(8)tztllmm}_\eff)$ 	\\
$H^{(7)(0B)}_{230}$ 	 & 	 $\frac{4}{5}\sqrt{\frac{\pi}{7}}(5\Ht^{(7)tzttzz}_\eff-\Ht^{(7)tzttmm}_\eff)-\frac{8}{5}\sqrt{\frac{\pi}{7}}\Ht^{(7)tmttmz}_\eff$ 	&&	$g^{(8)(0B)}_{230}$ 	 & 	 $\frac{4}{3}\sqrt{\frac{\pi}{7}}(5\gt^{(8)tztttzz}_\eff-\gt^{(8)tztttmm}_\eff)$ 	\\
$H^{(7)(0B)}_{430}$ 	 & 	 $\frac{4}{45}\sqrt{\frac{\pi}{7}}(3\Ht^{(7)tzmmzz}_\eff+2 \Ht^{(7)tmmzzz}_\eff)$ 	&&		 & 	 $\hskip 30pt -\frac{8}{3}\sqrt{\frac{\pi}{7}}\gt^{(8)tmtttmz}_\eff$ 	\\
                       	 & 	 $\hskip30pt -\frac{4}{75}\sqrt{\frac{\pi}{7}} (4\Ht^{(7)tmmllz}_\eff+\Ht^{(7)tzmmll}_\eff)$ 	&&	$g^{(8)(0B)}_{430}$ 	 & 	 $\frac{4}{9}\sqrt{\frac{\pi}{7}}(3\gt^{(8)tztmmzz}_\eff+ 2 \gt^{(8)tmtmzzz}_\eff$) 	\\
$H^{(7)(0B)}_{450}$ 	 & 	 $\frac{2}{15}\sqrt{\frac{\pi}{11}}(3\Ht^{(7)tzzzzz}_\eff-2\Ht^{(7)tzmmzz}_\eff)$ 	&&	                      	 & 	 $\hskip30pt -\frac{4}{15}\sqrt{\frac{\pi}{7}}(4\gt^{(8)tmtmllz}_\eff+\gt^{(8)tztmmll}_\eff)$ 	\\
                       	 & 	 $\hskip30pt +\frac{2}{105}\sqrt{\frac{\pi}{11}}(4\Ht^{(7)tmmllz}_\eff+\Ht^{(7)tzmmll}_\eff)$ 	&&	$g^{(8)(0B)}_{430}$ 	 & 	 $\frac{2}{3}\sqrt{\frac{\pi}{11}}(3\gt^{(7)tztzzzz}_\eff-2\gt^{(7)tztmmzz}_\eff)$ 	\\
                       	 & 	 $\hskip30pt -\frac{8}{45}\sqrt{\frac{\pi}{11}}\Ht^{(7)tmmzzz}_\eff$ 	&&	                      	 & 	 $\hskip30pt +\frac{2}{21}\sqrt{\frac{\pi}{11}}(4\gt^{(8)tmtmllz}_\eff+\gt^{(8)tztmmll}_\eff)$ 	\\
\hline\hline
\end{tabular}
\label{appa-2}
\end{table*}

\renewcommand\arraystretch{1.7}
\begin{table*}
\caption{Relations between spherical coefficients $\T_{kj0}^{(d)(1B)}$ 
and effective cartesian coefficients for $3\leq d \leq 8$.}
\setlength{\tabcolsep}{3pt}
\begin{tabular}{clccl}
\hline\hline
Spherical & \quad Cartesian & \hskip 10pt & Spherical & \quad Cartesian \\
\hline
$H^{(5)(1B)}_{210}$ 	  &  	 $\frac{2}{3}\sqrt{\frac{\pi}{3}}(3\Ht^{(5)nztn}_\eff-\Ht^{(5)tnnz}_\eff) +\frac{2}{3}\sqrt{\frac{\pi}{3}}\Ht^{(5)tznn}_\eff$ 	&&	$g^{(8)(1B)}_{210}$ 	  &  	 $\frac{5}{3}\sqrt{\frac{\pi}{3}}(4\gt^{(8)tztttmm}_\eff-4\gt^{(8)tmtttmz}_\eff)$ 	\\
$H^{(7)(1B)}_{210}$ 	  &  	 $4\sqrt{\frac{\pi}{3}}(\Ht^{(7)tzttnn}_\eff-\Ht^{(7)tnttnz}_\eff) +4\sqrt{\frac{\pi}{3}}\Ht^{(7)nztttn}_\eff$ 	&&	                      	  &  	 $\hskip30pt +5 \sqrt{\frac{\pi}{3}}\gt^{(8)mzttttm}_\eff$ 	\\
$H^{(7)(1B)}_{410}$ 	  &  	 $\frac{4}{25}\sqrt{3\pi}(\Ht^{(7)tzllnn}_\eff-\Ht^{(7)tnnllz}_\eff) +\frac{4}{5}\sqrt{3\pi}\Ht^{(7)nztnll}_\eff$ 	&&	$g^{(8)(1B)}_{410}$ 	  &  	 $\frac{4}{5}\sqrt{3\pi}(\gt^{(8)tztllmm}_\eff-\gt^{(8)tmtmllz}_\eff)$ 	\\
$H^{(7)(1B)}_{430}$ 	  &  	 $\frac{2}{5}\sqrt{\frac{6\pi}{7}}(5\Ht^{(7)nztnzz}_\eff+ \Ht^{(7)tznnzz}_\eff)$ 	&&	                       	  &  	 $\hskip30pt +2\sqrt{3\pi}\gt^{(8)mzttmll}_\eff$ 	\\
                       	  &  	 $\hskip30pt -\frac{2}{5}\sqrt{\frac{6\pi}{7}}(\Ht^{(7)tnnzzz}_\eff+\Ht^{(7)nztnll}_\eff)$ 	&&	$g^{(8)(1B)}_{610}$ 	  &  	 $\frac{1}{7}\sqrt{3\pi}\,\gt^{(8)mzmnnll}_\eff$ 	\\
                      	  &  	 $\hskip30pt +\frac{2}{25}\sqrt{\frac{6\pi}{7}}(\Ht^{(7)tnnllz}_\eff-\Ht^{(7)tznnll}_\eff)$ 	&&	$g^{(8)(1B)}_{430}$ 	  &  	 $\sqrt{\frac{6\pi}{7}}(5\gt^{(8)mzttmzz}_\eff+ 2 \gt^{(8)tztmmzz}_\eff)$ 	\\
$g^{(4)(1B)}_{210}$ 	  &  	 $\sqrt{\frac{\pi}{3}}\gt^{(4)nzn}_\eff$ 	&&	                      	  &  	 $\hskip30pt -\sqrt{\frac{6\pi}{7}}(2\gt^{(8)tmtmzzz}_\eff+\gt^{(8)mzttmll}_\eff)$ 	\\
$g^{(6)(1B)}_{210}$ 	  &  	 $\sqrt{\frac{4\pi}{3}}(\gt^{(6)tztnn}_\eff-\gt^{(6)tntnz}_\eff) +\sqrt{3\pi}\gt^{(6)nzttn}_\eff$ 	&&	                      	  &  	 $\hskip30pt +\frac{2}{5}\sqrt{\frac{6\pi}{7}}(\gt^{(8)tmtmllz}_\eff-\gt^{(8)tztmmll}_\eff)$ 	\\
$g^{(6)(1B)}_{410}$ 	  &  	 $\frac{1}{5}\sqrt{3\pi}\gt^{(6)nznmm}_\eff$ 	&&	$g^{(8)(1B)}_{630}$ 	  &  	 $\frac{1}{3}\sqrt{\frac{2 \pi}{21}}(5\gt^{(8)mzmllzz}_\eff-\gt^{(8)mzmllnn}_\eff)$ 	\\
$g^{(6)(1B)}_{430}$ 	  &  	 $\frac{1}{5}\sqrt{\frac{3\pi}{14}}(5\gt^{(6)lzlzz}_\eff-\gt^{(6)lzlnn}_\eff)$	&&	$g^{(8)(1B)}_{650}$ 	  &  	 $\frac{1}{3}\sqrt{\frac{5\pi}{33}}(3\gt^{(8)mzmzzzz}_\eff-2\gt^{(8)mzmllzz}_\eff)$ 	\\
	&		&&	                       	  &  	 $\hskip30pt +\frac{1}{21}\sqrt{\frac{5\pi}{33}} \gt^{(8)mzmllnn}_\eff$ 	\\
\hline\hline
\end{tabular}
\label{appa-3}
\end{table*}

Next, 
we consider the spin-dependent part of the single-particle hamiltonian.
For the component involving only the coefficients $\T_{kjm}^{(d)(0B)}$,
the relation between the cartesian and spherical terms is
\bea
\Tdual_{\eff}^{(d)\mu t} p_\mu &=&
\widetilde{\T}_{\eff}^{(d)\mu t\al_1\al_2\dots\al_{d-3}}
p_\mu p_{\al_1}p_{\al_2}\ldots p_{\al_{d-3}} 
\nonumber\\
&=&
\sum_{kjm} E_0^{d-3-k} \pmag^{k+1}(k+1)
\syjm{}{jm}(\phat) \T_{kjm}^{(d)(0B)}.
\nonumber\\
\label{app1}
\eea
Using orthonormality of the spherical harmonics then yields 
\bea
&&
\hskip-40pt
\int d\Om~ Y_{jm}^*(\phat) \Tdual_\eff^{(d)\mu t} p_\mu 
\nonumber \\
&&
\hskip20pt
=\sum_{k=j-1}^{d-3}E_0^{d-3-k} \pmag^{k+1} (k+1)\T^{(d)(0B)}_{kjm}
\label{app2}
\eea
between effective cartesian coefficients and spherical coefficients.
This result permits the extraction of explicit expressions
for the spin-dependent spherical coefficients $\T_{kjm}^{(d)(0B)}$
as linear combinations of effective cartesian coefficients.
Table \ref{appa-2} contains these expressions
for spherical coefficients $\T_{kjm}^{(d)(0B)}$ 
with $3\leq d\leq 8$, 
$m=0$, 
odd values of $j$ in the range $0\leq j\leq k+1$, 
and even values of $k$ in the range $0\leq k\leq d-3$. 
The structure of this table parallels that of Table \ref{appa-1}.

Determining the spherical coefficients $\T^{(d)(1B)}_{kjm}$ 
in terms of effective cartesian coefficients
requires more work because the relation 
containing $\T^{(d)(1B)}_{kjm}$ 
also incorporates the coefficients $\T^{(d)(0B)}_{kjm}$.
We find
\bea
&&
\hskip -30pt
\int d\Om~ \Tdual_\eff^{(d)j \nu}
\hat{\mbf{\ep}}_{+}^j\, p_\nu\, \syjm{1}{jm}^*(\phat) 
\nonumber \\
&&
\hskip -20pt
=\sum_{k=j-1}^{d-2}E_0^{d-2-k} \pmag^{k}
\Bigg(\sqrt{\dfrac{j(j+1)}{2}}\T^{(d)(0B)}_{kjm}
\nonumber\\
&&
\hskip 80pt
+\T^{(d)(1B)}_{kjm}
+i \T^{(d)(1E)}_{kjm}
\Bigg),
\label{app4}
\eea
where $\ephat_\pm = (\thhat \pm i\phhat)/\sqrt{2}$.
This result links three types 
of spherical coefficients with the effective cartesian coefficients. 
It can be disentangled 
first by eliminating the $\T^{(d)(0B)}_{kjm}$ via Eq.\ \rf{app2}
and then by grouping the remaining terms 
according to powers of the momentum magnitude. 
The point is that the $E$-type and $B$-type coefficients 
are proportional to distinct powers of the momentum when $j$ is fixed.
For example,
if $j$ is odd then the terms involving $B$-type and $E$-type coefficients 
can only contain even and odd powers of the momentum magnitude,
respectively.

For the particular case with $m=0$,
the spherical coefficients and the spin-weighted harmonics 
are all real numbers. 
It is therefore useful to separate 
the real and imaginary parts of Eq.\ \rf{app4}.
The real part is 
\bea
&&
\hskip -20pt
\int d\Om~ \Tdual^{(d)j \nu}_\eff 
\hat{\mbf{\th}}^j\, p_\nu\, \syjm{1}{j0}(\phat) 
\nonumber \\
&&
\hskip -20pt
= \sum_{k=j-1}^{d-2}E_0^{d-2-k} \pmag^{k}
\big(\sqrt{j(j+1)}\T^{(d)(0B)}_{kj0}
+\sqrt{2}\T^{(d)(1B)}_{kj0}\big),
\nonumber\\
\label{app5}
\eea
and it contains only $B$-type coefficients.
The imaginary part of Eq.\ \rf{app4} is given by
\bea
&& 
\hskip -40pt
\int d\Om ~ \Tdual^{(d)j \nu}_\eff 
\hat{\mbf{\ph}}^j\, p_\nu\, \syjm{1}{j0}(\phat)
\nonumber \\
&&
\hskip 40pt
=\sqrt{2}\sum_{k=j}^{d-2}E_0^{d-2-k} \pmag^{k} \T^{(d)(1E)}_{kj0}
\eea
and contains only $E$-type coefficients. 
By combining Eqs.\ \rf{app3} and \rf{app5},
we can extract explicit expressions 
for the coefficients $\T^{(d)(1B)}_{kjm}$
in terms of effective cartesian components.
Table \ref{appa-3} contains the results
for spherical coefficients with 
$4\leq d\leq 8$, 
$m=0$, 
odd values of $j$ in the range $0\leq j\leq k-1$, 
and even values of $k$ in the range $0\leq k\leq d-2$.
The structure of this table again follows that of Table \ref{appa-1}.

\section{Transformations to the Sun-centered frame}
\label{appSun}

\renewcommand\arraystretch{1.5}
\begin{table*}
\caption{Relations between spin-independent cartesian coefficients 
in laboratory and Sun-centered frames for $3\leq d \leq 8$.}
\setlength{\tabcolsep}{4pt}
\begin{tabular}{ccccccc}
\hline\hline
Laboratory & Factor & Sun-centered & $\hskip 30pt$
& Laboratory & Factor & Sun-centered \\
\hline
$a^{(3)0}_\eff$     	&	1	&	   $a^{(3)T}_\eff$	&&	$c^{(6)0000}_\eff$  	&	1	&	 $c^{(6)TTTT}_\eff$	\\
  	&	 $-\bevec^J $                                                              	&	 $a^{(3)J}_\eff$	&&	 	&	  $-4\bevec^J $                                 	&	 $c^{(6)TTTJ}_\eff$	\\
$a^{(5)000}_\eff$   	&	1	&	   $a^{(5)TTT}_\eff$	&&	$c^{(6)00jj}_\eff$  	&	1	&	 $c^{(6)TTJJ}_\eff$	\\
 	&	   $-3\bevec^J $                                                           	&	 $a^{(5)TTJ}_\eff$	&&	 	&	  $-2\bevec^J $                                  	&	 $c^{(6)TTTJ}_\eff$	\\
$a^{(5)0jj}_\eff$   	&	1	&	   $a^{(5)TKK}_\eff$	&&	 	&	  $-2\bevec^J $                                  	&	 $c^{(6)TKKJ}_\eff$	\\
 	&	   $-2\bevec^J $                                 	&	 $a^{(5)TTJ}_\eff$	&&	$c^{(6)0033}_\eff$  	&	  $\hat{B}^\Jj{1}\hat{B}^\Jj{2}$ 	&	 $c^{(6)TT\Jj{1}\Jj{2}}_\eff$	\\
 	&	   $-\bevec^J $                                 	&	 $a^{(5)KKJ}_\eff$	&&	 	&	  $-2\hat{B}^\Jj{1}\hat{B}^\Jj{2}\bevec^J$                                  	&	 $c^{(6)T\Jj{1}\Jj{2}J}_\eff$	\\
$a^{(5)033}_\eff$  	&	  $\hat{B}^\Jj{1}\hat{B}^\Jj{2}$                                                	&	  $a^{(5)T\Jj{1}\Jj{2}}_\eff$	&&	 	&	  $-2\hat{B}^\Jj{1} \hat{B}\cdot\bevec$                                	&	 $c^{(6)TTT\Jj{1}}_\eff$	\\
 	&	   $-\hat{B}^\Jj{1}\hat{B}^\Jj{2}\bevec^J $                               	&	 $a^{(5)\Jj{1}\Jj{2}J}_\eff$	&&	$c^{(6)jjkk}_\eff$  	&	1	&	 $c^{(6)KKLL}_\eff$	\\
 	&	   $-2\hat{B}^\Jj{1} \hat{B}\cdot\bevec$                            	&	 $a^{(5)TT\Jj{1}}_\eff$	&&	 	&	  $-4\bevec^J$                                                                 	&	 $c^{(6)TKKJ}_\eff$	\\
$a^{(7)00000}_\eff$  	&	1	&	  $a^{(7)TTTTT}_\eff$	&&	$c^{(6)33jj}_\eff$  	&	  $\hat{B}^\Jj{1}\hat{B}^\Jj{2}$ 	&	$c^{(6)KK\Jj{1}\Jj{2}}_\eff$	\\
   	&	   $-5\bevec^J $                                                                	&	  $a^{(7)TTTTJ}_\eff$	&&	 	&	  $-2\hat{B}^\Jj{1}\hat{B}^\Jj{2}\bevec^J$                                   	&	 $c^{(6)T\Jj{1}\Jj{2}J}_\eff$	\\
$a^{(7)000jj}_\eff$  	&	1	&	  $a^{(7)TTTKK}_\eff$	&&	 	&	  $-2\hat{B}^\Jj{1} \hat{B}\cdot\bevec$                                  	&	 $c^{(6)TKK\Jj{1}}_\eff$	\\
   	&	   $-2\bevec^J $                                                                	&	  $a^{(7)TTTTJ}_\eff$	&&	$c^{(6)3333}_\eff$  	&	  $\hat{B}^\Jj{1}\hat{B}^\Jj{2}\hat{B}^\Jj{3}\hat{B}^\Jj{4}$                    	&	 $c^{(6)\Jj{1}\Jj{2}\Jj{3}\Jj{4}}_\eff$	\\
   	&	   $-3\bevec^J $                                                                	&	  $a^{(7)TTKKJ}_\eff$	&&	  	&	  $-4\hat{B}^\Jj{1} \hat{B}^\Jj{2} \hat{B}^\Jj{3} \hat{B}\cdot\bevec$  	&	 $c^{(6)T\Jj{1}\Jj{2}\Jj{3}}_\eff$	\\
$a^{(7)00033}_\eff$  	&	 $\hat{B}^\Jj{1}\hat{B}^\Jj{2}$                                                   	&	 $a^{(7)TTT\Jj{1}\Jj{2}}_\eff$	&&	$c^{(8)000000}_\eff$ 	&	1	&	 $c^{(8)TTTTTT}_\eff$	\\
   	&	 $-3\hat{B}^\Jj{1}\hat{B}^\Jj{2} \bevec^J$ 	&	 $a^{(7)TT\Jj{1}\Jj{2}J}_\eff$	&&	 	&	  $-6\bevec^J $                                                                	&	$c^{(8)TTTTTJ}_\eff$	\\
   	&	 $-2\hat{B}^\Jj{1}\hat{B}\cdot\bevec$    	&	 $a^{(7)TTTT\Jj{1}}_\eff$	&&	$c^{(8)0000jj}_\eff$ 	&	1	&	 $c^{(8)TTTTKK}_\eff$	\\
$a^{(7)0jjkk}_\eff$  	&	1	&	  $a^{(7)TKKLL}_\eff$	&&	 	&	  $-2\bevec^J $                                                                	&	$c^{(8)TTTTTJ}_\eff$	\\
   	&	   $-4\bevec^J $                                                                	&	  $a^{(7)TTKKJ}_\eff$	&&	     	&	  $-4\bevec^J $                                                                	&	$c^{(8)TTTKKJ}_\eff$	\\
   	&	   $-\bevec^J $                                                                  	&	  $a^{(7)KKLLJ}_\eff$	&&	$c^{(8)000033}_\eff$ 	&	 $\hat{B}^\Jj{1}\hat{B}^\Jj{2}$  	&	 $c^{(8)TTTT\Jj{1}\Jj{2}}_\eff$	\\
$a^{(7)0jj33}_\eff$  	&	 $\hat{B}^\Jj{1}\hat{B}^\Jj{2}$                                                   	&	 $a^{(7)TKK\Jj{1}\Jj{2}}_\eff$	&&	 	&	  $-4\bevec^J \hat{B}^\Jj{1}\hat{B}^\Jj{2}$                               	&	$c^{(8)TTTJ\Jj{1}\Jj{2}}_\eff$	\\
   	&	 $-2\hat{B}^\Jj{1}\hat{B}^\Jj{2} \bevec^J$ 	&	 $a^{(7)TT\Jj{1}\Jj{2}J}_\eff$	&&	     	&	  $-2\hat{B}^\Jj{1}\hat{B}\cdot\bevec$                                	&	$c^{(8)TTTTT\Jj{1}}_\eff$	\\
   	&	 $-2\hat{B}^\Jj{1}\hat{B}\cdot\bevec$    	&	 $a^{(7)TTKK\Jj{1}}_\eff$	&&	$c^{(8)00jjkk}_\eff$ 	&	1	&	 $c^{(8)TTKKLL}_\eff$	\\
    	&	 $-\hat{B}^\Jj{1}\hat{B}^\Jj{2} \bevec^\Jj{3}$ 	&	 $a^{(7)KK\Jj{1}\Jj{2}\Jj{3}}_\eff$	&&	 	&	  $-4\bevec^J $                                                                	&	$c^{(8)TTTKKJ}_\eff$	\\
$a^{(7)03333}_\eff$  	&	 $\hat{B}^\Jj{1}\hat{B}^\Jj{2}\hat{B}^\Jj{3}\hat{B}^\Jj{4}$                     	&	 $a^{(7)T\Jj{1}\Jj{2}\Jj{3}\Jj{4}}_\eff$	&&	     	&	  $-2\bevec^J $                                                                	&	$c^{(8)TKKLLJ}_\eff$	\\
   	&	  $-\hat{B}^\Jj{1}\hat{B}^\Jj{2}\hat{B}^\Jj{3}\hat{B}^\Jj{4} \bevec^J$  	&	 $a^{(7)\Jj{1}\Jj{2}\Jj{3}\Jj{4}J}_\eff$	&&	$c^{(8)00jj33}_\eff$ 	&	 $\hat{B}^\Jj{1}\hat{B}^\Jj{2}$     	&	 $c^{(8)TTKK\Jj{1}\Jj{2}}_\eff$	\\
   	&	  $-4\hat{B}^\Jj{1}\hat{B}^\Jj{2}\hat{B}^\Jj{3}\hat{B}\cdot\bevec$  	&	   $a^{(7)TT\Jj{1}\Jj{2}\Jj{3}}_\eff$	&&	 	&	  $-2\bevec^J \hat{B}^\Jj{1}\hat{B}^\Jj{2}$                               	&	$c^{(8)TTTJ\Jj{1}\Jj{2}}_\eff$	\\
$c^{(4)00}_\eff$    	&	1	&	  $c^{(4)TT}_\eff$	&&	 	&	  $-2\bevec^J \hat{B}^\Jj{1}\hat{B}^\Jj{2}$                               	&	$c^{(8)TKK\Jj{1}\Jj{2}J}_\eff$	\\
 	&	 $-2\bevec^J $                                                             	&	  $c^{(4)TJ}_\eff$	&&	     	&	  $-2\hat{B}^\Jj{1} \hat{B}\cdot\bevec$                                 	&	$c^{(8)TTTKK\Jj{1}}_\eff$	\\
$c^{(4)jj}_\eff$    	&	1	&	  $c^{(4)KK}_\eff$	&&	$c^{(8)003333}_\eff$ 	&	 $\hat{B}^\Jj{1}\hat{B}^\Jj{2}\hat{B}^\Jj{3}\hat{B}^\Jj{4}$              	&	 $c^{(8)TT\Jj{1}\Jj{2}\Jj{3}\Jj{4}}_\eff$	\\
 	&	 $-2\bevec^J $                                                             	&	  $c^{(4)TJ}_\eff$	&&	     	&	  $-2\hat{B}^\Jj{1}\hat{B}^\Jj{2}\hat{B}^\Jj{3}\hat{B}^\Jj{4} \bevec^J$  	&	$c^{(8)T\Jj{1}\Jj{2}\Jj{3}\Jj{4}J}_\eff$	\\
$c^{(4)33}_\eff$    	&	 $\hat{B}^\Jj{1}\hat{B}^\Jj{2}$                                                	&	  $c^{(4)\Jj{1}\Jj{2}}_\eff$	&&	          	&	 $-4\hat{B}^\Jj{1} \hat{B}^\Jj{2} \hat{B}^\Jj{3} \hat{B}\cdot\bevec$   	&	$c^{(8)TKK\Jj{1}\Jj{2}\Jj{3}}_\eff$	\\
 	&	 $-2\hat{B}^\Jj{1} \hat{B}\cdot\bevec$                              	&	  $c^{(4)T\Jj{1}}_\eff$	&&						\\
\hline\hline
\end{tabular}
\label{appb-1}
\end{table*}

\renewcommand\arraystretch{1.6}
\begin{table}
\caption{The quantities $\Va{kj}{d}{JK\ldots M}$ for $5 \leq d \leq 8$.} 
\setlength{\tabcolsep}{5pt}
\begin{tabular}{ll}
\hline\hline
$\Va{kj}{d}{JK\ldots M}$ & Combination \\
\hline
$\Va{20}{5}{J}$ 	 & 	 $2a^{(5)TTJ}_\eff+a^{(5)KKJ}_\eff$	\\
$\Va{20}{6}{J}$ 	 & 	 $4c^{(6)TTTJ}_\eff+4c^{(6)TKKJ}_\eff$	\\
$\Va{20}{7}{J}$ 	 & 	 $\frac{10}{3}(2a^{(7)TTTTJ}_\eff+3a^{(7)TTKKJ}_\eff)$	\\
$\Va{20}{8}{J}$ 	 & 	 $10c^{(8)TTTTTJ}_\eff+20c^{(8)TTTKKJ}_\eff$	\\
$\Va{40}{7}{J}$ 	 & 	 $a^{(7)LLKKJ}_\eff+4 a^{(7)TTKKJ}_\eff$	\\
$\Va{40}{8}{J}$ 	 & 	 $6 c^{(8)TLLKKJ}_\eff+12 c^{(8)TTTKKJ}_\eff$	\\
$\Va{22}{5}{J\Jj{1}\Jj{2}}$ 	 & 	 $3 a^{(5)J\Jj{1}\Jj{2}}_\eff+6\de^{J\Jj{1}}a^{(5)TT\Jj{2}}_\eff$	\\
$\Va{22}{6}{J \Jj{1}\Jj{2}}$ 	 & 	 $12c^{(6)TJ\Jj{1}\Jj{2}}_\eff+12\de^{J\Jj{1}}c^{(6)TTT\Jj{2}}_\eff$	\\
$\Va{22}{7}{J \Jj{1}\Jj{2}}$ 	 & 	 $30a^{(7)TTJ\Jj{1}\Jj{2}}_\eff+20\de^{J\Jj{1}}a^{(7)TTTT\Jj{2}}_\eff$	\\
$\Va{22}{8}{J \Jj{1}\Jj{2}}$ 	 & 	 $60c^{(8)TTTJ\Jj{1}\Jj{2}}_\eff+30\de^{J\Jj{1}}c^{(8)TTTTT\Jj{2}}_\eff$	\\
$\Va{42}{7}{J \Jj{1}\Jj{2}}$ 	 & 	 $\frac{60}{7}(a^{(7)TTJ\Jj{1}\Jj{2}}_\eff+\de^{J\Jj{1}}a^{(7)TTLL\Jj{2}}_\eff)$	\\
                              	 & 	 $\hskip 30pt +\frac{30}{7}a^{(7)LLJ\Jj{1}\Jj{2}}$	\\
$\Va{42}{8}{J \Jj{1}\Jj{2}}$ 	 & 	 $\frac{180}{7}(c^{(8)TTTJ \Jj{1}\Jj{2}}_\eff+c^{(8)TKKJ \Jj{1}\Jj{2}}_\eff)$	\\
                             	 & 	 $\hskip 30pt +\frac{180}{7}\de^{J \Jj{1}}c^{(8)TTTKK\Jj{2}}_\eff$	\\
$\Va{44}{7}{J \Jj{1}\Jj{2}\Jj{3}\Jj{4}}$ 	 & 	 $5(\de^{J \Jj{1}}a^{(7)TT\Jj{2}\Jj{3}\Jj{4}}_\eff+a^{(7)J \Jj{1}\Jj{2}\Jj{3}\Jj{4}}_\eff)$	\\
$\Va{44}{8}{J \Jj{1}\Jj{2}\Jj{3}\Jj{4}}$ 	 & 	 $60 \de^{J \Jj{1}} c^{(8)TTT\Jj{2}\Jj{3}\Jj{4}}_\eff+30c^{(8)TJ \Jj{1}\Jj{2}\Jj{3}\Jj{4}}_\eff$	\\
\hline\hline
\end{tabular}
\label{appb-2}
\end{table}

\renewcommand\arraystretch{1.5}
\begin{table*}
\caption{Relations between spin-dependent cartesian coefficients 
in laboratory and Sun-centered frames for $3\leq d \leq 6$.}
\setlength{\tabcolsep}{4pt}
\begin{tabular}{ccccccc}
\hline\hline
Laboratory & Factor & Sun-centered & $\hskip 30pt$ & 
Laboratory & Factor & Sun-centered \\
\hline
$\Ht^{(3)03}_\eff$ 	&	 $\hat{B}^\Jj{1}$ 	&	 $\Ht^{(3)T\Jj{1}}_\eff$	&&	$\gt^{(6)03000}_\eff$ 	&	 $\hat{B}^\Jj{1}$ 	&	 $\gt^{(6)T\Jj{1}TTT}_\eff$	\\
 	&	 $\hat{B}^\Jj{1} \bevec^J $ 	&	 $\Ht^{(3)\Jj{1}J}_\eff$	&&	 	&	 $4\hat{B}^\Jj{1} \bevec^J $ 	&	 $\gt^{(6)\Jj{1}(TTTJ)}_\eff$	\\
$\Ht^{(5)0300}_\eff$ 	&	 $\hat{B}^\Jj{1}$ 	&	 $\Ht^{(5)T\Jj{1}TT}_\eff$	&&	$\gt^{(6)0j0j3}_\eff$ 	&	 $\hat{B}^\Jj{1}$ 	&	 $\gt^{(6)TKTK\Jj{1}}_\eff$	\\
 	&	 $3\hat{B}^\Jj{1} \bevec^J $ 	&	 $\Ht^{(5)\Jj{1}(TTJ)}_\eff$	&&	 	&	 $\hat{B}^\Jj{1} \bevec^J $ 	&	 $\gt^{(6)JTTT\Jj{1}}_\eff$	\\
$\Ht^{(5)j30j}_\eff$ 	&	 $\hat{B}^\Jj{1}$ 	&	 $\Ht^{(5)K\Jj{1}TK}_\eff$	&&	 	&	 $2\hat{B}^\Jj{1} \bevec^J $ 	&	 $\gt^{(6)K(JTK)\Jj{1}}_\eff$	\\
 	&	 $2\hat{B}^\Jj{1} \bevec^J $ 	&	 $\Ht^{(5)\Jj{1}(JT)T}_\eff$	&&	 	&	 $\hat{B}\cdot\bevec $ 	&	 $\gt^{(6)KTTTK}_\eff$	\\
 	&	 $\hat{B}^\Jj{1} \bevec^J $ 	&	 $\Ht^{(5)\Jj{1}KKJ}_\eff$	&&	$\gt^{(6)030jj}_\eff$ 	&	 $\hat{B}^\Jj{1}$ 	&	 $\gt^{(6)T\Jj{1}TKK}_\eff$	\\
 	&	 $\hat{B}\cdot\bevec $ 	&	 $\Ht^{(5)TKTK}_\eff$	&&	 	&	 $2\hat{B}^\Jj{1} \bevec^J $ 	&	 $\gt^{(6)\Jj{1}TTTJ}_\eff$	\\
$\Ht^{(5)0jj3}_\eff$ 	&	 $\hat{B}^\Jj{1}$ 	&	 $\Ht^{(5)TKK\Jj{1}}_\eff$	&&	 	&	 $2\hat{B}^\Jj{1} \bevec^J $ 	&	 $\gt^{(6)\Jj{1}(TJ)KK}_\eff$	\\
 	&	 $\hat{B}^\Jj{1} \bevec^J $ 	&	 $\Ht^{(5)JTT\Jj{1}}_\eff$	&&	$\gt^{(6)j300j}_\eff$ 	&	 $\hat{B}^\Jj{1}$ 	&	 $\gt^{(6)K\Jj{1}TTK}_\eff$	\\
 	&	 $\hat{B}^\Jj{1} \bevec^J $ 	&	 $\Ht^{(5)KJK\Jj{1}}_\eff$	&&	 	&	 $2\hat{B}^\Jj{1} \bevec^J $ 	&	 $\gt^{(6)\Jj{1}KTJK}_\eff$	\\
 	&	 $\hat{B}\cdot\bevec $ 	&	 $\Ht^{(5)KTKT}_\eff$	&&	 	&	 $2\hat{B}^\Jj{1} \bevec^J $ 	&	 $\gt^{(6)\Jj{1}(JT)TT}_\eff$	\\
$\Ht^{(5)03jj}_\eff$ 	&	 $\hat{B}^\Jj{1}$ 	&	 $\Ht^{(5)T\Jj{1}KK}_\eff$	&&	 	&	 $\hat{B}\cdot\bevec $ 	&	 $\gt^{(6)TKTTK}_\eff$	\\
 	&	 $2\hat{B}^\Jj{1} \bevec^J $ 	&	 $\Ht^{(5)\Jj{1}TTJ}_\eff$	&&	$\gt^{(6)j3jkk}_\eff$ 	&	 $\hat{B}^\Jj{1}$ 	&	 $\gt^{(6)J\Jj{1}JKK}_\eff$	\\
 	&	 $\hat{B}^\Jj{1} \bevec^J $ 	&	 $\Ht^{(5)\Jj{1}JKK}_\eff$	&&	 	&	 $4\hat{B}^\Jj{1} \bevec^J $ 	&	 $\gt^{(6)\Jj{1}(KTKJ)}_\eff$	\\
$\Ht^{(5)0333}_\eff$ 	&	 $\hat{B}^\Jj{1}\hat{B}^\Jj{2}\hat{B}^\Jj{3}$ 	&	 $\Ht^{(5)T\Jj{1}\Jj{2}\Jj{3}}_\eff$	&&	 	&	 $\hat{B}\cdot\bevec $ 	&	 $\gt^{(6)TKKLL}_\eff$	\\
 	&	 $2\hat{B}^\Jj{1}\hat{B}^\Jj{2}\hat{B}\cdot\bevec $ 	&	 $\Ht^{(5)\Jj{1}TT\Jj{2}}_\eff$	&&	$\gt^{(6)03033}_\eff$ 	&	 $\hat{B}^\Jj{1}\hat{B}^\Jj{2}\hat{B}^\Jj{3}$ 	&	 $\gt^{(6)T\Jj{1}T\Jj{2}\Jj{3}}_\eff$	\\
 	&	 $\hat{B}^\Jj{1}\hat{B}^\Jj{2}\hat{B}^\Jj{3} \bevec^J $ 	&	 $\Ht^{(5)\Jj{1}J\Jj{2}\Jj{1}}_\eff$ 	&&	 	&	 $2 \hat{B}^\Jj{1}\hat{B}^\Jj{2}\hat{B}\cdot\bevec $ 	&	 $\gt^{(6)\Jj{1}TTT\Jj{2}}_\eff$	\\
$\gt^{(4)030}_\eff$ 	&	 $\hat{B}^\Jj{1}$ 	&	 $\gt^{(4)T\Jj{1}T}_\eff$	&&	 	&	 $2\hat{B}^\Jj{1}\hat{B}^\Jj{2}\hat{B}^\Jj{3} \bevec^J $ 	&	 $\gt^{(6)\Jj{1}(TJ)\Jj{2}\Jj{3}}_\eff$	\\
 	&	 $2\hat{B}^\Jj{1} \bevec^J $ 	&	 $\gt^{(4)\Jj{1}(TJ)}_\eff$	&&	$\gt^{(6jl3j33}_\eff$ 	&	 $\hat{B}^\Jj{1}\hat{B}^\Jj{2}\hat{B}^\Jj{3}$ 	&	 $\gt^{(6)K\Jj{1}K\Jj{2}J}_\eff$	\\
$\gt^{(4)j3j}_\eff$ 	&	 $\hat{B}^\Jj{1}$ 	&	 $\gt^{(4)K\Jj{1}K}_\eff$	&&	 	&	 $-3 \hat{B}^\Jj{1}\hat{B}^\Jj{2}\hat{B}\cdot\bevec $ 	&	 $\gt^{(6)K(\Jj{1}T\Jj{2})K}_\eff$	\\
 	&	 $2\hat{B}^\Jj{1} \bevec^J $ 	&	 $\gt^{(4)\Jj{1}(TJ)}_\eff$	&&	 	&	 $2\hat{B}^\Jj{1}\hat{B}^\Jj{2}\hat{B}^\Jj{3} \bevec^J $ 	&	 $\gt^{(6)\Jj{1}(TJ)\Jj{2}\Jj{3}}_\eff$	\\
 	&	 $\hat{B}\cdot\bevec $ 	&	 $\gt^{(4)TKK}_\eff$	&&		&		&		\\
\hline\hline
\end{tabular}
\label{appb-3a}
\end{table*}

\renewcommand\arraystretch{1.4}
\begin{table*}
\caption{Relations between spin-dependent cartesian coefficients 
in laboratory and Sun-centered frames for $7\leq d \leq 8$.}
\setlength{\tabcolsep}{4pt}
\begin{tabular}{ccccccc}
\hline\hline
Laboratory & Factor & Sun-centered & $\hskip 30pt$ & 
Laboratory & Factor & Sun-centered \\
\hline
$\Ht^{(7)030000}_\eff$ 	&	 $\hat{B}^\Jj{1}$ 	&	 $\Ht^{(7)T\Jj{1}TTTT}_\eff$	&&	$\gt^{(8)0300000}_\eff$ 	&	 $\hat{B}^\Jj{1}$ 	&	 $\gt^{(8)T\Jj{1}TTTTT}_\eff$	\\
 	&	 $ 5 \hat{B}^\Jj{1} \bevec^J$ 	&	 $\Ht^{(7)\Jj{1}(JTTTT)}_\eff$	&&	 	&	 $6 \hat{B}^\Jj{1} \bevec^J$ 	&	 $\gt^{(8)\Jj{1}(TTTTTJ)}_\eff$ 	\\
$\Ht^{(7)0j00j3}_\eff$ 	&	 $\hat{B}^\Jj{1}$ 	&	 $\Ht^{(7)TKTTK\Jj{1}}_\eff$	&&	$\gt^{(8)0j000j3}_\eff$ 	&	 $\hat{B}^\Jj{1}$ 	&	 $\gt^{(8)TKTTTK\Jj{1}}_\eff$	\\
 	&	 $\hat{B}^\Jj{1}\bevec^J$ 	&	 $\Ht^{(7)JTTTT\Jj{1}}_\eff$	&&	 	&	 $\hat{B}\cdot\bevec$ 	&	 $\gt^{(8)KTTTTTK}_\eff$	\\
 	&	 $3\hat{B}^\Jj{1}\bevec^J$ 	&	 $\Ht^{(7)K(JTT)K\Jj{1}}_\eff$	&&	 	&	 $\hat{B}^\Jj{1} \bevec^J$ 	&	 $\gt^{(8)JTTTTT\Jj{1}}_\eff$	\\
 	&	 $\hat{B}\cdot\bevec$ 	&	 $\Ht^{(7)KTTTTK}_\eff$	&&	 	&	 $4\hat{B}^\Jj{1} \bevec^J$ 	&	 $\gt^{(8)K(JTTT)\Jj{1}K}_\eff$	\\
$\Ht^{(7)0300jj}_\eff$ 	&	 $\hat{B}^\Jj{1}$ 	&	 $\Ht^{(7)T\Jj{1}TTKK}_\eff$	&&	$\gt^{(8)03000jj}_\eff$ 	&	 $\hat{B}^\Jj{1}$ 	&	 $\gt^{(8)T\Jj{1}TTTKK}_\eff$ 	\\
 	&	 $2\hat{B}^\Jj{1} \bevec^J$ 	&	 $\Ht^{(7)\Jj{1}TTTTJ}_\eff$	&&	 	&	 $2\hat{B}^\Jj{1}\bevec^J$ 	&	 $\gt^{(8)\Jj{1}TTTTTJ}_\eff$	\\
 	&	 $3\hat{B}^\Jj{1} \bevec^J$ 	&	 $\Ht^{(7)\Jj{1}(JTT)KK}_\eff$	&&	 	&	 $4\hat{B}^\Jj{1}\bevec^J$ 	&	 $\gt^{(8)\Jj{1}(JTTT)KK}_\eff$	\\
$\Ht^{(7)j3000j}_\eff$ 	&	 $\hat{B}^\Jj{1}$ 	&	 $\Ht^{(7)K\Jj{1}TTTK}_\eff$	&&	$\gt^{(8)j30000j}_\eff$ 	&	 $\hat{B}^\Jj{1}$ 	&	 $\gt^{(8)K\Jj{1}TTTTK}_\eff$	\\
 	&	 $2\hat{B}^\Jj{1} \bevec^J$ 	&	 $\Ht^{(7)\Jj{1}(JT)TTT}_\eff$	&&	 	&	 $2\hat{B}^\Jj{1}\bevec^J$ 	&	 $\gt^{(8)\Jj{1}(JT)TTTT}_\eff$	\\
 	&	 $3\hat{B}^\Jj{1} \bevec^J$ 	&	 $\Ht^{(7)\Jj{1}KTTJK}_\eff$	&&	 	&	 $4\hat{B}^\Jj{1}\bevec^J$ 	&	 $\gt^{(8)\Jj{1}KTTTKJ}_\eff$	\\
 	&	 $\hat{B}\cdot\bevec$ 	&	 $\Ht^{(7)TKTTTK}_\eff$	&&	 	&	 $\hat{B}\cdot\bevec$ 	&	 $\gt^{(8)TKTTTTK}_\eff$	\\
$\Ht^{(7)j30jkk}_\eff$ 	&	 $\hat{B}^\Jj{1}$ 	&	 $\Ht^{(7)J\Jj{1}TJKK}_\eff$	&&	$\gt^{(8)j300jkk}_\eff$ 	&	 $\hat{B}^\Jj{1}$ 	&	 $\gt^{(8)J\Jj{1}TTJKK}_\eff$	\\
 	&	 $2\hat{B}^\Jj{1}\bevec^J$ 	&	 $\Ht^{(7)\Jj{1}(JT)TKK}_\eff$	&&	 	&	 $\hat{B}\cdot\bevec$ 	&	 $\gt^{(8)TKTTKLL}_\eff$	\\
 	&	 $\hat{B}^\Jj{1}\bevec^J$ 	&	 $\Ht^{(7)\Jj{1}KJKLL}_\eff$	&&	 	&	 $4\hat{B}^\Jj{1} \bevec^J$ 	&	 $\gt^{(8)\Jj{1}(JTKK)TT}_\eff$	\\
 	&	 $2\hat{B}^\Jj{1}\bevec^J$ 	&	 $\Ht^{(7)\Jj{1}KTTKJ}_\eff$	&&	 	&	 $2\hat{B}^\Jj{1} \bevec^J$ 	&	 $\gt^{(8)\Jj{1}KTJKLL}_\eff$	\\
 	&	 $\hat{B}\cdot\bevec$ 	&	 $\Ht^{(7)TKTKJJ}_\eff$	&&	$\gt^{(8)030jjkk}_\eff$ 	&	 $\hat{B}^\Jj{1}$ 	&	 $\gt^{(8)T\Jj{1}TKKJJ}_\eff$	\\
$\Ht^{(7)03jjkk}_\eff$ 	&	 $\hat{B}^\Jj{1}$ 	&	 $\Ht^{(7)T\Jj{1}KKLL}_\eff$	&&	 	&	 $4\hat{B}^\Jj{1} \bevec^J$ 	&	 $\gt^{(8)\Jj{1}TTTJKK}_\eff$	\\
 	&	 $\hat{B}^\Jj{1}\bevec^J$ 	&	 $\Ht^{(7)\Jj{1}JLLKK}_\eff$	&&	 	&	 $2\hat{B}^\Jj{1} \bevec^J$ 	&	 $\gt^{(8)\Jj{1}(JT)LLKK}_\eff$	\\
 	&	 $4\hat{B}^\Jj{1}\bevec^J$ 	&	 $\Ht^{(7)\Jj{1}TTJKK}_\eff$	&&	$\gt^{(8)0j0jkk3}_\eff$ 	&	 $\hat{B}^\Jj{1}$ 	&	 $\gt^{(8)TKTKLL\Jj{1}}$	\\
$\Ht^{(7)0jjkk3}_\eff$ 	&	 $\hat{B}^\Jj{1}$ 	&	 $\Ht^{(7)TKKJJ\Jj{1}}_\eff$	&&	 	&	 $\hat{B}\cdot\bevec$ 	&	 $\gt^{(8)KTTTKLL}_\eff$	\\
 	&	 $\hat{B}^\Jj{1} \bevec^J$ 	&	 $\Ht^{(7)LJLKK\Jj{1}}_\eff$	&&	 	&	 $2\hat{B}^\Jj{1} \bevec^J$ 	&	 $\gt^{(8)K(JT)KLL\Jj{1}}_\eff$	\\
 	&	 $-3\hat{B}^\Jj{1} \bevec^J$ 	&	 $\Ht^{(7)T(KKJ)\Jj{1}T}_\eff $	&&	 	&	 $-3\hat{B}^\Jj{1} \bevec^J$ 	&	 $\gt^{(8)T(KKJ)TT\Jj{1}}_\eff$	\\
 	&	 $\hat{B}\cdot\bevec$ 	&	 $\Ht^{(7)KTTKLL}_\eff $ 	&&	$\gt^{(8)j300j33}_\eff$ 	&	 $\hat{B}^\Jj{1}\hat{B}^\Jj{2}\hat{B}^\Jj{3}$ 	&	 $\gt^{(8)K\Jj{1}TTK\Jj{2}\Jj{3}}_\eff$	\\
$\Ht^{(7)030033}_\eff$ 	&	 $\hat{B}^\Jj{1}\hat{B}^\Jj{2}\hat{B}^\Jj{3}$ 	&	 $\Ht^{(7)T\Jj{1}TT\Jj{2}\Jj{3}}_\eff$	&&	 	&	 $2\hat{B}^\Jj{1}\hat{B}^\Jj{2}\hat{B}^\Jj{3}\bevec^J$ 	&	 $\gt^{(8)\Jj{1}KT\Jj{2}K\Jj{3}}_\eff$	\\
&	 $3\hat{B}^\Jj{1}\hat{B}^\Jj{2}\hat{B}^\Jj{3}\bevec^J$ 	&	 $\Ht^{(7)\Jj{1}(JTT)\Jj{2}\Jj{3}}_\eff$	&&	 	&	 $-3\hat{B}^\Jj{1}\hat{B}^\Jj{2}\hat{B}\cdot\bevec$ 	&	 $\gt^{(8)K(\Jj{1}\Jj{2}T)TTK}_\eff$	\\	
 	&	 $2\hat{B}^\Jj{1}\hat{B}^\Jj{2}\hat{B}\cdot\bevec $ 	&	 $\Ht^{(7)\Jj{1}TTTT\Jj{2}}_\eff$	&&	 	&	 $2\hat{B}^\Jj{1}\hat{B}^\Jj{2}\hat{B}^\Jj{3}\bevec^J$ 	&	 $\gt^{(8)\Jj{1}(JT)TT\Jj{2}\Jj{3}}_\eff$	\\
$\Ht^{(7)j30j33}_\eff$ 	&	 $\hat{B}^\Jj{1}\hat{B}^\Jj{2}\hat{B}^\Jj{3}$ 	&	 $\Ht^{(7)K\Jj{1}TK\Jj{2}\Jj{3}}_\eff$	&&	$\gt^{(8)0j0j333}_\eff$ 	&	 $\hat{B}^\Jj{1}\hat{B}^\Jj{2}\hat{B}^\Jj{3}$ 	&	 $\gt^{(8)TKTK\Jj{1}\Jj{2}\Jj{3}}_\eff$	\\
 	&	 $2\hat{B}^\Jj{1}\hat{B}^\Jj{2}\hat{B}^\Jj{3}\bevec^J$ 	&	 $\Ht^{(7)\Jj{1}(JT)T\Jj{2}\Jj{3}}_\eff$	&&	 	&	 $3\hat{B}^\Jj{1}\hat{B}^\Jj{2}\hat{B}\cdot\bevec$ 	&	 $\gt^{(8)KTTTK\Jj{1}\Jj{2}}_\eff$	\\
 	&	 $\hat{B}^\Jj{1}\hat{B}^\Jj{2}\hat{B}^\Jj{3}\bevec^J$ 	&	 $\Ht^{(7)\Jj{1}KJK\Jj{2}\Jj{3}}_\eff$	&&	 	&	 $\hat{B}^\Jj{1}\hat{B}^\Jj{2}\hat{B}^\Jj{3}\bevec^J$ 	&	 $\gt^{(8)JTTT\Jj{1}\Jj{2}\Jj{3}}_\eff$	\\
 	&	 $-3\hat{B}^\Jj{1}\hat{B}^\Jj{2}\hat{B}\cdot\bevec$ 	&	 $\Ht^{(7)K(T\Jj{1}\Jj{2})TK}_\eff$	&&	 	&	 $2\hat{B}^\Jj{1}\hat{B}^\Jj{2}\hat{B}^\Jj{3}\bevec^J$ 	&	 $\gt^{(8)K(JT)K\Jj{1}\Jj{2}\Jj{3}}_\eff$	\\
$\Ht^{(7)0jj333}_\eff$ 	&	 $\hat{B}^\Jj{1}\hat{B}^\Jj{2}\hat{B}^\Jj{3}$ 	&	 $\Ht^{(7)TKK\Jj{1}\Jj{2}\Jj{3}}_\eff$	&&	$\gt^{(8)030jj33}_\eff$ 	&	 $\hat{B}^\Jj{1}\hat{B}^\Jj{2}\hat{B}^\Jj{3}$ 	&	 $\gt^{(8)T\Jj{1}TKK\Jj{2}\Jj{3}}_\eff$	\\
 	&	 $3\hat{B}^\Jj{1}\hat{B}^\Jj{2}\hat{B}\cdot\bevec$ 	&	 $\Ht^{(7)KTTK\Jj{1}\Jj{2}}_\eff$	&&	 	&	 $2\hat{B}^\Jj{1}\hat{B}^\Jj{2}\hat{B}\cdot\bevec$ 	&	 $\gt^{(8)\Jj{1}TTT\Jj{2}KK}_\eff$	\\
 	&	 $\hat{B}^\Jj{1}\hat{B}^\Jj{2}\hat{B}^\Jj{3}\bevec^J$ 	&	 $\Ht^{(7)JTT\Jj{1}\Jj{2}\Jj{3}}_\eff$	&&	 	&	 $2\hat{B}^\Jj{1}\hat{B}^\Jj{2}\hat{B}^\Jj{3}\bevec^J$ 	&	 $\gt^{(8)\Jj{1}TTT\Jj{2}\Jj{3}J}_\eff$	\\
 	&	 $\hat{B}^\Jj{1}\hat{B}^\Jj{2}\hat{B}^\Jj{3}\bevec^J$ 	&	 $\Ht^{(7)KJK\Jj{1}\Jj{2}\Jj{3}}_\eff$	&&	 	&	 $2\hat{B}^\Jj{1}\hat{B}^\Jj{2}\hat{B}^\Jj{3}\bevec^J$ 	&	 $\gt^{(8)\Jj{1}(JT)\Jj{2}\Jj{3}KK}_\eff$	\\
$\Ht^{(7)03jj33}_\eff$ 	&	 $\hat{B}^\Jj{1}\hat{B}^\Jj{2}\hat{B}^\Jj{3}$ 	&	 $\Ht^{(7)T\Jj{1}KK\Jj{2}\Jj{3}}_\eff$	&&	$\gt^{(8)0300033}_\eff$ 	&	 $\hat{B}^\Jj{1}\hat{B}^\Jj{2}\hat{B}^\Jj{3}$ 	&	 $\gt^{(8)T\Jj{1}TTT\Jj{2}\Jj{3}}_\eff$	\\
 	&	 $2\hat{B}^\Jj{1}\hat{B}^\Jj{2} \hat{B}\cdot\bevec$ 	&	 $\Ht^{(7)\Jj{1}TT\Jj{2}KK}_\eff$	&&	 	&	 $4\hat{B}^\Jj{1}\hat{B}^\Jj{2}\hat{B}^\Jj{3} \bevec^J$ 	&	 $\gt^{(8)\Jj{1}(TTTJ)\Jj{2}\Jj{3}}_\eff$	\\
 	&	 $2\hat{B}^\Jj{1}\hat{B}^\Jj{2}\hat{B}^\Jj{3}\bevec^J$ 	&	 $\Ht^{(7)\Jj{1}TTJ\Jj{2}\Jj{3}}_\eff$	&&	 	&	 $2\hat{B}^\Jj{1}\hat{B}^\Jj{2}\hat{B}\cdot\bevec$ 	&	 $\gt^{(8)\Jj{1}TTTTT\Jj{2}}_\eff$	\\
	&	 $\hat{B}^\Jj{1}\hat{B}^\Jj{2}\hat{B}^\Jj{3}\bevec^J$ 	&	 $\Ht^{(7)\Jj{1}JKK\Jj{2}\Jj{3}}_\eff$	&&	$\gt^{(8)0303333}_\eff$ 	&	 $\hat{B}^\Jj{1}\hat{B}^\Jj{2}\hat{B}^\Jj{3}\hat{B}^\Jj{4}\hat{B}^\Jj{5}$ 	&	 $\gt^{(8)T\Jj{1}T\Jj{2}\Jj{3}\Jj{4}\Jj{5}}_\eff$	\\
	&		&		&&	 	&	 $2\hat{B}^\Jj{1}\hat{B}^\Jj{2}\hat{B}^\Jj{3}\hat{B}^\Jj{4}\hat{B}^\Jj{5} \bevec^J$ 	&	 $\gt^{(8)\Jj{1}(JT)\Jj{2}\Jj{3}\Jj{4}\Jj{5}}_\eff$	\\
	&		&		&&	 	&	 $4\hat{B}^\Jj{1}\hat{B}^\Jj{2}\hat{B}^\Jj{3} \hat{B}^\Jj{4}\hat{B}\cdot\bevec$ 	&	 $\gt^{(8)\Jj{1}TTT\Jj{2}\Jj{3}\Jj{4}}_\eff$ 	\\
\hline\hline
\end{tabular}
\label{appb-3b}
\end{table*}

\renewcommand\arraystretch{1.6}
\begin{table*}
\caption{The quantities $\Tg{0B,kj}{d}{JK\ldots M}$ 
and $\Tg{1B,kj}{d}{JK\ldots M}$ for $3 \leq d \leq 8$.} 
\setlength{\tabcolsep}{2pt}
\begin{tabular}{llcll}
\hline\hline
$\Tg{0B,kj}{d}{JK\ldots M}$ & Combination & $\hskip5pt$ & 
$\Tg{1B,kj}{d}{JK\ldots M}$ & Combination \\
\hline
$\Tg{0B,01}{3}{J\Jj{1}}$	 & 	$\Ht^{(3)\Jj{1}J}_\eff$	&&	$\Tg{1B,21}{5}{J\Jj{1}}$ 	 & 	 $\frac{1}{5}(15\Ht^{(5)\Jj{1}(JTT)}_\eff+2\Ht^{(5)T(\Jj{1}J)T}_\eff$	\\
$\Tg{0B,01}{4}{J\Jj{1}}$	 & 	$2\gt^{(4)\Jj{1}(JT)}_\eff$	&&	                                 	 & 	  $+6\Ht^{(5)\Jj{1}(JKK)}_\eff-2\Ht^{(5)K(\Jj{1}J)K}_\eff$	\\
$\Tg{0B,01}{5}{J\Jj{1}}$	 & 	$3\Ht^{(5)\Jj{1}(JTT)}_\eff$	&&		 & 	 $\hskip30pt +6\Ht^{(5)TKTK}_\eff\de_{J\Jj{1}})$	\\
$\Tg{0B,01}{6}{J\Jj{1}}$	 & 	$4\gt^{(6)\Jj{1}(JTTT)}_\eff$	&&	$\Tg{1B,21}{6}{J\Jj{1}}$ 	 & 	 $\frac{1}{10}(60\gt^{(6)\Jj{1}(JTTT)}_\eff+12\gt^{(6)T(\Jj{1}J)TT}_\eff$	\\
$\Tg{0B,01}{7}{J\Jj{1}}$	 & 	$5\Ht^{(7)\Jj{1}(JTTTT)}_\eff$	&&	                                 	 & 	  $+48\gt^{(6)\Jj{1}(TJKK)}_\eff-18\gt^{(6)K(\Jj{1}JT)K}_\eff$	\\
$\Tg{0B,01}{8}{J\Jj{1}}$	 & 	$6\gt^{(8)\Jj{1}(JTTTTT)}_\eff$	&&		 & 	 $\hskip30pt +21\gt^{(6)TKTTK}_\eff\de_{J\Jj{1}})$	\\
$\Tg{0B,21}{5}{J\Jj{1}}$	 & 	$-\frac{1}{5}(3\Ht^{(5)J(\Jj{1}KK)}_\eff+4\Ht^{(5)T(\Jj{1}J)T}_\eff$	&&	$\Tg{1B,21}{7}{J\Jj{1}}$ 	 & 	 $\frac{2}{5}(25\Ht^{(7)\Jj{1}(TTTTJ)}_\eff+6\Ht^{(7)T(\Jj{1}J)TTT}_\eff$	\\
	 & 	$\hskip30pt +2\Ht^{(5)TKTK}_\eff\de_{J\Jj{1}})$	&&	                                 	 & 	  $+30\Ht^{(7)\Jj{1}(TTJKK)}_\eff-12\Ht^{(7)K(TTJ\Jj{1})K}_\eff$	\\
$\Tg{0B,21}{6}{J\Jj{1}}$	 & 	$-\frac{3}{5}(3\gt^{(6)J(\Jj{1}KK)T}_\eff+4\gt^{(6)T(\Jj{1}J)TT}_\eff$	&&		 & 	 $\hskip30pt +8\Ht^{(7)TKTTTK}_\eff\de_{J\Jj{1}})$	\\
	 & 	$+3\gt^{(6)T(\Jj{1}KK)J}_\eff+2\gt^{(6)TKTTK}_\eff\de_{J\Jj{1}})$	&&	$\Tg{1B,21}{8}{J\Jj{1}}$ 	 & 	 $\frac{1}{2}(30\gt^{(8)\Jj{1}(JTTTTT)}_\eff+8\gt^{(8)T(\Jj{1}J)TTTT}_\eff$	\\
$\Tg{0B,21}{7}{J\Jj{1}}$	 & 	$-\frac{6}{5}(3\Ht^{(7)J(\Jj{1}KK)TT}_\eff+6\Ht^{(7)T(\Jj{1}KK)TJ}_\eff$	&&	                                 	 & 	  $+48\gt^{(8)\Jj{1}(TTTJKK)}_\eff-20\gt^{(8)K(\Jj{1}JTTT)K}_\eff$	\\
	 & 	$+4\Ht^{(7)T(\Jj{1}J)TTT}_\eff+2\Ht^{(7)TKTTTK}_\eff\de_{J\Jj{1}})$	&&		 & 	 $\hskip30pt +9\gt^{(8)TKTTTTK}_\eff\de_{J\Jj{1}})$	\\
$\Tg{0B,21}{8}{J\Jj{1}}$	 & 	$-2(3\gt^{(8)J(\Jj{1}KK)TTT}_\eff+9\gt^{(8)T(\Jj{1}KK)TTJ}_\eff$	&&	$\Tg{1B,41}{7}{J\Jj{1}}$ 	 & 	 $\frac{3}{35}(70\Ht^{(7)\Jj{1}(JTTKK)}_\eff-4\Ht^{(7)K(J\Jj{1})KLL}_\eff$	\\
	 & 	$+4\gt^{(8)T(\Jj{1}J)TTTT}_\eff+2\gt^{(8)TKTTTTK}_\eff\de_{J\Jj{1}})$	&&	                                 	 & 	  $+15\Ht^{(7)\Jj{1}(JLLKK)}_\eff+8\Ht^{(7)T(J\Jj{1}LL)T}_\eff$	\\
$\Tg{0B,41}{7}{J\Jj{1}}$	 & 	$-\frac{3}{35}(16\Ht^{(7)T(\Jj{1}JKK)T}_\eff+5\Ht^{(7)J(\Jj{1}KKLL)}_\eff$	&&		 & 	 $\hskip30pt +16\Ht^{(7)TKKLLT}_\eff\de^{J\Jj{1}})$	\\
	 & 	$\hskip30pt +4\Ht^{(7)TKKLLT}_\eff\de_{J\Jj{1}})$	&&	$\Tg{1B,41}{8}{J\Jj{1}}$ 	 & 	 $\frac{3}{7}(18\gt^{(8)\Jj{1}(JTKKLL)}_\eff-2\gt^{(8)K(J\Jj{1}T)KLL}_\eff$	\\
$\Tg{0B,41}{8}{J\Jj{1}}$	 & 	$-\frac{3}{7}(16\gt^{(8)T(\Jj{1}JKK)TT}_\eff+5\gt^{(8)J(\Jj{1}KKLL)T}_\eff$	&&	                                 	 & 	  $+8\gt^{(8)T(KK\Jj{1}J)TT}_\eff+42\gt^{(8)\Jj{1}(JLLTTT)}_\eff$	\\
	 & 	$+5\gt^{(8)T(\Jj{1}KKLL)J}_\eff+4\gt^{(8)TKKLLTT}_\eff\de_{J\Jj{1}})$	&&		 & 	 $\hskip30pt +9\gt^{(8)TKKLLTT}_\eff\de^{J\Jj{1}})$	\\
$\Tg{0B,23}{5}{J\Jj{1}\Jj{2}\Jj{3}}$	 & 	$-(\Ht^{(5)J\Jj{1}\Jj{2}\Jj{3}}_\eff+2\de^{J\Jj{1}}\Ht^{(5)T\Jj{2}T\Jj{3}}_\eff)$	&&	$\Tg{1B,43}{7}{J\Jj{1}}$ 	 & 	 $-\frac{\sqrt{6}}{15}(15\Ht^{(7)\Jj{1}(JKKTT)}_\eff$	\\
$\Tg{0B,23}{6}{J\Jj{1}\Jj{2}\Jj{3}}$	 & 	$6(\gt^{(6)\Jj{1}(JT)\Jj{2}\Jj{3}}_\eff-\de^{J\Jj{1}}\gt^{(6)T\Jj{2}\Jj{3}TT}_\eff)$	&&		 & 	 $-2\Ht^{(7)K(J\Jj{1})KLL}_\eff+5\Ht^{(7)\Jj{1}(JLLKK)}_\eff$	\\
$\Tg{0B,23}{7}{J\Jj{1}\Jj{2}\Jj{3}}$	 & 	$18\Ht^{(7)\Jj{1}(JTT)\Jj{2}\Jj{3}}_\eff-12\de^{J\Jj{1}}\Ht^{(7)T\Jj{2}\Jj{3}TTT}_\eff$	&&		 & 	 $-4\Ht^{(7)T(J\Jj{1}LL)T}_\eff+2\Ht^{(7)TKKLLT}_\eff\de^{J\Jj{1}})$	\\
$\Tg{0B,23}{8}{J\Jj{1}\Jj{2}\Jj{3}}$	 & 	$40\gt^{(8)\Jj{1}(JTTT)\Jj{2}\Jj{3}}_\eff$	&&	$\Tg{1B,43}{8}{J\Jj{1}}$ 	 & 	 $-\frac{1}{\sqrt{6}}(12\gt^{(8)\Jj{1}(JTKKLL)}_\eff$	\\
	 & 	$\hskip30pt -20\de^{J\Jj{1}}\gt^{(8)T(\Jj{2}\Jj{3})TTTT}_\eff$	&&		 & 	 $+6\gt^{(8)K(J\Jj{1}T)KLL}_\eff-8\gt^{(8)T(KK\Jj{1}J)TT}_\eff$	\\
$\Tg{0B,43}{7}{J\Jj{1}\Jj{2}\Jj{3}}$	 & 	$-\frac{2}{9}(8\Ht^{(7)T(J\Jj{1}\Jj{2}\Jj{3})T}_\eff+5\Ht^{(7)J(\Jj{1}\Jj{2}\Jj{3}KK)}_\eff$	&&		 & 	 $+18\gt^{(8)\Jj{1}(JLLTTT)}_\eff+\gt^{(8)TKKLLTT}_\eff\de^{J\Jj{1}})$	\\
	 & 	$\hskip30pt +12\de^{J\Jj{1}}\Ht^{(7)T(\Jj{2}\Jj{3}KK)T}_\eff)$	&&	$\Tg{1B,43}{7}{J\Jj{1}\Jj{2}\Jj{3}}$ 	 & 	 $-\frac{\sqrt{6}}{9}(12\Ht^{(7)K(J\Jj{1}\Jj{2}\Jj{3})K}_\eff$	\\
$\Tg{0B,43}{8}{J\Jj{1}\Jj{2}\Jj{3}}$	 & 	$-\frac{10}{9}(8\gt^{(8)T(J\Jj{1}\Jj{2}\Jj{3})TT}_\eff$	&&		 & 	 $+5\Ht^{(7)J(\Jj{1}\Jj{2}\Jj{3}KK)}_\eff-27\Ht^{(7)\Jj{1}(JTT)\Jj{2}\Jj{3}}_\eff$	\\
	 & 	$+5\gt^{(8)J(\Jj{1}\Jj{2}\Jj{3}KK)T}_\eff+5\gt^{(8)T(\Jj{1}\Jj{2}\Jj{3}KK)J}_\eff$	&&		 & 	 $\hskip30pt -4\Ht^{(7)T(J\Jj{1}\Jj{2}\Jj{3})T}_\eff$	\\
	 & 	$\hskip30pt +12\de^{J\Jj{1}}\gt^{(7)T(\Jj{2}\Jj{3}KK)TT}_\eff)$	&&		 & 	 $\hskip30pt +12\de^{J\Jj{1}}\Ht^{(7)T(\Jj{2}\Jj{3}KK)T}_\eff$	\\
$\Tg{0B,45}{7}{J\Jj{1}\Jj{2}\Jj{3}\Jj{4}\Jj{5}}$	 & 	$-(\Ht^{(7)J\Jj{1}\Jj{2}\Jj{3}\Jj{4}\Jj{5}}_\eff+\de^{J\Jj{1}}\Ht^{(7)T\Jj{2}\Jj{3}\Jj{4}\Jj{5}T}_\eff)$	&&		 & 	 $\hskip30pt +36\de^{J\Jj{1}}\Ht^{(7)K(\Jj{2}\Jj{3}TT)K}_\eff$)	\\
$\Tg{0B,45}{8}{J\Jj{1}\Jj{2}\Jj{3}\Jj{4}\Jj{5}}$	 & 	$-5(\gt^{(8)J\Jj{1}\Jj{2}\Jj{3}\Jj{4}\Jj{5}T}_\eff+\gt^{(8)T\Jj{1}\Jj{2}\Jj{3}\Jj{4}\Jj{5}J}_\eff$	&&	$\Tg{1B,43}{8}{J\Jj{1}\Jj{2}\Jj{3}}$ 	 & 	 $-\frac{5}{3\sqrt{6}}( 10\gt^{(8)K(J\Jj{1}\Jj{2}\Jj{3}T)K}_\eff$	\\
	 & 	$\hskip30pt +4\de^{J\Jj{1}}\gt^{(8)T\Jj{2}\Jj{3}\Jj{4}\Jj{5}TT}_\eff)$	&&		 & 	 $-8\gt^{(8)T(J\Jj{1}\Jj{2}\Jj{3})TT}_\eff-24\gt^{(8)\Jj{1}(JKKT)\Jj{2}\Jj{3}}_\eff$	\\
	 & 		&&		 & 	 $\hskip30pt -36\gt^{(8)\Jj{1}(JTTT)\Jj{2}\Jj{3}}_\eff$	\\
	 & 		&&		 & 	 $\hskip30pt +24\de^{J\Jj{1}}\gt^{(8)T(\Jj{2}\Jj{3}KK)TT}_\eff$	\\
	 & 		&&		 & 	 $\hskip30pt +45\de^{J\Jj{1}}\gt^{(8)K(\Jj{2}\Jj{3}TTT)K}_\eff$)	\\
\hline\hline
\end{tabular}
\label{appb-45}
\end{table*}

Constraints on the coefficients for Lorentz violation 
are commonly reported in the Sun-centered frame
\cite{tables}. 
This appendix describes the conversion 
of coefficients for Lorentz violation in a laboratory frame 
into combinations of coefficients in the Sun-centered frame,
including effects at zeroth and linear boost order.
The primary focus here is on effective cartesian coefficients,
which are better suited for boost analyses. 
We use Greek indices to denote spacetime indices
and Latin indices to represent spatial components. 
Generic indices in the laboratory frame
are represented by lowercase letters,
while indices in the Sun-centered frame are represented 
by uppercase ones.
For definiteness,
we label cartesian components in the laboratory frame by 0,1,2,3
and assume that the Lorentz transformation is given by Eq.\ \rf{LTlinear} 
with $\mathcal{R}^{3}{}_{J}=\hat{B}^J$.
Cartesian components in the Sun-centered frame 
are denoted by $T$, $X$, $Y$, $Z$,
and contractions of spatial uppercase indices 
imply summation over components in the Sun-centered frame. 

Consider first the effective cartesian coefficients associated
with spin-independent Lorentz and CPT violation.
The expressions for these effective cartesian coefficients 
in the laboratory frame
in terms of effective cartesian coefficients in the Sun-centered frame
can be reconstructed at linear boost order
from the information contained in Table \ref{appb-1}.
The table limits attention to coefficients in the laboratory frame
that contribute to the spherical coefficients with $3\leq d \leq 8$
discussed in Appendix \ref{sphtocar},
which are the ones relevant to the clock-comparison experiments
analyzed in this work.
The table contains two triplets of columns.
In each triplet,
the first column lists the cartesian components of interest
in the laboratory frame.
Entries in the second column are factors 
involving the boost $-\bevec^J$
and the direction $\hat{B}^J$ of the magnetic field.
The third column lists the relevant cartesian components
in the Sun-centered frame.
The expression converting a given coefficient
from the laboratory frame to the Sun-centered frame
is obtained by multiplying the entries in the second and third columns
and adding the associated rows.
For example,
the first two rows of the table generate the equation
$a^{(3)t}_\eff=a^{(3)T}_\eff-\bevec^J a^{(3)J}_\eff$.
Using the contents of this table and the results 
in Appendix \ref{sphtocar},
it is straightforward to convert 
spin-independent spherical coefficients in the laboratory frame
to effective cartesian coefficients in the Sun-centered frame 
at linear boost order.

To analyze the experiments discussed in this work,
it is useful to find analogous expressions 
converting the nonrelativistic coefficients for Lorentz violation
to the Sun-centered frame.
The nonrelativistic coefficients are 
combinations of spherical coefficients for Lorentz violation 
of arbitrary mass dimension,
as illustrated in Eqs.\ (111) and (112) of Ref.\ \cite{km13}.
All the spherical coefficients contributing
to a particular nonrelativistic coefficient 
behave the same way under rotations,
so at zeroth boost order the conversion between frames
is given by the comparatively simple result \rf{ltos}. 
However,
the spherical coefficients transform differently under boosts,
so converting nonrelativistic coefficients at linear boost order 
becomes involved.
In contrast, 
the effective cartesian coefficients 
have comparatively simple transformations under boosts
and so are better suited for studying boost effects. 

To circumvent this issue,
we limit attention here to terms involving effective cartesian coefficients 
that contribute at zeroth order in $\pmag/m_\f$,
which yields the dominant contributions at linear boost order
and suffices for the experimental analyses of interest.
With this assumption,
the spin-independent nonrelativistic coefficients $\Vnr{kjm}$
in the laboratory frame 
are expressed in terms of spherical coefficients as 
\beq
\Vnr{kjm} \approx \sum_d \m^{d-3-k} \V^{(d)}_{kjm}.
\label{b1}
\eeq
The spherical coefficients can then be translated 
into effective cartesian coefficients in the laboratory frame 
using the results in Appendix \ref{sphtocar}.
To perform the conversion between the laboratory frame
and the Sun-centered frame,
we note that any nonrelativistic coefficient $\K^\nr_{kjm}$
can be expanded to linear boost order as
\beq
\K^\nr_{kjm}\approx
\K^\nr_{kjm}\Big|_{\bevec^J=0}
+\Bigg(\dfrac{\partial\K_{kjm}^\nr}{\partial\bevec^J}
\Big|_{\bevec^J=0}\Bigg)\bevec^J.
\label{NRtoSF}
\eeq
For all coefficients,
the zeroth-order term is given by Eq.\ \rf{ltos}. 

At linear boost order,
we are interested in the contribution $\Vnr{kj0}(\O({\be}))$
to the nonrelativistic spin-independent coefficients with $m=0$.
Decomposing this contribution as a polynomial in 
the unit vector $\hat{B}^J$ along the magnetic field
yields the result
\bea
\Vnr{kj0}(\O({\be}))&=&
\Bigg(\dfrac{\partial\Vnr{kj0}}{\partial\bevec^J}
\Big|_{\bevec^J=0}\Bigg)\bevec^J
\nonumber\\
&=&
-\sqrt{\dfrac{4\pi}{2j+1}} m^{d-k-3}
\Bigg[\sum_{d=3}^8\Va{kj}{d}{J}\bevec^J
\nonumber\\
&&
+\sum_{d=5}^8\Va{kj}{d}{J\Jj{1}\Jj{2}}
\bevec^J\hat{B}^\Jj{1}\hat{B}^\Jj{2}
\nonumber\\
&&
+\sum_{d=7}^8\Va{kj}{d}{J\Jj{1}\Jj{2}\Jj{3}\Jj{4}}
\bevec^J\hat{B}^\Jj{1}\hat{B}^{\Jj{2}}\hat{B}^\Jj{3}\hat{B}^{\Jj{4}}
\Bigg].
\nonumber\\
\eea
The quantities $\Va{kj}{d}{J \Jj{1}\dots \Jj{n}}$ 
with $d-3-k<0$ vanish. 
For $3\leq d \leq 8$,
Table \ref{appb-2} provides explicit expressions for 
many nonvanishing $\Va{kj}{d}{J \Jj{1}\dots \Jj{n}}$ 
in terms of combinations of effective cartesian coefficients
in the Sun-centered frame.
The other quantities of relevance can be obtained 
from entries in this table using the relations
\bea
\Va{22}{d}{J}&=&-\Va{20}{d}{J},
\quad 
\Va{42}{d}{J}=-\tfrac{10}{7}\Va{40}{d}{J},
\nonumber\\
\Va{44}{d}{J}&=&\tfrac{3}{7}\Va{40}{d}{J},
\quad
\Va{44}{d}{J\Jj{1}\Jj{2}}=-\Va{42}{d}{J\Jj{1}\Jj{2}}.
\quad
\label{relboostsi}
\eea

With the above results for spin-independent coefficients in hand,
we next consider spin-dependent effects.
The relations connecting the sets 
of spin-dependent effective cartesian coefficients
in the laboratory frame and the Sun-centered frame
up to linear boost order can be found using the information 
in Tables \ref{appb-3a} and \ref{appb-3b}.
These tables restrict attention to coefficients
with $3\leq d \leq 6$ and $7\leq d \leq 8$,
respectively,
which are the ones of relevance 
to our analysis of clock-comparison experiments.
Each table contains two triplets of columns,
and each triplet has the same structure
as that of Table \ref{appb-1}.
Taking products of the second and third entries in a row
and summing over rows relevant 
to the chosen laboratory-frame coefficient 
yields the desired equation converting
the effective cartesian coefficients
from the laboratory to the Sun-centered frame,
as before.

In parallel with the above discussion for spin-independent effects,
the analysis of experiments is facilitated
by translating nonrelativistic coefficients for Lorentz violation
in the laboratory frame 
to expressions involving effective cartesian coefficients 
in the Sun-centered frame. 
Adopting the assumptions leading to Eq.\ \rf{b1},
the spin-dependent nonrelativistic coefficients
can be approximated in terms of spherical coefficients as
\bea
\TzBnr{kjm} &\approx& \sum_{d} \m^{d-3-k} 
(k+1) \T^{(d)(0B)}_{kjm} ,
\nonumber\\
\ToBnr{kjm} &\approx& \sum_d \m^{d-3-k} 
\Big(\T^{(d)(1B)}_{kjm}+\sqrt{\tfrac{j(j+1)}{2}} 
\T^{(d)(0B)}_{kjm} \Big).
\nonumber\\
\eea
The spherical coefficients can then in turn
be converted to effective cartesian coefficients
using the results in Appendix \ref{sphtocar}.
The conversion can be implemented to linear boost order
via Eq.\ \rf{NRtoSF},
where the zeroth-order term is again given by Eq.\ \rf{ltos}.

At linear boost order,
the relevant spin-dependent nonrelativistic coefficients 
$\TqBnr{kj0}(\cO(\be))$ have $m=0$.
Expanding them in powers of the unit vector $\hat{B}^J$
along the magnetic field,
we obtain 
\bea
\TqBnr{kj0}(\cO(\be))&=&
\Bigg(\dfrac{\partial\TqBnr{kj0}}{\partial\bevec^J}
\Big|_{\bevec^J=0}\Bigg)\bevec^J
\nonumber\\
&&
\hskip-60pt
=\sqrt{\dfrac{4\pi}{2j+1}} m^{d-k-3}
\Bigg[
\sum_{d=3}^8\Tg{sB, kj}{d}{J \Jj{1}}
\bevec^J\hat{B}^\Jj{1}
\nonumber\\
&&
\hskip-60pt
+\sum_{d=5}^8
\Tg{sB, kj}{d}{J\Jj{1}\Jj{2}\Jj{3}}
\bevec^J\hat{B}^\Jj{1}\hat{B}^\Jj{2}\hat{B}^\Jj{3}
\nonumber\\
&&
\hskip-60pt
+\sum_{d=7}^8
\Tg{sB, kj}{d}{J\Jj{1}\Jj{2}\Jj{3}\Jj{4}\Jj{5}}
\bevec^J\hat{B}^\Jj{1}\hat{B}^\Jj{2}\hat{B}^\Jj{3}
\hat{B}^\Jj{4}\hat{B}^\Jj{5}\Bigg],
\nonumber\\
\eea
where the quantities $\Tg{sB,kj}{d}{J\Jj{1}\dots \Jj{n}}$ 
with $d-3-k<0$ vanish. 
Explicit expressions 
for nonvanishing $\Tg{0B,kj}{d}{J\Jj{1}\dots \Jj{n}}$ 
in terms of effective cartesian coefficients in the Sun-centered frame
can be found in the first two columns of Table \ref{appb-45} 
and by using the relations
\bea
\Tg{0B,23}{d}{J\Jj{1}}&=&-\Tg{0B,21}{d}{J\Jj{1}},
\quad
\Tg{0B,43}{d}{J\Jj{1}}=-\tfrac{14}{9}\Tg{0B,41}{d}{J\Jj{1}},
\nonumber\\
\Tg{0B,45}{d}{J\Jj{1}\Jj{2}\Jj{3}}&=&-\Tg{0B,43}{d}{J\Jj{1}\Jj{2}\Jj{3}},
\quad
\Tg{0B,45}{d}{J\Jj{1}}=\tfrac{5}{9}\Tg{0B,41}{d}{J\Jj{1}}.
\nonumber\\
\eea
The nonvanishing quantities $\Tg{1B,kj}{d}{J\Jj{1}\dots \Jj{n}}$
are compiled in the second pair of columns
of Table \ref{appb-45}.

In working with these results,
the reader is cautioned that
the coefficients $\TzBnr{kjm}$ and $\ToBnr{kjm}$ with $j=k+1$ 
are linearly dependent at zeroth order in $\pmag_\f/m_\f$
because the spherical coefficients $\T^{(d)(1B)}_{kjm}$ vanish for $j=k+1$. 
One implication of this,
for instance,
is the existence of the relationships 
\bea
\Tg{1B,k(k+1)}{d}{J\Jj{1}}&=&
\sqrt{\dfrac{k+2}{2(k+1)}}\Tg{0B,k(k+1)}{d}{J\Jj{1}},
\nonumber\\
\Tg{1B,k(k+1)}{d}{J\Jj{1}\Jj{2}\Jj{3}}&=&
\sqrt{\dfrac{k+2}{2(k+1)}}\Tg{0B,k(k+1)}{d}{\Jj{1}\Jj{2}\Jj{3}},
\nonumber\\
\Tg{1B,k(k+1)}{d}{J\Jj{1}\Jj{2}\Jj{3}\Jj{4}\Jj{5}}&=&
\sqrt{\dfrac{k+2}{2(k+1)}}\Tg{0B,k(k+1)}{d}{\Jj{1}\Jj{2}\Jj{3}\Jj{4}\Jj{5}}
\quad
\eea
that link the quantities with subscripts $0B$ and $1B$.

\end{document}